\documentclass[rmp, aps, twocolumn, floatfix]{revtex4-1}
\usepackage{amssymb,amsmath,array,delarray,amsfonts,graphicx}

\newcommand{\eop}{\hspace*{\fill}{\footnotesize $\blacksquare$}}

\newcommand{\bt}{\begin{theorem}}
\newcommand{\et}{\end{theorem}}
\newcommand{\bc}{\begin{corollary}}
\newcommand{\bl}{\begin{lemma}}
\newcommand{\ec}{\end{corollary}}
\newcommand{\el}{\end{lemma}}
\newcommand{\bo}{\begin{observation}}
\newcommand{\eo}{\end{observation}}
\newcommand{\bp}{\begin{proposition}}
\newcommand{\ep}{\end{proposition}}
\newcommand{\br}{\begin{remark}}
\newcommand{\er}{\end{remark}}
\newcommand{\brt}{\begin{result}}
\newcommand{\ert}{\end{result}}
\newcommand{\bq}{\begin{question}}
\newcommand{\eq}{\end{question}}

\newtheorem{theorem}{Theorem}[section]
\newtheorem{observation}[theorem]{Observation}
\newtheorem{corollary}[theorem]{Corollary}
\newtheorem{lemma}[theorem]{Lemma}
\newtheorem{proposition}[theorem]{Proposition}
\newtheorem{remark}[theorem]{Remark}
\newtheorem{result}[theorem]{Result}
\newtheorem{question}[theorem]{\textsc{Question}}

\newcommand{\zv}{\ensuremath{\mathbf{0}}}

\newcommand{\PGL}{\ensuremath{\mathbf{PGL}}}
\newcommand{\PSL}{\ensuremath{\mathbf{PSL}}}

\newcommand{\PG}{\ensuremath{\mathbf{PG}}}

\newcommand{\id}{\ensuremath{\mathbf{id}}}
\newcommand{\1}{\ensuremath{\mathbf{1}}}

\newcommand{\M}{\ensuremath{\mathfrak{M}}}
\newcommand{\mL}{\ensuremath{\mathbf{L}}}

\newcommand{\A}{\ensuremath{\mathbf{A}}}
\newcommand{\K}{\ensuremath{\mathbb{K}}}
\newcommand{\hA}{\mathbf{A}}
\newcommand{\spec}{\ensuremath{\mathrm{spec}}}
\newcommand{\B}{\ensuremath{\mathbf{B}}}
\newcommand{\II}{\ensuremath{\mathbb{I}}}
\newcommand{\JJ}{\ensuremath{\mathbb{J}}}
\newcommand{\mP}{\ensuremath{\mathcal{P}}}
\newcommand{\mB}{\ensuremath{\mathcal{B}}}
\newcommand{\I}{\ensuremath{\mathbf{I}}}

\newcommand{\vv}{\mathfrak{v}}

\newcommand{\cA}{\mathcal{A}} %blocking sets
\newcommand{\cC}{\mathcal{C}}
\newcommand{\cD}{\mathcal{D}}

\newcommand{\cK}{\mathcal{K}} %arcs, partial ovoids
\newcommand{\cL}{\mathcal{L}}
\newcommand{\cN}{\mathcal{N}}
\newcommand{\cNo}{\overline{\mathcal{N}}}
\newcommand{\cT}{\mathcal{T}}

\newcommand{\cV}{\mathcal{V}}

\def\im{{\mathcal I}m\ }

\def\scl{\mathrm{s.c.}}
\def\cc{\textrm{c.c.}}

\def\bfa{{\bf a}}
\def\bfq{{\bf q}}

\begin{document}

\author{Olivier Giraud}
\email{olivier.giraud@lptms.u-psud.fr}
\affiliation{\mbox{Univ. Paris-Sud, CNRS, LPTMS, UMR 8626, Orsay, F-91405, France}}
\affiliation{\mbox{Laboratoire de Physique Th\'{e}orique (IRSAMC), CNRS and UPS,}}
\affiliation{\mbox{Universit\'{e} Paul Sabatier, F-31062 Toulouse, France}}
\author{Koen Thas}
\email{kthas@cage.UGent.be}
\affiliation{\mbox{Ghent University, Department of Pure Mathematics
and Computer Algebra,}}
\affiliation{\mbox{Krijgslaan 281, S25, B-9000 Ghent, Belgium}}

\title{Hearing  shapes of drums | mathematical and physical aspects of isospectrality}
\date{August 6, 2010}
%\date{\today}
\begin{abstract}
In a celebrated paper ``Can one hear the shape of a drum?'' M. Kac [{\em
  Amer. Math. Monthly}  {\bf 73}, 1  (1966)] asked his
  famous question about the existence of nonisometric billiards having
  the same spectrum of the Laplacian. This question was eventually
answered positively in 1992 by the construction of noncongruent planar isospectral pairs.
This review highlights mathematical and physical aspects of isospectrality.
\end{abstract}

\maketitle

\tableofcontents 

\vspace{2cm}
%%%%%%%%%%%%%%%%%%%%%%%%%%%%%%%%%%%%%%%%%%%%%%%%%%%%%%%%%%%%%%%%%%%%%%%%%%%%%%
\section{Introduction}
%%%%%%%%%%%%%%%%%%%%%%%%%%%%%%%%%%%%%%%%%%%%%%%%%%%%%%%%%%%%%%%%%%%%%%%%%%%%%%

Elastic plates are probably some of the oldest supports of sound
production. They were used by most human cultures. Clay drums dated from
the Chalcolithic have been found in graves in central Europe, and bronze drums dated from
the second millenary B.C. have been discovered in Sweden and Hungary. However, it is usually
acknowledged that the scientific study of the vibration of elastic plates goes
back only to the end of the 18th century, when the German researcher Ernst
Chladni carried out the first systematic investigations on the
production of sound by plates \cite{Chl, SmiSto}. When the plate was fixed in its middle and
struck with a bow, it was set into vibration. The mode that
was being excited was physically visualized by pouring sand on the
plate: the sand accumulates at nodal lines, that is lines along which
the plate does not oscillate. Some insight was brought into the
mathematical theory of vibrating plates by the French mathematician 
Sophie Germain, who published  {\it Recherches sur la th\'eorie des surfaces
  \'elastiques} in 1821. In the course of the 19th century,
Poisson, Kirchhoff, Lam\'e, Mathieu, and Clebsch, devised analytic
expressions for the description of the oscillation for elementary shapes such as
the rectangle, the triangle, the circle, and the ellipse. 

The motivation
for studying this problem was mainly that the wave phenomenon at the heart of
membrane oscillations is in fact quite general. The
stationary wave equation describing the problem arises in a variety of
situations. In many fields of physics, such as acoustics, seismology, hydrodynamics,
and heat propagation, the mathematical formulation of the problem
involves partial differential equations, and general solutions of
these equations can be found as superpositions of solutions of the
so-called Helmholtz equation. In a $d$-dimensional space a stationary
solution to the wave equation is an unknown function of $d$ variables
describing the problem, and the Helmholtz equation reads
\begin{equation}
\label{helmholtz}
\Delta f+E f=0,
\end{equation}
where $\Delta$ is the $d$-dimensional Laplacian. Under suitable approximations,
numerous problems can be cast in that form. For instance, in a certain
regime the oscillations of the height $f=f(x,y)$ of a thin vibrating plate 
at point $(x,y)$ can be described by \eqref{helmholtz}.

At the end of the 19th century,
James C. Maxwell showed that the electric and the magnetic field
behave like waves and established equations governing the time
evolution of the electromagnetic field. From Maxwell's equations it is
easy to prove that the electric and the magnetic field components also
obey the same wave equation \eqref{helmholtz}. Further interest developed in this equation
when the wave-like behavior of matter was discovered in the early
years of quantum mechanics. The Schr\"odinger equation was established in 1926
by Erwin Schr\"odinger to describe the spacetime evolution of a
quantum system. The behavior of a particle can be described, in the 
framework of quantum mechanics, by a wave function $\psi$, which is a
function of the position of the particle, and which characterizes the 
probability amplitude $\psi(x)$ that the particle be located at a 
position $x$. If the system is described by the Hamiltonian $H$, the
wave function satisfies the stationary Schr\"odinger equation
$H\Psi=E\Psi$, where $E$ is the energy of the particle.
For a particle of mass $m$ and momentum $p$ evolving in a box defined by its
contour $\partial B$, the Hamiltonian describing the free
motion inside the box reads $H=p^2/2m$ inside the box enclosure 
$\partial B$ and $\infty$ outside, and the time-independent Schr\"odinger
equation takes the form \eqref{helmholtz}.\\

Mathematically, solutions of the Helmholtz equation are readily obtained
in dimension $d=1$. The problem of vibrating strings had been solved
in the 18th century by Jean Le Rond d'Alembert. For a string of
length $L$ fixed at its two ends, solutions are simply given by
$f(x)=\sin(n\pi x/L)$, where $n$ is an integer. The sound produced by
the string has the possible frequencies $n \nu_0$, with the
fundamental frequency given by $\nu_0=c/(2L)$.

Just as the one-dimensional case | which can describe a variety of
physical situations | can be seen as a problem of
vibrating strings, the two-dimensional case is usually studied from
the perspective of 
its simplest mathematical equivalent, namely billiards.
Billiards (in the mathematical sense) are two-dimensional compact
domains of the Euclidean plane $\mathbb{R}^2$. 
For instance, in quantum mechanics, the billiard models the
behavior of a particle moving freely in a box  whose dimensions are
such that it can be approximated by a two-dimensional
enclosure. 
The billiard problem is solved by looking for eigenfunctions $\psi$ and
eigenvalues $E$ that are solutions of Eq.~\eqref{helmholtz}
inside the billiard, imposing boundary conditions on the boundary
$\partial B$ of the billiard. Physical problems impose specific
boundary conditions. For instance hard wall domains in quantum
mechanics impose that the wave function vanishes on the
boundary. In acoustics, clamping an elastic membrane imposes that the
oscillations and their derivative along the boundary vanish. The
billiard problem usually considers the two following boundary conditions: 
Dirichlet boundary conditions $\psi_{|\partial B}=0$, for which the function vanishes on the boundary, or Neumann
boundary conditions $\partial_{\bf n}\psi_{|\partial B}=0$, for which the normal derivative vanishes on the boundary. 
If such boundary conditions are imposed there is an infinite but countable
number of solutions to Eq.~\eqref{helmholtz}. We denote eigenfunctions
of the operator $-\Delta$ by $\psi_n$ and eigenvalues by $E_n$, $n\in\mathbb{N}$, 
with $0<E_1\leq E_2\leq E_3\cdots$.
Of course any combination of the above boundary conditions yields a
different spectral problem. In this review however, we will be mainly
concerned with Dirichlet boundary conditions.

In the second half of the
20th century, quantum billiards were
studied in the framework of quantum chaos. Quantum properties of classical systems were investigated, and different behaviors were found according to the properties of integrability or chaoticity of the underlying classical dynamics. This quantum-classical correspondence led to various conjectures
for integrable systems \cite{BerTab77a} and chaotic systems \cite{BohGiaSch84}.
These conjectures rest on  powerful mathematical tools that allow insight into the properties of solutions of
the Helmholtz equation \eqref{helmholtz}. 
For instance, the Weyl formula (see section \ref{meandensity}), or 
semiclassical trace formulas (see section \ref{scgreen}), provide a
connection between the density of energy levels and classical features
of the domains such as area, perimeter or properties of classical
trajectories in the domain.
The existence of such formulas and the conjectures on the
quantum-classical correspondence indicate that the spectrum of a
billiard contains a certain amount of information about the shape of the billiard. Therefore
it is natural to ask how much information about the billiard
can be retrieved from knowledge of the eigenvalue spectrum.
For rectangular or triangular billiards, it is known that a
finite number of eigenvalues suffices to entirely specify the shape of
the billiard (see e.g. \cite{ChaDet}), but is this true for more
complicated shapes?\\

In 1966, in a celebrated paper \cite{Kac}, Mark Kac
formulated the famous question ``Can one hear the shape of a
drum?''. This provocative question is of course to be understood
mathematically as follows: Is it possible to find two (or more)
non-isometric Euclidean simply connected domains for which
the sets $\{ E_n \parallel n \in\mathbb{N}\}$ of solutions of 
(\ref{helmholtz}) with $\Psi_{\vert \mbox{Boundary}} = 0$ are
identical? More broadly, the question raises the issue of the inverse problem of retrieving information about a drum from knowledge of its spectral properties. As the spectroscopist A.~Schuster put it in an 1882 report to the British Association for the Advancement of Science: ''To find out the different tunes sent out by a vibrating system is a problem which may or may not be solvable in certain special cases, but it would baffle the most skillful mathematicians to solve the inverse problem and to find out the shape of a bell by means of the sounds which it is capable of sending out. And  
this is the problem which ultimately spectroscopy hopes to solve in the case of light. In the meantime we must welcome 
with delight even the smallest step in the desired direction.'' \cite{MehRec00}.
Actually, it was known very early, from
Weyl's formula, that one can ``hear'' the area of a drum and the
length of its perimeter (see section \ref{meandensity}, and \cite{VaaKocBlu} for a historical account of the problem). But could the
shape itself be retrieved from the spectrum?  That is, what kind of information on the geometry is it possible to gather from the knowledge of the spectrum, for instance, using semiclassical methods that allow investigation of the quantum-classical correspondence? And what kind of sufficient conditions allow the geometry to be entirely specified from the spectrum?

Formally, an answer ``no'' to Kac's question amounts to finding
{\em isospectral billiards}, that is non-isometric billiards having
exactly the same eigenvalue spectrum. 
Since the appearance of Kac's paper \cite{Kac}, far more than 500
papers have been written on the subject, and innumerable variations on
``hearing the shape of something'' can be found in the literature. 
Early examples of flat tori sharing the same
eigenvalue spectrum were found in 1964 by Milnor in $\mathbb{R}^{16}$ from
nonisometric lattices of rank $16$ in $\mathbb{R}^{16}$ (see section
\ref{milnor}).  Other examples of isospectral Riemannian manifolds
were constructed later, for example on lens spaces \cite{Ike} or on surfaces
with constant negative curvature \cite{Vig}. In 1982, H. Urakawa produced the
first examples of isospectral domains in $\mathbb{R}^{n}$, $n\geq 4$ 
\cite{Ura}.
(These examples are also described by \cite{Pro}.)
 More specifically, it is proved that there exist domains $C$ 
and $C'$ in the unit sphere $\mathbb{S}^{n - 1}$
in $\mathbb{R}^n$, $n \geq 4$, which are Dirichlet and Neumann 
isospectral but not congruent in $\mathbb{S}^{n - 1}$.
This existence follows from the observation that there are finite 
reflection groups $W$ and $W'$ that act on
the same Euclidean space $\mathbb{R}^n$, $n \geq 4$, for which the 
sets of exponents coincide, and the intersections
($C$ and $C'$) of their chambers with $\mathbb{S}^{n - 1}$ are not 
congruent in $\mathbb{S}^{n - 1}$. Then work
of \textcite{BerBes} is applied.\\
In the late 1980s, various other papers appeared, giving necessary conditions that any family of
billiards sharing the same spectrum should satisfy (\cite{Mel},
\cite{OsgPhiSar}, \cite{OsgPhiSar2}), and necessary conditions given as
inequalities on the eigenvalues were reviewed in \cite{Pro}.\\

\begin{figure}[ht]
\begin{center}
\includegraphics[width=0.8\linewidth]{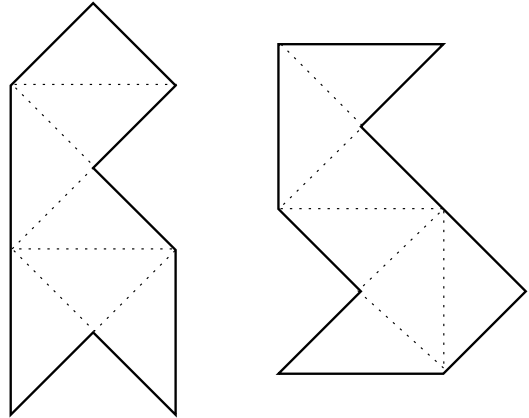}
\end{center}
\caption{Paradigmatic pair of isospectral billiards with seven
  half-square shaped base tiles. The dotted lines are just for the eye.
\label{celebrated}}
\end{figure}

But it was almost 30 years after Kac's paper that the first example of
two-dimensional billiards having exactly the same spectrum was finally
exhibited in 1992. The pair was found by C.~Gordon, D.~ Webb and S.~ Wolpert in their paper
``Isospectral plane domains and surfaces via Riemannian orbifolds''
\cite{GorWebWol}. They gave a no as a final answer to Kac's
question, and as a reply to Kac's paper, they published a paper titled ``One cannot hear the shape of a
drum'' \cite{GorWebWol2}. The most popularized example is shown in Fig.~\ref{celebrated}.
Crucial for finding the example was a theorem by Sunada
(see section \ref{sunada}) asserting that when two subgroups are
``almost conjugate'' in a group that acts by isometries on a Riemannian manifold, the quotient
manifolds are isospectral. In fact, the other examples which were 
constructed after 1992 all used Sunada's method. 
Later, the so-called transplantation technique was used, giving
an easier way for detecting isospectrality of planar billiards. Still,
essentially only 17 families of examples that say no to Kac's
question were constructed in a 40 year period. \\

Since the literature on isospectrality is large, and covers a
broad spectrum of mathematical topics, we have chosen here to
put the focus on isospectral billiards, that is, two-dimensional
isospectral domains of the Euclidean plane, with Dirichlet boundary
conditions. It is worth noting that simple examples of isospectral domains
can be constructed in the case of mixed Dirichlet-Neumann boundary conditions. 
Such constructions were proposed by \cite{LevParPol} (see section
\ref{levitin}). We now review some results on related topics, to
which we will not return in this paper.

First we mention several fundamental results on isospectrality 
that will be omitted. 
\textcite{Zeld98} proved that isospectral simple analytic 
surfaces of revolution are isometric.
That is, he considered the moduli space $\mathcal{R}$ of metrics of 
revolution $(\mathbb{S}^2,g)$ with the 
following properties. Suppose that there is an effective action of 
$\mathbb{S}^1$ by isometries of $(\mathbb{S}^2,g)$.
The two fixed points are $N$ and $S$. Denote by $(r,\theta)$ geodesic 
polar coordinates centered at $N$, with $\theta = 0$
being some fixed meridian $\gamma_M$ from $N$ to $S$. The metric 
$g$ can then be written as $g = dr^2 + a(r)d\theta^2$,
where $a: [0,L] \mapsto \mathbb{R}^+$ is defined by 
$a(r) = \vert S_r(N)\vert/(2\pi)$, with $\vert S_r(N)\vert$ 
the length of the distance circle of radius $r$ centered at $N$.              
The properties now are as follows: (i) $g$ is real analytic, 
(ii) $a$ has precisely one critical point $r_0 \in ]0,L[$, with $a''(r_0) < 0$, corresponding
to an equatorial geodesic $\gamma_E$, and (iii) the nonlinear Poincar\'{e} map $\mathcal{P}_{\gamma_E}$ for $\gamma_E$ is of twist type.

Denote by $\mathcal{R}^* \subset \mathcal{R}$ the subset of metrics with simple length spectra in the sense 
of \cite{Zeld98}.
Then Zelditch proved that $\mathrm{Spec}: \mathcal{R}^* 
\mapsto \mathbb{R}^{\mathbb{N}}_{+}$ is 1-1.
Furthermore, in \cite{Zeld99} | see also \cite{Zeld00}, Zelditch showed that real plane domains $\Omega$ that (1) are simply connected and real analytic,
(2) are $\mathbb{Z}_2\times\mathbb{Z}_2$-symmetric (i.e., have the symmetry of an ellipse), and (3) have at least one 
axis that is a nondegenerate bouncing ball orbit, the length of which has multiplicity 1 in the length spectrum $Lsp(\Omega)$, are 
indeed determined by their spectrum. 
In recent work \textcite{Zeld04} pursued his goal of eventually solving the inverse spectral
problem for general  real analytic plane domains. We will return to this issue in more detail in section \ref{analyticdomains}.\\

Concerning the known counterexamples in the plane, it should be remarked
that the constructed domains are not convex (see e.g.~Appendix \ref{gallery}).    
The objective of \textcite{GorWeb} is to exhibit pairs of convex domains in the
hyperbolic plane $\mathbb{H}^2$ that are  
both Dirichlet and Neumann isospectral. They are obtained from nonconvex
examples in the real plane by modifying the shape of a fundamental tile.
Other interesting variations on the problem include the construction of
a pair of isospectral (nonisometric) compact three-manifolds, called ``Tetra'' and
``Didi'', which have different closed geodesics \cite{DoyRoss04}.\\

The related question of graph isospectrality has also attracted much
interest. We mention here a few results.
A {\em quantum graph} is a metric graph equipped with a differential
operator (typically the negative
Laplacian) and homogeneous differential boundary conditions at the vertices.
(Recall that a {\em metric graph} is a graph such that to each edge $e$ is
assigned a finite (strictly positive) length $\ell_e \in \mathbb{R}$, so
that it can be identified with the closed interval $[0,\ell_e] \subset
\mathbb{R}$. Without the boundary conditions, the graph ``consists of''
edges with functions defined separately on each edge.) So there is a
natural spectral theory associated with quantum graphs. Many results exist,
and we just mention a few striking ones.
One of the main results in that spectral theory can be found in
\cite{GutkSmil}, where
the trace formula is used to show that (under certain conditions) a
quantum graph can be recovered from the
spectrum of its Laplacian. (Necessary conditions include the graph being
simple and the edges having rationally independent lengths.) Using a
spectral trace formula, \textcite{Roth} in an early paper constructed 
isospectral quantum graphs. \textcite{vonBel}, on
the other hand, used the connection between spectra of discrete graphs and
spectra of (equilateral) quantum graphs to transform isospectral discrete
graphs into isospectral quantum graphs. Finally, we note that \textcite{ParzBand}
presented a method for
constructing isospectral quantum graphs, based on linear
representations of finite groups. 
Note that a different notion of graph isospectrality was considered by
\textcite{KT3} based on the spectrum of the adjacency matrix of the graph. We
return to this point in section \ref{isograph}.\\

To end this section, we give a short description of the contents of
the paper.

To familiarize the reader with the notions involved, we start by
presenting a simple proof of isospectrality for the seminal example of
\textcite{GorWebWol} in section
\ref{pedestrian}. Then the first historical examples of higher-dimensional
isospectral pairs of flat tori are constructed (section
\ref{milnor}). (Much more work has been done on isospectrality for the Laplace-Beltrami operator on flat tori in higher dimensions than just the material we cover in section II. We refer to that section for more commentaries on that matter.)
Section \ref{transplantation} is devoted to the
mathematical aspects lying behind the construction of the known examples of
isospectral pairs. Then we review various
aspects of the properties of isospectral pairs (section
\ref{properties}), as well as experimental implementations and numerical
checks of isospectrality (section \ref{expandnum}). As the first examples
  of isospectral billiards were produced by applying Sunada theory, a
  review of this theory is given in section \ref{sunadasection}.
In the last section we examine questions related to Kac's problem.

%\newpage
%%%%%%%%%%%%%%%%%%%%%%%%%%%%%%%%%%%%%%%%%%%%%%%%%%%%%%%%%%%%%%%%%%%%%%%%%%%%%%
\section{A Pedestrian Proof of Isospectrality}
%%%%%%%%%%%%%%%%%%%%%%%%%%%%%%%%%%%%%%%%%%%%%%%%%%%%%%%%%%%%%%%%%%%%%%%%%%%%%%
\label{pedestrian}
The first examples of isospectral billiards in the Euclidean plane
were constructed using powerful mathematical tools. We 
postpone these historical constructions to section \ref{ST}. 
The present section aims at illustrating the main ideas involved in
isospectrality, so that the reader can acquire some intuition about it.
More rigorous mathematical grounds will be provided in the next sections.

\subsection{Paper-folding proof}
\label{paperfoldingproof}
We 
start with a simple construction method that was proposed by \textcite{Cha}. 
It is based on the so-called ''paper-folding'' method. To illustrate it we follow
\cite{Tha}, where the method is illustrated on a simple example.\\
\begin{figure}[ht]
\begin{center}
\includegraphics[width=0.98\linewidth]{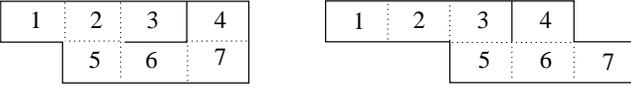}
\end{center}
\caption{The pair $7_3$ (see Appendix \ref{gallery})
of isospectral billiards with a rectangular base shape.
\label{squarebase}}
\end{figure}
Consider the two billiards in Fig.~\ref{squarebase}. Each billiard is made of
seven identical rectangular building blocks. The solid lines are hard
wall boundaries, the dotted lines are just a guide to the eye marking the building blocks. 
Let  $\phi$ be
an  eigenfunction of the left billiard with eigenvalue $E$.
The goal is to construct 
an eigenfunction of the right billiard with the same eigenvalue, that
is a function which:
\begin{itemize}
\item verifies the Helmholtz equation \eqref{helmholtz};
\item vanishes on the boundary of the billiard;
\item has a continuous normal derivative inside the billiard.
\end{itemize}
The idea is to define a function $\psi$ on the right billiard as a superposition of
translations of the function $\phi$.
Since the Helmholtz equation \eqref{helmholtz} satisfied by $\phi$ is linear, any linear combination
of translations of $\phi$ will be a solution of the Helmholtz equation 
with the same eigenvalue $E$ in the interior of each building block of the second billiard. 
The problem reduces to finding a linear combination that vanishes on the boundary
and has the correct continuity properties inside the billiard. The paper-folding method 
allows to satisfy all these conditions simultaneously.
\begin{center}
\begin{figure*}[ht]
\includegraphics[width=0.8\linewidth]{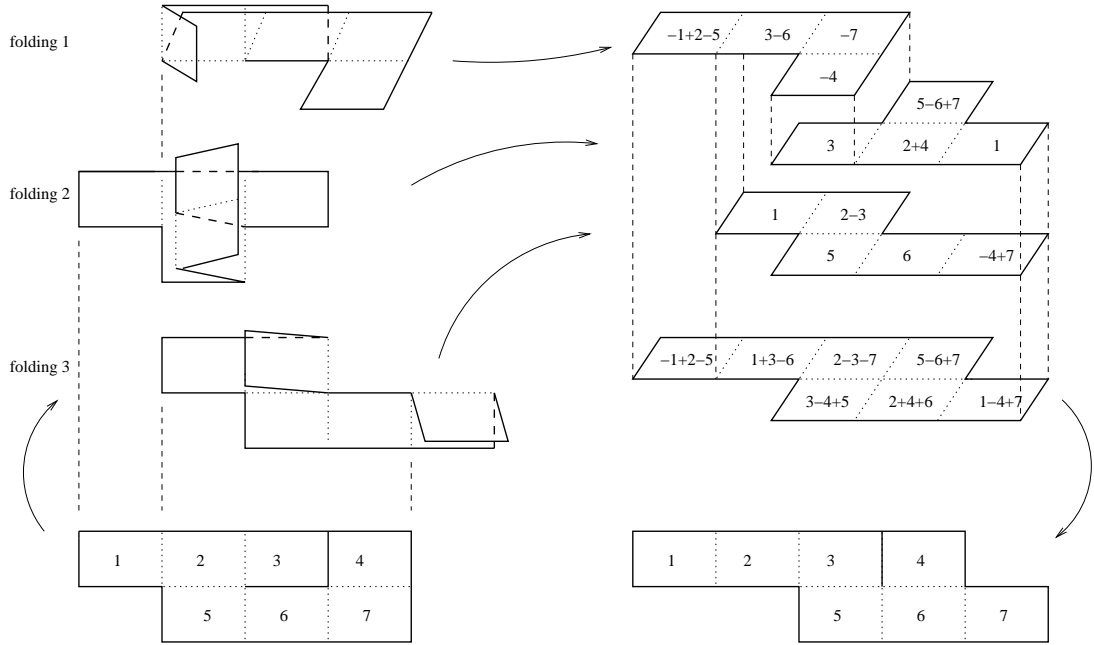}
\caption{Pictorial representation of the paper-folding method.\label{paperfolding}}
\end{figure*}
\end{center}
Take three copies of the left billiard of 
Fig.~\ref{squarebase}. Fold each copy in a different way, as shown
in Fig.~\ref{paperfolding} (left column). Then the three-times folded billiards are
stacked on top of each other as indicated in the right column of 
Fig.~\ref{paperfolding}; note that the first shape (folding 1) has been translated on
the left before being stacked, and that the second shape (folding 2) 
has been rotated by $\pi$ in the plane of the figure. Once superposed, these three billiard yield
the shape on the bottom right, which is the right billiard of Fig.~\ref{squarebase}. 

Now we make a correspondence between stacking two sheets of paper and adding the
functions defined on these sheets; moreover, stacking the
reverse of a sheet corresponds to assigning a minus sign to the function. For 
instance, in folding $3$, a minus sign is associated in the right column with tiles 3 
and 4, since they are folded back, and a plus sign is assigned to the
other tiles since they are not folded.
The function $\psi$ is defined by this ``folding and stacking'' procedure. For instance
it is defined in the tile numbered 1 in the right billiard of Fig.~\ref{squarebase} by
\begin{equation}
\label{exvanish}
\psi|_{\textrm{tile } 1}=-\phi|_{\textrm{tile } 1}
+\phi|_{\textrm{tile } 2}-\phi|_{\textrm{tile } 5}.
\end{equation}
The procedure above ensures that $\psi$ vanishes on the boundary and has a continuous
derivative across the tile boundaries. Indeed, consider for instance the leftmost vertical boundary
of the right billiard (i.e. the left edge of tile 1). On this boundary
we have $\phi|_{\textrm{tile } 5}=0$  (since it is
at the boundary of the left billiard), and $\phi|_{\textrm{tile } 1}=\phi|_{\textrm{tile } 2}$
since tiles 1 and 2 are glued together. Thus, $\psi$ given by
Eq.~\eqref{exvanish} indeed vanishes on the leftmost vertical boundary
of the right billiard. After we have checked by inspection all (inner
and outer) boundaries, we have proved that the two billiards of
Fig.~\ref{squarebase} are isospectral.\\

\begin{figure}[b]
\begin{center}
\includegraphics[width=0.96\linewidth]{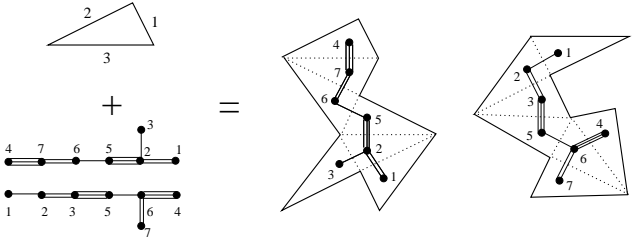}
\end{center}
\caption{Graphs corresponding to a pair of isospectral billiards: 
If we label the sides of the triangle by $\mu=1,2,3$, the unfolding rule
by symmetry with respect to side $\mu$ can be represented by edges made 
of $\mu$ braids in the graph. {F}rom a given pair of graphs, one can construct
infinitely many pairs of isospectral billiards by applying the unfolding
rules to any shape.\label{graphs_review}}
\end{figure}

With the paper-folding method, 
it is clear that what matters is the way the building blocks (the elementary
rectangles in our example) are glued to each other, irrespective of
their shape. We now show how the paper-folding proof generalizes to
other shapes. 
Suppose we denote by 1, 2, and 3 respectively the left, right, and bottom edge of 
tile 4 in the left billiard of Fig.~\ref{squarebase}. To obtain the whole
billiard one unfolds tile 4 with respect to its side number 3, getting tile 7. Then
tile 7 is unfolded with respect to its side number 2, yielding tile 6, and so on.
The unfolding rules can be summed up in a graph specifying the way we unfold the 
building block. The graphs in Fig.\ref{graphs_review} correspond to
the unfoldings yielding the billiards of Fig.~\ref{squarebase} when
applied to a rectangular building block.
The vertices of the graph represent the building blocks,
and the edges of the graph are ``colored'' according to the unfolding rule, that is,
depending on which of its sides the building block is unfolded. The graphs can
alternatively be encoded by permutations $a^{(\mu)}, b^{(\mu)}$, $1\leq\mu\leq 3$.
For instance for the first graph we have $a^{(1)}=(2 3)(5 6)$, $a^{(2)}=(1 2)(6 7)$, and
$a^{(3)}=(2 5)(4 7)$.
In fact, only three sides of the rectangle are involved in the unfolding.
So we can start with any triangular-shaped building block, and unfold it with respect to
its sides just as the billiards in Fig.~\ref{squarebase} are obtained
from the rectangular building block. This leads to billiard pairs whose isospectrality
is granted by the paper-folding proof given above.
For example, starting from the triangle in Fig.~\ref{graphs_review}
and following the same 
unfolding rules, we get the pair of isospectral billiards shown in Fig.~\ref{graphs_review} 
right. Taking a building block in the form of a
half-square, we recover the example of
Fig.~\ref{celebrated} when the same unfolding rules are applied.
\begin{figure}[hbt!]
\begin{center}
\includegraphics[width=0.96\linewidth]{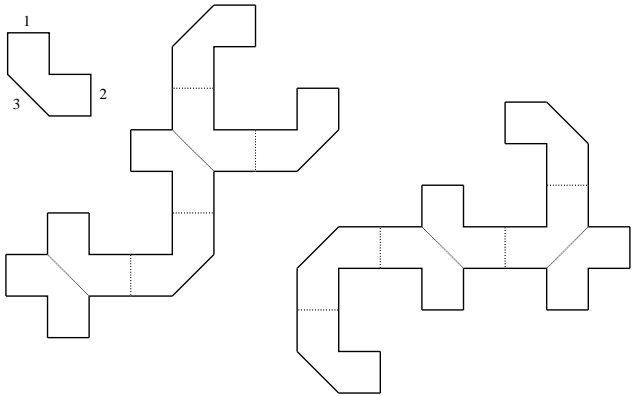}
\end{center}
\caption{Isospectral billiards. The top left figure is the seven-edged building block. From \cite{GorWebWol}.\label{gordonwebwolpert}}
\end{figure}

The  building block is in fact not even required to be a triangle or a rectangle.
Any building block possessing three edges around which to 
unfold leads to a different pair of isospectral billiards. 
Another interesting example is obtained by taking a heptagon and unfolding it
with respect to three of its sides, following the unfolding rules
of Fig.~\ref{graphs_review}. This yields the first example produced
by~\textcite{GorWebWol, GorWebWol2} (see  Fig.~\ref{gordonwebwolpert}).

\begin{figure}[h]
\begin{center}
\includegraphics[width=0.98\linewidth]{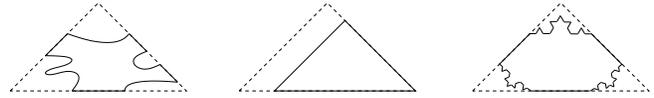}
\end{center}
\caption{Examples of building blocks yielding isospectral pairs.\label{buildingblocks}}
\end{figure}

\textcite{Cha} produced more involved examples, following the same procedure.
Starting from the building block of Fig.~\ref{buildingblocks} left, one obtains
an example of a pair of chaotic billiards with holes. Similarly 
\textcite{DhaMadUdaSri} constructed chaotic isospectral billiards 
based on the same idea: scattering circular disks were added inside the
base triangular shape in a way consistent with the unfolding.

The central building block of Fig.~\ref{buildingblocks}
yields a simple disconnected pair where each billiard consists
of a disjoint rectangle and triangle. In this case, isospectrality can be checked 
directly by calculating the eigenvalues, since the eigenvalue problem can be solved exactly
for triangles and half-squares. 

\textcite{SleHua} considered a
building block with piecewise fractal boundary: starting
from a $(\pi/2, \pi/3, \pi/6)$ base triangle they cut each side into three
pieces and remove the three triangular corners. Along the freshly made
cuts a Koch curve is constructed, while the untouched sides still allow
the Chapman unfolding (Fig.~\ref{buildingblocks} right). This yields a pair of isospectral billiards with fractal
boundary of dimension $\ln 4/\ln 3$.\\

A separate problem that will not be presented here is to find
inhomogeneous vibrating membranes isospectral to a homogeneous
membrane with the same shape (see, e.g., \cite{Got} for circular
membranes). \textcite{KnoMcC} used the isospectrality of the billiards of
Fig.~\ref{celebrated} to construct a pair of  isospectral
 circular membranes by a conformal mapping.

\subsection{Transplantation proof}
\label{transplantationproof}
The paper-folding proof can be made more formal be means of
the so-called ``transplantation'' method. This
method was introduced in \textcite{Bera, Berard, Berard2}, and
discussed by \textcite{BusConDoySem} and \textcite{OkaShu}.
It will be presented in more detail in section 
\ref{transplantation}. Here we sketch the main ideas using a simple
example.\\

Consider the isospectral pair of Fig.~\ref{squarebase}. Let $\phi$ be an eigenstate of the
first billiard. Any point in the billiard can be specified by its coordinates
$\bfa=(x,y)$ inside a building block, and a number $i$ arbitrarily associated with
the building block (for example $1\leq i\leq 7$ in our example of Fig.~\ref{squarebase}). 
Thus $\phi$ is a function of the variable $(\bfa,i)$.
According to the paper-folding proof, a building block $i$ of the second billiard
is constructed from a superposition of three building blocks $j$ obtained by folding the first
billiard. We can code the result of the folding-and-stacking procedure in a matrix $T$, as
\begin{equation}
\label{matriceT}
T=\left(
\begin{array}{ccccccc} -1&1&0&0&1&0&0\cr 1&0&1&0&0&-1&0\cr 0&1&-1&0&0&0&-1\cr
0&0&0&0&1&-1&1\cr 0&0&1&-1&1&0&0\cr 0&1&0&1&0&1&0\cr 1&0&0&-1&0&0&1
\cr\end{array}\right).
\end{equation}
The paper-folding proof consists in showing that one can construct 
an eigenstate $\psi$ of the second billiard as
\begin{equation}
\label{psiphi}
\psi(\bfa, i)=\cN\sum_j T_{i j}\phi(\bfa,j),
\end{equation}
where $\cN$ is some normalization factor. That is, one can ''transplant''
the eigenfunction of the first billiard to the second one. The matrix
$T$ is called a ``transplantation matrix''. The proof of
isospectrality reduces to checking that $\psi$ given by 
\eqref{matriceT}-\eqref{psiphi} vanishes on the 
boundary and has a continuous derivative inside the billiard.\\

Let us first transform the
problem into an equivalent one on translation surfaces. Translation surfaces 
\cite{GutJud00}, also called planar structures, are manifolds of
zero curvature with a finite number of singular points
(see \cite{Vor96} for a more rigorous mathematical definition).
A construction by \textcite{ZemKat76} allows to construct 
a planar structure on rational polygonal billiards, that is polygonal billiards
whose angles at the vertices are of the form  $\alpha_i=\pi m_i/n_i$, 
with $m_i,n_i$ positive integers. 
This planar structure is obtained by ``unfolding'' the polygon, that is by 
gluing to the initial polygon its images obtained by mirror reflection with 
respect to each of its sides, and repeating this process on the images. 
For polygons with angles $\alpha_i=\pi m_i/n_i$,  this process terminates
and $2n$ copies of the
initial polygon are required, where $n$ is the gcd of the $n_i$. Identifying 
parallel sides, one gets a planar structure of genus in general greater than 1.
This structure has singular points corresponding to vertices of the initial 
polygon where the angle $\alpha_i=\pi m_i/n_i$ is such that $m_i\neq 1$.
The genus of the translation surface thus obtained is given by \cite{BerRic81}
\begin{equation}
g=1+\frac{n}{2}\sum_i\frac{m_i-1}{n_i}.
\end{equation}
A very simple example of a translation surface is the flat torus,
obtained by identifying the opposite sides of a square. Such a
translation surface corresponds to four copies of a square billiard
glued together.

\begin{figure}[ht]
\begin{center}
\includegraphics[width=0.98\linewidth]{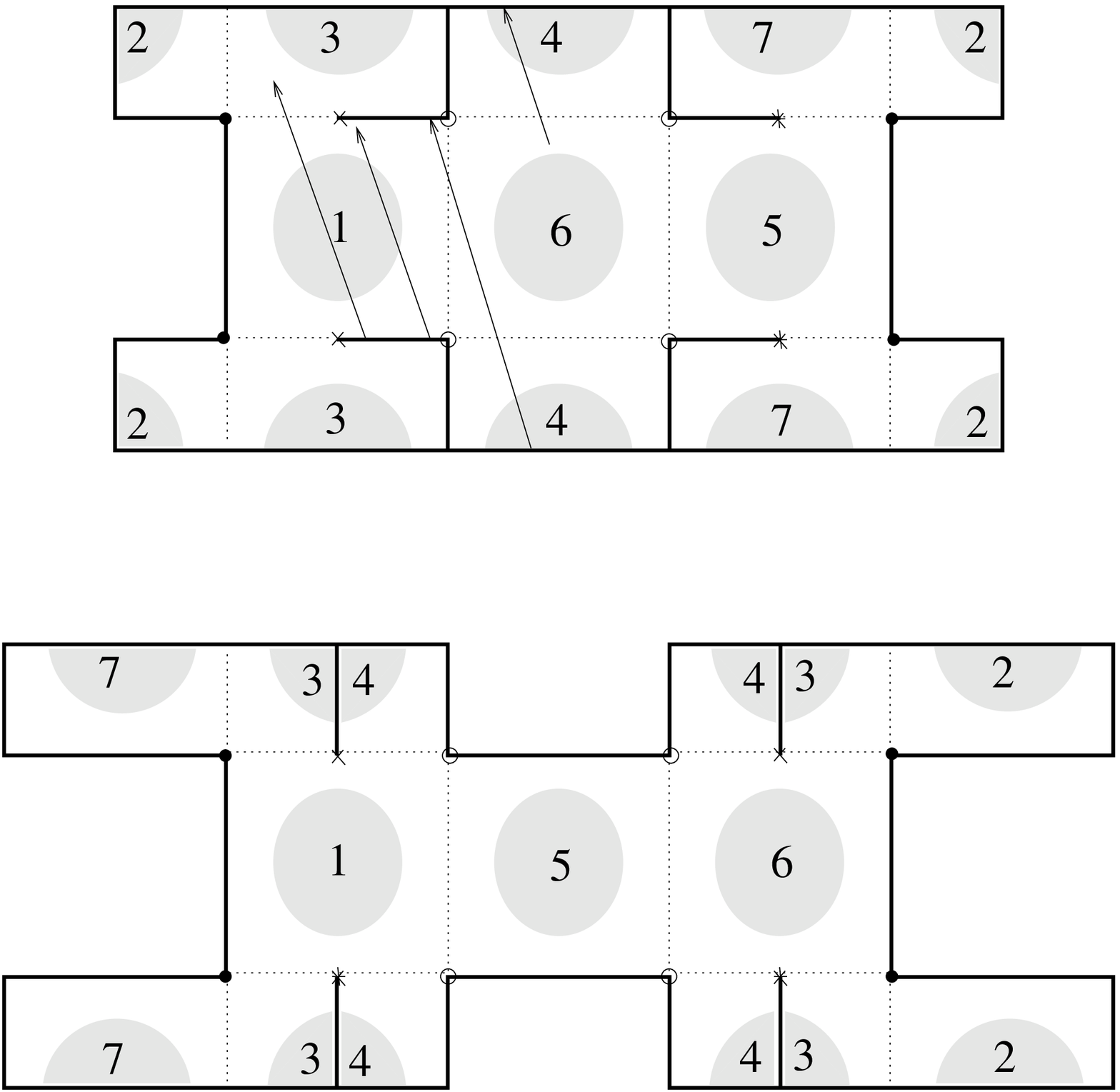}
\end{center}
\caption{The pair $7_3$ of isospectral billiards with a rectangular base shape
unfolded to a translation surface (i.e. flat billiard with opposite
sides identified).\label{squarebaseunfolded}}
\end{figure}

The billiards of Fig.~\ref{squarebase} possess one $2\pi$-angle, two $3\pi/2$-angles
and eight $\pi/2$-angles each. The translation surfaces associated to these billiards 
are obtained by gluing together $2n=4$ copies of the billiards, yielding planar surfaces 
of genus $4$. They are shown in Fig.~\ref{squarebaseunfolded}. 
Opposite sides are identified (e.g. in the first surface, the left edge of tile 1 
is identified with the right edge of tile 5). Each surface has four
singular points. The symbols $\circ$ and $\bullet$
represent a $6\pi$-angle, while the $\times$ and $\ast$ symbols denote a $8\pi$
angle. An example of a straight line drawn on the first surface is shown on
Fig.~\ref{squarebaseunfolded}. The eigenvalue problem on these
surfaces is equivalent to the problem on the billiards. It is however
simpler to handle since the translation surfaces have no
boundary. Thus, only the continuity properties of the eigenfunctions
have to be checked.

Each translation surface is tiled by seven rectangles. Again, any point on the surface can
be specified by its coordinates $(\bfa, i)$.
Each tile on the translation surface has six neighboring tiles, attached  
at its left, upper left, upper right, right, lower right and lower left edge, and
numbered from 1 to 6 respectively. For instance tile 1 is surrounded by: tile 5 on its
left edge, tile 6 on its right edge, tile 3 on its upper left edge, tile 1 itself on its
upper right edge (because of the identification of opposite sides), tile 3 on its
lower left edge  and tile 1 on its lower right edge. 
The way the tiles are glued together can be specified 
by permutation matrices $A^{(\nu)}$, $1\leq \nu\leq 6$, such that
$A^{(\nu)}_{ij}=1$ if and only if the edge number $\nu$ of $i$ glues
tile $i$ to tile $j$.
For instance for the first translation surface, the matrix
specifying which tile is on the right of which is  
\begin{equation}
A^{(2)}=\left(
\begin{array}{ccccccc} 0&0&0&0&0&1&0\cr 0&0&1&0&0&0&0\cr 0&0&0&0&0&0&1\cr
0&0&0&1&0&0&0\cr 1&0&0&0&0&0&0\cr 0&0&0&0&1&0&0\cr 0&1&0&0&0&0&0\cr
\end{array}\right)
\end{equation}
(tile 6 is on the right of tile 1, therefore $A^{(1)}_{1,6}=1$, and so on).
In a similar way, matrices $B^{(\nu)}$, $1\leq \nu\leq 6$, can be defined for the 
second translation surface. Now suppose there exists a matrix $T$ such that 
\begin{equation}
\label{commute}
\forall \nu,\ 1\leq \nu\leq 6,\ A^{(\nu)}T=T B^{(\nu)}.
\end{equation}
Then for any given eigenstate $\phi$ 
of the first translation surface we can construct an eigenstate $\psi$ for the
second translation surface, defined by Eq. \eqref{psiphi}. In order to prove isospectrality
we only have to check for continuity properties at each edge.
Suppose tiles $i$ and $j$ are neighbors. This means that
there exists a $\nu$, $1\leq \nu\leq 6$, such that
$A^{(\nu)}_{ij}=1$.
To prove the continuity of $\psi$ between tiles $i$ and $j$, we have to show
that the quantity
\begin{equation}
\label{conditionpsipsi1}
\cC=\psi(\bfa,i)-\psi(\bfa,j)
\end{equation}
is equal to zero for all $\bfa$ belonging to the edge between $i$ and
$j$. By definition of $\nu$ we have $A^{(\nu)}_{ik}=1$ if and only if
$k=j$. Therefore
\begin{equation}
\label{dvppsi}
\psi(\bfa, j)=\sum_{k}A^{(\nu)}_{ik}\psi(\bfa, k),
\end{equation}
and $\cC$ is given by
\begin{equation}
\label{conditionpsipsi2}
\cC=\psi(\bfa,i)-\sum_{k}A^{(\nu)}_{ik}\psi(\bfa, k).
\end{equation}
Using Eq.~\eqref{psiphi}, we get
\begin{equation}
\cC=\cN\sum_k T_{i k}\phi(\bfa,k)-\cN\sum_{k, k'}A^{(\nu)}_{ik} T_{k k'}\phi(\bfa,k').
\end{equation}
The sum over $k$ on the right-hand side yields a term $(A^{(\nu)}T)_{ik'}$. 
According to the commutation relation \eqref{commute}, it is equal to $(T B^{(\nu)})_{ik'}$,
which gives
\begin{equation}
\cC=\sum_k T_{i k}\left(\phi(\bfa,k)-\sum_{k'}B^{(\nu)}_{k k'}\phi(\bfa,k')\right).
\end{equation}
Now the continuity of the function $\phi$ ensures that all the terms between parentheses
vanish. Thus $\cC=0$, and continuity of $\psi$ is proved. Continuity of partial
derivatives is proved in the same way.\\

The proof rests entirely on the fact that we assumed the 
existence of a transplantation matrix $T$ satisfying
the commutation properties \eqref{commute}. It turns out that such a  matrix exists. 
One can check that given the matrix
\begin{equation}
\label{matriceTperiodic}
T=\left(
\begin{array}{ccccccc} 1&0&0&1&0&0&1\cr 0&1&0&0&1&0&1\cr 0&0&1&0&0&1&1\cr
1&0&0&0&1&1&0\cr 0&1&0&1&0&1&0\cr 0&0&1&1&1&0&0\cr 1&1&1&0&0&0&0\cr\end{array}\right),
\end{equation}
the commutation relations \eqref{commute} are satisfied for all $\nu$, $1\leq \nu\leq 6$. Thus
the proof of isospectrality is completed.
We return in section \ref{transplantation} 
on this transplantation proof
of isospectrality.\\

A natural question is to know how one can find a suitable matrix $T$ 
and permutation matrices $A^{(\nu)}$, $B^{(\nu)}$
verifying all commutation equations \eqref{commute}. Historically these matrices were obtained
by the construction of Sunada triples, as will be explained in section \ref{sunada}.
In fact, it turns out that the matrix $T$ is just the incidence matrix of the graph 
associated with a certain finite projective space (the Fano plane in our example), 
as will be explained in detail in section \ref{transplantation}.

%\newpage
%%%%%%%%%%%%%%%%%%%%%%%%%%%%%%%%%%%%%%%%%%%%%%%%%%%%%%%%%%%%%%%%%%%%%%%%%%%%%%%%%%%%5
\section{Further Examples in Higher Dimensions}
%%%%%%%%%%%%%%%%%%%%%%%%%%%%%%%%%%%%%%%%%%%%%%%%%%%%%%%%%%%%%%%%%%%%%%%%%%%%%%%%%%%%5
\label{milnor}

\textcite{milnor} showed that from two nonisometric lattices
of rank $16$ in $\mathbb{R}^{16}$  discovered by \textcite{witt},
one can  construct a pair of flat tori that have the same spectrum of
eigenvalues (all relevant terms are defined below).

In this section, we describe a simple criterion for the
construction of nonisometric  flat tori with the same eigenvalues for
the Laplace operator, from  certain lattices (which was used by Milnor
for the particular case  mentioned above),
and then we construct, for  each integer $n \geq 17$,  a pair of
lattices of rank $n$ in $\mathbb{R}^n$  that match the criterion.
Furthermore, we describe results of S. Wolpert and  M. Kneser
on  the moduli space of flat tori. An interesting survey paper focused on the (elementary) construction
theory of isospectral manifolds has been given by \textcite{Brooks88}.\\

\subsection{Lattices and flat tori}

A lattice (that is, a discrete additive subgroup) can be prescribed as $A\mathbb{Z}^n$ with $A$ a fixed matrix.
For example, set
\begin{equation}  A = \left(\begin{array}{cc}1 &0\\1 &1\\ \end{array} \right);            \end{equation}
then the lattice $A\mathbb{Z}^2$ consists of the points of the form
\begin{equation} 
a(1,1) + b(0,1),\ \ a,b \in \mathbb{Z}.
\end{equation}
An  {\em $n$-dimensional (flat) torus} $T$ is $\mathbb{R}^n$ factored
by a lattice $\mL = A\mathbb{Z}^n$ with  $A \in
\mathbf{GL}(n,\mathbb{R})$. The torus is thus determined by  identifying
points that differ by an element of the lattice.

If we return to the planar example above, the torus topologically
is a donut | one may see this by cutting out the  parallelogram
determined by $(1,1)$ and $(0,1)$, and then gluing  opposite sides together.\\

With $A,B \in \mathbf{GL}(n,\mathbb{R})$ are associated the lattices
$A\mathbb{Z}^n$ and $B\mathbb{Z}^n$. The tori $\mathbb{R}^n/A\mathbb{Z}^n$ and
$\mathbb{R}^n/B\mathbb{Z}^n$, $B \in \mathbf{GL}(n,\mathbb{R})$, are
isometric if and only  if $A\mathbb{Z}^n$ and $B\mathbb{Z}^n$ are
isometric by left multiplication  by an element of
$\mathbf{O}(n,\mathbb{R})$. The matrices $A$ and $B$  are associated
with the same lattice if and only if  they are equivalent by
multiplication on the right by an element of
$\mathbf{GL}(n,\mathbb{Z})$. So the tori $\mathbb{R}^n/A\mathbb{Z}^n$
and $\mathbb{R}^n/B\mathbb{Z}^n$ are isometric  if and only if $A$ and $B$ are equivalent in

\begin{equation}   \mathbf{O}(n,\mathbb{R})\setminus\mathbf{GL}(n,\mathbb{R})/\mathbf{GL}(n,\mathbb{Z}).\footnote{Let $H, K$ be subgroups of the group $G$. Then the {\em space of double cosets}
$H\setminus G /K$ consists of the subsets (``double cosets'') of the form $HgK$, with $g \in G$. (It is clear that $G$ can be partitioned in these double cosets, and each such 
double coset itself can be partitioned in right cosets of $H$, and also in left cosets of $K$.) 
So in $H \setminus G/K$, $x \sim y$ if and only if there are 
$h \in H$ and $k \in K$ such that $hxk = y$.}
                \end{equation}
Here, $\mathbf{O}(n,\mathbb{R})$ is the orthogonal group in $n$ dimensions.

The metric structure of $\mathbb{R}^n$ projects to $T$, and volume$(T) = \vert \mbox{det}A\vert$; $T$ carries a Laplace operator
\begin{equation} \Delta = -\sum_i\partial^2/\partial x_i^2,         
\end{equation}
which is just the projection of the Laplacian of $\mathbb{R}^n$.
The lengths of closed geodesics of $T$ are given by $\parallel a
\parallel$ for $a$ arbitrary in $\mL$,  $\parallel \cdot \parallel$
being the Euclidean  norm.\\

Let $P$ be a symmetric matrix that defines a quadratic form on $\mathbb{R}^n$. The {\em spectrum} of $P$ is defined to be the sequence (with multiplicities) of values $\gamma = N^TPN$ for $N \in \mathbb{Z}^n$. The sequence of squares of lengths of closed geodesics of $\mathbb{R}^n/A\mathbb{Z}^n$ is the spectrum of $A^TA= Q$; the sequence of eigenvalues of the Laplacian is the spectrum of $4\pi^2(A^{-1})(A^{-1})^T = 4\pi^2Q^{-1}$. The Jacobi inversion formula yields for positive $\tau$,
\begin{eqnarray}  
 \sum_{N \in \mathbb{Z}^n} \mbox{exp}(-4\pi^2\tau
  N^TQ^{-1}N)\hspace{2cm}\nonumber\\ 
= \frac{\mbox{volume}(T)}{(4\pi\tau)^{n/2}}\sum_{M \in
    \mathbb{Z}^n}\mbox{exp}(\frac{-1}{4\tau}M^TQM).
\end{eqnarray}

This equation therefore relates the eigenvalue spectrum of the torus
to its length spectrum. We will see in section \ref{formuledetraces}  other examples
of this connection between the spectrum of the Laplacian and the
length spectrum.

\subsection{Construction of examples}

\bigskip
If $\mL$ is a lattice of $\mathbb{R}^n$, $\mL^*$ denotes its dual lattice, which
consists of all $y \in \mathbb{R}^n$ for which $\langle x,y \rangle \in \mathbb{Z}$ for all $x \in \mL$; here, $\langle \cdot,\cdot \rangle$ is the usual scalar product on $\mathbb{R}^n \times \mathbb{R}^n$.
Clearly, ${(\mL^*)}^* = \mL$, and two lattices $\mL$ and $\mL'$ are isometric if and only if $\mL^*$ and ${\mL'}^*$ are.

Recall that two flat tori of the form $\mathbb{R}^n/\mL_i$, $i \in \{1,2\}$,
are isometric if and only if the lattices $\mL_1$ and $\mL_2$ are isometric.
The following theorem gives a criterion for isospectrality of flat tori.

\bt
\label{balls}
Let $\mL_1$ and $\mL_2$ be two nonisometric lattices of rank $n$ in $\mathbb{R}^n$, $n \geq 2$, and suppose that for each $r > 0$ in $\mathbb{R}$, the ball of radius $r$ about the origin contains the same number of points of $\mL_1$ and $\mL_2$. Then
the flat tori $\mathbb{R}^n/\mL_1^*$ and $\mathbb{R}^n/\mL_2^*$ are nonisometric while having the same spectrum for
the Laplace operator.
\et

{\em Proof}.\quad
Suppose $x \ne \zv$ is an element of $\mL_1$ of length $\alpha$.
Then  there is an $\alpha' < \alpha$ such that the ball of radius $\alpha'$ centered at $\zv$ contains all elements
of $\mL_1$ with length strictly smaller than $\alpha$ (since $\mL_1$ is discrete). For any $\alpha' \leq \alpha'' < \alpha$, the ball of radius $\alpha''$ centered at $\mathbf{0}$ contains that same
number of elements.
This ball contains as many elements of $\mL_2$ as of $\mL_1$, and since the ball centered at $\mathbf{0}$ with radius $\alpha$ contains strictly more elements of $\mL_1$,
it follows easily that $\mL_2$ also
contains vectors of length $\alpha$.

Each  element $z \in \mL_i$, $i \in \{1,2\}$, determines an eigenfunction $f(x) = e^{2\pi\langle x,z \rangle i}$ for the
Laplace operator on $\mathbb{R}^n/\mL_i^*$, with corresponding eigenvalue $\lambda = (2\pi)^2\langle z,z\rangle$, so
the number of eigenvalues less than or equal to $(2\pi r)^2$ is equal to the number of points of $\mL_i$ contained in
the ball centered at $\zv$ with radius $r$.

We conclude that $\mathbb{R}^n/\mL_1^*$ and $\mathbb{R}^n/\mL_2^*$ have the same spectrum of eigenvalues, while not being isometric.
\eop \\

\bigskip
\textsc{Milnor's Construction}.\quad
By using the Witt nonisometric lattices in $\mathbb{R}^{16}$ \cite{witt}, \textcite{milnor} essentially used the aforementioned criterion to construct the first example of nonisometric isospectral flat tori.\\

Starting from these two  nonisometric lattices $\mL_1^{16}$ and
$\mL_2^{16}$ of rank $16$ in $\mathbb{R}^{16}$ as described 
in \textcite{witt}, one can in fact construct examples of isospectral flat
tori in  $\mathbb{R}^{n}$ for all $n$, $n\geq 16$, as follows. The
lattices  $\mL_1^{16}$ and $\mL_2^{16}$ satisfy the condition of Theorem
\ref{balls} \cite[p. 324]{witt}. Now embed  $\mathbb{R}^{16}$
in $\mathbb{R}^{17}$ in the canonical way. Denote the coordinate axes
of the latter by $X_1,X_2,\ldots,X_{17}$,  such that
$\langle X_1,X_2,\ldots,X_{16}\rangle = \mathbb{R}^{16}$.
Suppose $\ell \ne \zv$ is a vector on the $X_{17}$-axis which has
length strictly smaller than any non-zero vector  of
$\mL_1$ (and $\mL_2$).  Define two new lattices $\mL_i^{17}$ (of rank
$17$)  generated  by $\mL_i^{16}$ and $\ell$,  $i = 1,2$.
Since $X_{17} \perp \mathbb{R}^{16}$, it follows easily that for any
$r > 0$, the ball centered at the origin with  radius
$r$ contains the same number of elements of $\mL_1^{17}$ as of
$\mL_2^{17}$. One observes that these lattices are  nonisometric. Thus, by Theorem \ref{balls},
we obtain two nonisometric flat tori $\mathbb{R}^{17}/{\mL_i^{17}
}^*$, $i = 1,2$, which have the same spectrum  of eigenvalues for the Laplace operator.

Inductively, we can now define, in a similar way, the nonisometric
lattices $\mL_1^{n}$ and $\mL_2^n$ of rank  $n$, $n \geq 17$, satisfying the
condition of Theorem \ref{balls}, and thus leading to  nonisometric
flat tori $\mathbb{R}^{n}/{\mL_i^{n} }^*$,  $i = 1,2$, which have the
same spectrum of eigenvalues for the  Laplace operator. \\

\subsection{The four-parameter family of Conway and Sloane}

Let $\Lambda$ be a positive-definite lattice. The {\em theta function}\index{theta function} of $\Lambda$ is:
\begin{equation}
\label{thetaseries}
 \Theta_{\Lambda}(\tau) = \sum_{x \in \Lambda}e^{i\pi\tau\parallel x\parallel^2}                 = \sum_{x \in \Lambda}q^{\parallel x\parallel^2} = \sum_{m = 0}^{\infty}N_mq^m,
\end{equation}
where $\mathrm{Im}(\tau) > 0$, and $N_m$ is the number of vectors $x \in \Lambda$ of norm $m$.
$\Theta_{\Lambda}$ can be thought of as a formal power series in the indeterminate $q$, although sometimes one takes $q = e^{i\pi\tau}$ for further investigation, with $\tau$ a complex variable. In that case, $\Theta_{\Lambda}(\tau)$ is a holomorphic function of $\tau$ for $\mathrm{Im}(\tau) \geq 0$.\\

\textcite{ConSlo} construct a four-parameter family of pairs of four-dimensional lattices that
are isospectral (equivalently, that have the same theta series (\ref{thetaseries})). In a similar way as before, such lattice pairs yield isospectral flat tori. 
The main construction of \cite{ConSlo} is given by the next result.

\bt[Conway and Sloane, 1992]
Let $e_{\infty}$, $e_0$, $e_1$, $e_2$ be orthogonal vectors satisfying

\[ e_{\infty}\cdot e_{\infty} = a/12,\ \ e_0\cdot e_0 = b/12, \ \ e_1\cdot e_1 = c/12, \ \ e_2\cdot e_2 = d/12,       \]

where $a,b,c,d > 0$, and let $[w,x,y,z]$ denote the vector $we_{\infty} + xe_0 + ye_1 + ze_2$.
Let $v_{\infty}^{\pm} = [\pm 3,-1,-1,-1], v_0^{\pm} = [1,\pm 3,1,-1], v_1^{\pm} = [1,-1,\pm 3,1],
v_2^{\pm} = [1,1,-1,\pm 3]$.
Then the lattices $\mathbf{L}^+(a,b,c,d)$ spanned by $v_{\infty}^+,v_0^+,v_1^+,v_2^+$ and 
$\mathbf{L}^-(a,b,c,d)$ spanned by $v_{\infty}^-,v_0^-,v_1^-,v_2^-$ are isospectral.
\et

Some small values of $a, b, c, d$ give examples which were first found by \textcite{Schiemann}.
Substituting $(a,b,c,d) = (7,13,19,49)$, one obtains the pair of \textcite{EarNipp}.\\

\subsection{The eigenvalue spectrum as moduli for flat tori}

We now discuss some interesting results on the eigenvalue spectrum for flat tori.
We already saw that there exist nonisometric isospectral flat tori. A
natural question is now how such tori are distributed. 

The following
theorem gives an insight into this question by considering the case of a continuous family of isospectral flat tori.

\bt[\textcite{Wolp}]
Let $T_s$ be a continuous family of isospectral tori defined for $s \in [0,1]$. Then the tori $T_s$, $s \in [0,1]$, are isometric.
\et

An interesting result by M. Kneser is the following (see \cite{Wolp} for a proof). It states that, given an eigenvalue spectrum of some torus, only a finite number of nonisometric tori can be isospectral to it.

\bt[M. Kneser]
The total number of nonisometric tori with a given eigenvalue spectrum is finite.
\et

The following result is rather technical. Its main message is that given two tori $\mathbb{R}^n/A\mathbb{Z}^n$ and $\mathbb{R}^n/B\mathbb{Z}^n$ with the same eigenvalue spectrum, then either these two tori are isometric, or the quadratic forms
$(A^TA)$ and $(B^TB)$ lie on a certain subvariety in the space of positive definite quadratic forms.
A more precise statement is as follows. Denote the space of positive
definite symmetric $n\times n$-matrices by $\wp(n,\mathbb{R})$, and
observe that the map
\begin{equation}
  A \in \mathbf{GL}(n,\mathbb{R}) \mapsto A^TA \in \wp(n,\mathbb{R})  
\end{equation}
determines a bijection from $\mathbf{O}(n,\mathbb{R})\setminus
\mathbf{GL}(n,\mathbb{R})$ to $\wp(n,\mathbb{R})$. Then the following
theorem holds.

\bt[\textcite{Wolp}]
There is a properly discontinuous group $G_n$ acting on $\wp(n,\mathbb{R})$ containing the transformation group induced by the 
$\mathbf{GL}(n,\mathbb{Z})$ action 
\begin{equation} 
S \mapsto A[\mathcal{Z}],           
\end{equation}

where $S \in \wp(n,\mathbb{R})$ and $\mathcal{Z} \in \mathbf{GL}(n,\mathbb{Z})$. Given $P, S \in \wp(n,\mathbb{R})$ with the same spectrum, either $g(P) = S$ for some $g \in G_n$, or $P, S \in \mathfrak{V}_n$, where the latter is a subvariety of $\wp(n,\mathbb{R})$. Moreover,
\begin{itemize}
\item[{\rm(i)}]
$\mathfrak{V}_n = \{ Q \in \wp(n,\mathbb{R}) \parallel$ {\rm spec}(Q) $=$ {\em spec}(R), $R \in \wp(n,\mathbb{R})$ with $R \ne g(Q)$  for all  $g \in G_n\}$, and
\item[{\rm(ii)}]
$\mathfrak{V}_n$ is the intersection of $\wp(n,\mathbb{R})$ and a countable union of subspaces of $\mathbb{R}^m$ for some $m$.
\end{itemize}
\et

In this section we have seen that is essentially ``easy'' to construct (nonisometric) isospectral flat tori. The Milnor example
was exhibited in 1964. But it has taken about 30 years to find counterexamples to Kac's question in the real plane \ldots

%\newpage
%%%%%%%%%%%%%%%%%%%%%%%%%%%%%%%%%%%%%%%%%%%%%%%%%%%%%%%%%%%%%%%%%%%%%%%%%%%%%%%%%%%%%%%%%%%%%%%%%%%%%%%%
\section{Transplantation}
%%%%%%%%%%%%%%%%%%%%%%%%%%%%%%%%%%%%%%%%%%%%%%%%%%%%%%%%%%%%%%%%%%%%%%%%%%%%%%%%%%%%%%%%%%%%%%%%%%%%%%%%
\label{transplantation}

The aim of this section is to describe the idea of transplantation
in a more mathematical way than in section \ref{pedestrian}.
 This concept was presumably first introduced by \textcite{Berard,Berard2}.
There is in fact a deep connection between transplantation theory and
the mathematical field of finite geometries. First we review some
elementary facts about finite geometries. Application of these tools to
transplantation theory sheds light on the reasons for the existence of isospectrality.

\subsection{Tiling}

\subsubsection{Graphs and billiards by tiling}

In this section, we follow \textcite{OkaShu}.\\

{\bf Tiling}.\quad
All known isospectral billiards can be obtained by unfolding
polygonal-shaped tiles. As the unfolding is done along only three sides of
the polygon we can essentially consider triangles. We call
such examples isospectral {\em Euclidean TI-domains}. The known ones are
listed in Appendix \ref{gallery}.
The way the tiles are unfolded can be specified by three permutation
$d\times d$-matrices $M^{(\mu)}$, $1 \leq \mu \leq 3$ and $d \in
\mathbb{N}$, associated with the three sides of the triangle and defined in
the following way: $M^{(\mu)}_{ij}  =  1$ if tiles $i$ and $j$ are glued by their side $\mu$;
$M^{(\mu)}_{ii}  =  1$ if the side $\mu$ of tile $i$ is the boundary of the billiard, and
$0$ otherwise. The number of tiles is, of course, $d$. Call the matrices $M^{(\mu)}$
``adjacency matrices''. 

One can sum up the action of the $M^{(\mu)}$ in a graph with colored edges:
each copy of the base tile is associated with a vertex, and vertices $i$ and
$j$, $i \ne j$, are joined by an edge of color $\mu$ if and only if
$M^{(\mu)}_{ij}  =  1$. In the same way, in the second member of the pair,
the tiles are unfolded according to permutation matrices $N^{(\mu)}$, $1
\leq \mu \leq 3$. We call such a colored graph an {\em involution graph}
for reasons to be explained later in this section. 
An example of such graphs is given in Fig.~\ref{graphs_review}.
If $D$ is a Euclidean TI-domain with base tile a triangle, and $\M = \{M^{(\mu)} \parallel \mu \in \{1,2,3\}\}$ is the set of
associated permutation matrices (or, equivalently, the associated
coloring), denote by $\Gamma(D,\M)$ the corresponding involution
graph.\\

The following proposition is easy but rather useful \cite{KT3}.

\bp
\label{adjmat}
Let $D$ be a Euclidean TI-domain with base tile a triangle, and let
$\M = \{M^{(\mu)} \parallel \mu \in \{1,2,3\}\}$ be the set of
associated permutation matrices. Then the matrix
\begin{equation}
\Delta_{ij}=\sum_{\mu = 1}^3\left(M^{(\mu)}_{ij} - M^{(\mu)}_{ii}\delta_{ij}\right),
\end{equation}
where $\delta_{ij}$ is the Kronecker symbol, is the adjacency matrix
of $\Gamma(D,\M)$.\eop \\
\ep

{\bf Transplantability}.\quad
Two billiards are said to be {\em transplantable} if there exists an invertible matrix $T$ | the {\em transplantation matrix} | such that
\begin{equation} 
TM^{(\mu)}  =  N^{(\mu)}T \ \ \mbox{for all}\ \ \mu. 
\end{equation}
If the matrix $T$ is a permutation matrix, the two domains would just have the same shape.  One can show that transplantability implies isospectrality, as seen in section \ref{pedestrian}.\\

We now discuss an example exhibited by \textcite{BusConDoySem}, and first found by \textcite{GorWebWol}.

\subsubsection{The example of Gordon et al.}

\textcite{Buser} constructed a pairs of isospectral flat surfaces $\M_1$ and $\M_2$ as 
covers of a certain surface $\M_0$, using a pair of almost conjugate subgroups of $\mathbf{SL}(3,2)$.
\textcite{GorWebWol} similarly constructed orbifolds $O_1$ and $O_2$, respectively being 
the quotient by an involutive isometry of $\M_1$ and $\M_2$.\footnote{{\em Orbifolds} are generalizations
of manifolds; they are locally modeled on quotients of open subsets of $\mathbb{R}^n$ by finite group actions.
We refer to \cite{Bull} for a formal introduction.}
$O_1$ and $O_2$ have a common orbifold cover | it is the quotient by an involutive isometry of the common 
cover of $\M_1$ and $\M_2$.
The Neumann orbifold spectrum of $O_i$ is 
precisely the Neumann spectrum of the underlying manifold $\M(O_i)$, and  these latter underlying spaces 
are simply connected real plane domains. Furthermore, Dirichlet isospectrality of $\M(O_1)$ and $\M(O_2)$ is obtained 
by exploiting the Dirichlet isospectrality of $\M_1$ and $\M_2$.\\

We now analyze this pair of isospectral but non-congruent Euclidean domains. We
follow the very transparent approach of 
\textcite{BusConDoySem} to show isospectrality. As the reader will
notice, this will in fact be an easy approach to (and example of) transplantability.\\

\begin{figure}
\includegraphics[width=0.98\linewidth]{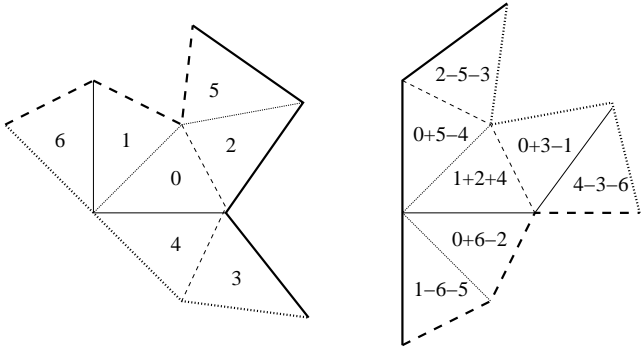}
\caption{Two isospectral billiards with a triangular base shape on seven tiles.\label{figBus}}
\end{figure}

\textsc{Setting}.\quad
Let  $f$ be an eigenfunction of the Laplacian with eigenvalue $\lambda \in \mathbb{R}$ for the Dirichlet problem corresponding to the left-hand
billiard  in Fig.~\ref{figBus}. Let $f_0,f_1,\ldots,f_6$ denote the
functions obtained by restriction of $f$ to each of the seven tiles of the
left-hand billiard, as indicated on the left in Fig.~\ref{figBus}. For the
sake of convenience, we write $\mathbf{i}$ for $f_i$.\\ 

The Dirichlet boundary condition is that $f$ must vanish on each boundary segment.
This is equivalent to the assertion that $f$ goes into $-f$ if continued as a smooth eigenfunction across any boundary segment; in fact,  it goes into $f \circ \sigma$ where $\sigma$ is the reflection on the
boundary segment.

%\begin{figure}
%\includegraphics[width=13cm]{Propeller.jpg}
%\end{figure}

On the right in Fig.~\ref{figBus}, we show how to obtain from $f$ another eigenfunction of eigenvalue $\lambda$  for the right-hand domain. We define the function $\mathbf{1} + \mathbf{2} + \mathbf{4}$
which is actually the function
\begin{equation}
f_1 \circ \tau_1 + f_2\circ \tau_2 + f_4\circ \tau_4,
\end{equation}
where for $k = 1,2,4$, $\tau_k$ is the isometry from the central triangle of the right-hand billiard to the triangle labeled $k$ on the left-hand one.
Now we see from the left-hand side that the functions $\mathbf{1},\mathbf{2},\mathbf{4}$ continue smoothly across dotted lines into copies of the functions $\mathbf{0},\mathbf{5},-\mathbf{4}$
respectively, so that their sum continues into $\mathbf{0} + \mathbf{5}- \mathbf{4}$ as shown.  Similarly way one observes that this continues across a solid line to $\mathbf{4} - \mathbf{5} - \mathbf{0}$
(its negative), and across a dashed line to $\mathbf{2} - \mathbf{5} - \mathbf{3}$, which continues
 across either a solid or dotted line to its own negative. These assertions, together with the similar ones obtained by  cyclic permutation of the arms of the billiards, suffice to show that the transplanted function is an
 eigenfunction of the eigenvalue $\lambda$ that vanishes along each boundary segment of the right-hand domain.

We have defined a linear map which for each $\lambda$ transforms the $\lambda$-eigenspace for the left-hand billiard into the $\lambda$-eigenspace for the right-hand one.
This is a non-singular map (the corresponding matrix is non-singular), and so the dimension of the eigenspace on the right-hand side is larger than or equal to the dimension on the left-hand side.
By symmetry, it follows that the dimensions are equal.  Since $\lambda$ was arbitrary,  the two billiards are Dirichlet isospectral.

\subsubsection{The other known examples}

A similar technique as in the previous section allowed \textcite{BusConDoySem} to show that the series of billiard pairs they
produced are indeed isospectral. All these pairs are listed in Appendix \ref{gallery}; they
were first found by searching for suitable Sunada triples, and then verified to be isospectral (in the plane) by the transplantation method (see also \cite{OkaShu} for a further discussion about the subject of this section).

\subsubsection{Euclidean TI-domains and their involution graphs}
\label{isograph}
To conclude this section, we address a related problem, namely
isospectrality of the involution graphs associated with the isospectral billiards.
We say that two (undirected) graphs are {\em isospectral} if their adjacency matrices have the same multiset of eigenvalues. Note that this definition of graph isospectrality is different from the definition introduced in e.g.~\cite{GutkSmil}, where the spectrum of a metric graph is defined as the spectrum of the Laplacian on the graph whose edges are assigned a given length.

The following question was posed by \textcite{KT3}: Let $(D_1,D_2)$ be a pair of nonisometric isospectral Euclidean TI-domains, and let $\Gamma(D_1) = \Gamma(D_1,\{ M^{(\mu)} \parallel \mu \in \{1,2,3\}\})$ and $\Gamma(D_2) = \Gamma(D_2,\{ N^{(\mu)} \parallel \mu \in \{1,2,3\}\})$ be the
corresponding involution graphs. Are $\Gamma(D_1)$ and $\Gamma(D_2)$ isospectral?
Note that one does not require the domains to be transplantable. (The term ``cospectrality'' is also sometimes used in graph theory, instead of ``isospectrality''.)

We now show that the answer is ``yes'' when the domains are transplantable. The proof is taken from \cite{KT3}.

\bt
Let $(D_1,D_2)$ be a pair of nonisometric isospectral Euclidean TI-domains, and let $\Gamma(D_1) = \Gamma(D_1,\{ M^{(\mu)} \parallel \mu \in \{1,2,3\}\})$ and $\Gamma(D_2) = \Gamma(D_2,\{ N^{(\mu)} \parallel \mu \in \{1,2,3\}\})$ be the
corresponding involution graphs. Then $\Gamma(D_1)$ and $\Gamma(D_2)$ are isospectral.
\et

{\em Proof}.\quad
Define, for $\mu = 1,2,3$, $M^{(\mu)}_*$ as the matrix which has the same entries as $M^{(\mu)}$, except on the diagonal,
where it has only zeros. Define matrices $N^{(\mu)}_*$ analogously.  Suppose that $TM^{(\mu)}T^{-1} = N^{(\mu)}$ for all $\mu$.
Note the following properties:
\begin{itemize}
\item
$M^{(\mu)}_*$ and $N^{(\mu)}_*$, $\mu = 1,2,3$, are symmetric $(0,1)$-matrices, with at most one $1$ entry on each row;
\item
$[M^{(\mu)}_*]^m = M^{(\mu)}_*$ if the natural number $m$ is odd and $[M^{(\mu)}_*]^m = \mathbb{I}_M^{(\mu)}$,
where $[\mathbb{I}_M^{(\mu)}]_{ii} = 1$ if there is a $1$ on the $i$-th row of $M_*^{(\mu)}$, and $0$ otherwise,
if $m$ is even, $\mu = 1,2,3$, and similar properties hold for the $N^{(\mu)}_*$;
\item
$\mbox{Tr}(M^{(i)}_*M^{(j)}_*) = \mbox{Tr}(M^{(j)}_*M^{(i)}_*) = 0$ for $i \ne j$ and $\mbox{Tr}(N^{(i)}_*N^{(j)}_*) = \mbox{Tr}(N^{(j)}_*N^{(i)}_*) = 0$
for $i \ne j$;
\item
$\mbox{Tr}(M^{(i)}_*M^{(j)}_*M^{(k)}_*)$ and $\mbox{Tr}(N^{(i)}_*N^{(j)}_*N^{(k)}_*)$ are independent of the permutation $(ijk)$ of $(123)$
(this is because the individual matrices are symmetric);
\item
the value of all traces in the previous property is $0$ (note that, if
$\{i,j,k\} = \{1,2,3\}$, such a trace equals $0$ since the existence of a nonzero diagonal entry of  $M^{(i)}_*M^{(j)}_*M^{(k)}_*$, respectively $N^{(i)}_*N^{(j)}_*N^{(k)}_*$, implies $\Gamma(D_1)$, respectively $\Gamma(D_2)$,
to have closed circuits of length $3$).\\
\end{itemize}

{F}rom Proposition \ref{adjmat} it follows that $A = \sum_{\mu = 1}^3M^{(\mu)}_*$ is the adjacency matrix of $\Gamma(D_1)$, and 
$B = \sum_{\mu = 1}^3N^{(\mu)}_*$, the adjacency matrix of $\Gamma(D_2)$.

Consider a natural number $n \in \mathbb{N}_0$. Then, with the previous
properties in mind, it follows that 
\begin{equation} 
\mbox{Tr}(A^n) = \mbox{Tr}(B^n). 
\end{equation}
Thus by the following lemma (cf. \cite[Lemma 1]{VDH})  the adjacency matrices of $\Gamma(D_1)$ and $\Gamma(D_2)$ have the same spectrum.
\bl
Two $k\times k$-matrices $K$ and $K'$ are isospectral if and only if {\rm Tr}$(K^l) =$ {\rm Tr}$({K'}^l)$ for $l = 1,2,\ldots,k$.
\el
\eop \\
In section \ref{RQ} we will see that other graph theoretical problems turn
up in Kac theory.

\subsection{Some projective geometry}
There is a fascinating relation between the structure of isospectral billiards and the geometry of vector spaces over finite fields. In section \ref{pedestrian} we constructed pairs of isospectral billiards using unfolding rules. These unfolding rules can be encoded into graphs, like the ones in Fig.~\ref{graphs_review}. 
Thus the structure of a pair of isospectral billiards is entirely encoded
into a pair of graphs that have certain specific properties.
The graphs of Fig.~\ref{graphs_review} are ''colored'' according to a certain set of
permutations. It turns out that the group generated by these permutations is
precisely the automorphism group of a projective space over a
finite field, the so-called Fano plane. The Fano plane has many beautiful
properties and appears in various places, such as combinatorial problems or
the multiplication table of the octonions. A representation of this finite
projective plane is given in Fig.~\ref{fano}.
Here we will see that the adjacency matrix of the graph representing the
Fano plane is nothing but the transplantation matrix between the two
isospectral billiards of  Fig.~\ref{graphs_review}. 

In order to understand this deep connection, basic notions of finite geometries
and design theory are required. In this section we provide the necessary
tools. More details about the notions considered here can be found in \cite{Hirsch}. Note that some remarks about
isospectrality, projective geometry and groups are made in \cite{VoroStep},
however the results there are not fully mathematically rigorous.

\subsubsection{Finite projective geometry}

Let $\mathbb{F}_q$ be the finite field with $q$ elements, $q$ a prime power,
and denote by $V(n,q)$ the $n$-dimensional
vector space over $\mathbb{F}_q$, $n$ a nonzero natural number. 
Define the {\em $(n - 1)$-dimensional projective geometry $\PG(n - 1,q)$} over
$\mathbb{F}_q$ as the set of all subspaces of $V(n,q)$.
Note that $\PG(n - 1,q)$ is often called the ``Desarguesian'' or
``classical'' projective space. The projective space $\PG(-1,q)$ is the
empty set, and has dimension $-1$. 

Points in $\PG(n,q)$ correspond to one-dimensional subspaces of $V(n,q)$,
lines in $\PG(n,q)$ correspond to two-dimensional subspaces of $V(n,q)$,
and so on.
Any $d$-dimensional subspace of $\PG(n,q)$ contains $(q^{d+1}-1)/(q-1)$
points. In particular, $\PG(n,q)$ itself has $(q^{n+1}-1)/(q-1)$ points. It also has
$(q^{n+1}-1)/(q-1)$ hyperplanes (i.e.~$(n - 1)$-dimensional subspaces).\\

\begin{figure}
\centering 
\includegraphics[width=0.5\linewidth]{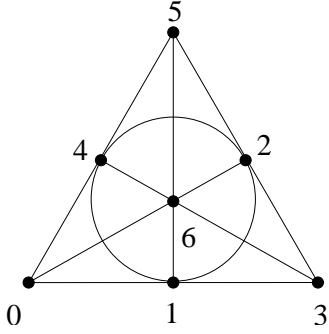}
\caption{The Fano plane.\label{fano}}
\end{figure}

{\bf Example}.\quad
The Fano plane $\PG(2,2)$ shown in Fig.~\ref{fano} has seven points and seven
hyperplanes or lines (one of which is represented as a circle in
Fig.~\ref{fano}).
Any line contains three points (we say that three points are ``incident''
with each line) and any point belongs to three lines (we say that three
lines are ``incident'' with each point). The use of the word `'incident'' in
both cases enhances the symmetry between points and lines in this
geometry. It is precisely this geometry that lies at the root of isospectrality.

\subsubsection{Automorphism groups}
The automorphism groups of finite projective spaces play a key role in
isospectrality as the generators of these groups allow us to construct the
graphs that encode the unfolding rules for the billiard construction. We
 now define these groups and mention some of their properties.
For group theoretical notions not explained here, we refer to the beginning
of section \ref{sunadasection}.

An {\em automorphism} or {\em collineation} of a finite projective space is
a bijection of the points that preserves the type of each subspace
(i.e. lines are mapped to lines, and more generally $d$-dimensional spaces
to $d$-dimensional spaces) and preserves incidence properties (i.e. intersecting
lines are transformed into intersecting lines, etc...).
It can be shown that any automorphism of a $\PG(n,q)$, $n \geq 3$,
necessarily has the following form: 
\begin{equation} 
\theta: \mathbf{x}^T \mapsto A(\mathbf{x}^{\sigma})^T,          
\end{equation}
where $A \in \mathbf{GL}(n + 1,q)$, $\sigma$ is a field automorphism of $\mathbb{F}_q$, the homogeneous coordinate $\mathbf{x} = (x_0,x_1,\ldots,x_n)$ represents a point of the space (which is determined up to a scalar), and
$\mathbf{x}^{\sigma} = (x_0^{\sigma},x_1^{\sigma},\ldots,x_n^{\sigma})$ (recall that $x_i^{\sigma}$ is the image of $x_i$ under $\sigma$).

The set of automorphisms of a projective space naturally forms a group, and in case of $\PG(n,q)$, $n \geq 3$,
this group  is denoted by $\mathbf{P\Gamma L}(n + 1,q)$.
The normal subgroup of $\mathbf{P\Gamma L}(n + 1,q)$ which consists of all automorphisms for which the
companion field automorphism $\sigma$ is the identity, is the {\em projective general linear group}, and
denoted by $\PGL(n + 1,q)$. So $\PGL(n + 1,q) = \mathbf{GL}(n + 1,q)/Z(\mathbf{GL}(n + 1,q))$, where
$Z(\mathbf{GL}(n + 1,q))$ is the central subgroup of all scalar matrices of $\mathbf{GL}(n + 1,q)$.
Similarly one defines $\PSL(n + 1,q) = \mathbf{SL}(n + 1,q)/Z(\mathbf{SL}(n + 1,q))$, where
$Z(\mathbf{SL}(n + 1,q))$ is the central subgroup of all scalar matrices of $\mathbf{SL}(n + 1,q)$ with unit determinant. Recall that $\mathbf{SL}(n + 1,q)$ consists of the elements of $\mathbf{GL}(n + 1,q)$ with unit determinant.

An {\em elation} of $\PG(n,q)$ is an automorphism of which the fixed points structure precisely is a hyperplane, or the space itself. A {\em homology} either is the identity, or it is an automorphism that fixes a hyperplane pointwise, and one further point not contained in that hyperplane.

\subsubsection{Involutions in finite projective space}

Let $\PG(n,q)$, $n \in \mathbb{N} \cup \{-1\}$, be  the $n$-dimensional projective space over the finite field
$\mathbb{F}_q$ with $q$ elements, so that $q$ is a prime power; we have $\vert \PG(n,q)\vert = \frac{q^{n + 1} - 1}{q - 1}$.
(Note again that $\PG(-1,q)$ is  the empty space.)

We discuss the different types of involutions that can occur in the
automorphism group of a finite projective space  \cite{Segre}. 
\begin{itemize}
\item
\textsc{Baer Involutions}.\quad
A {\em Baer involution} is an involution that is not contained in the linear automorphism group of the space so that $q$ is a square, and it fixes an $n$-dimensional subspace over $\mathbb{F}_{\sqrt{q}}$ pointwise.
\item
\textsc{Linear Involutions in Even Characteristic}.\quad
If $q$ is even, and $\theta$ is an involution that is not of Baer type, $\theta$ must fix an $m$-dimensional subspace of $\PG(n,q)$ pointwise, with $1 \leq m \leq n \leq 2m + 1$.
In fact, to avoid trivialities, one assumes that $m \leq n - 1$.
\item
\textsc{Linear Involutions in Odd Characteristic}.\quad
If $\theta$ is a linear involution of $\PG(n,q)$, $q$ odd, the set of fixed points is the union of
two disjoint complementary subspaces. Denote these by $\PG(k,q)$ and $\PG(n - k -1,q)$, $k \geq n - k - 1 > -1$.\footnote{We do not consider the possibility of involutions without fixed points, as they are not relevant for our purpose.}
\end{itemize}

We are now ready to explore a connection between Incidence Geometry and Kac Theory.

\subsection{Projective isospectral data}

\subsubsection{Transplantation matrices, projective spaces and isospectral data}

Suppose one wants to construct a pair of isospectral billiards, starting
from a planar polygonal base shape. The idea described in \cite{Giraud} is to start from the
transplantation matrix $T$, and choose it in such a way that the existence
of commutation relations
\begin{equation} 
T M^{(\mu)}= N^{(\mu)} T 
\end{equation}
for some
permutation matrices $M^{(\mu)}, N^{(\mu)}$ will be known a priori.
This is the case if $T$ is taken to be
the incidence matrix of a finite projective space; the matrices $M^{(\mu)}$ and
$N^{(\mu)}$ are then permutations of the points and the hyperplanes
of the finite projective space.\\

An $(N,k, \lambda)-${\it symmetric balanced incomplete block design}
(SBIBD) is a rank $2$  incidence geometry, defined on a set of
$N$ points, each belonging to $N$ subsets (called {\it blocks}) such that
each block is incident with $k$ points, any two distinct points are contained in exactly $\lambda$ blocks, and
each point is incident with $k$ different blocks.

{\bf Example}.\quad
The points and hyperplanes of an $n$-dimensional projective space
$\PG(n,q)$ defined over $\mathbb{F}_q$ is an $(N,k, \lambda)$-SBIBD with $N=(q^{n+1}-1)/(q-1)$,
$k=(q^{n}-1)/(q-1)$ and $\lambda=(q^{n-1}-1)/(q-1)$.\\
So the Fano plane is a $(7,3,1)-$SBIBD.\\

Let $\Gamma$ be an $(N,k,\lambda)$-SBIBD.
The points and the blocks can be labeled from $0$ to $N - 1$.
One can define an $N\times N$-{\it incidence matrix}
$T$ describing to which block each point belongs. The entries $T_{ij}$
of the matrix are $T_{ij}=1$ if the point $j$ belongs to the block $i$, and $0$
otherwise. It is easy to see that the matrix $T$ verifies the relation
 \begin{equation} 
T T^{T}=\lambda \JJ+(N-k)\lambda/(k-1)\II, 
\end{equation}
where $\JJ$ is the $N\times N$-matrix with all entries equal to $1$
and $\II$ the $N\times N$ identity matrix.
In particular, the incidence matrix of $\PG(n,q)$ verifies
\begin{equation}
\label{tt}
T T^{T}=\lambda \JJ+(k-\lambda)\II
\end{equation}
with $k$ and $\lambda$ as given above.\\

%{\bf Example}.\quad
%The incidence matrix
%of the Fano plane corresponds to a labeling of the lines such that line the $0$
%contains the points $0,1,3$, and the  line $1$ contains the points $1,2,4$, etc.\\

Any permutation $\sigma$ of the points of a finite projective space
can be written as a $d\times d$ permutation matrix $M$ defined by
$M_{i\sigma(i)}=1$ and the other entries equal to zero.
Here $d$ is the number of points. If $M$ is a
permutation matrix associated with an automorphism of the space, then there exists a
permutation matrix $N$ such that
\begin{equation}
\label{commutation}
TM=NT.
\end{equation}
In other words, (\ref{commutation}) means that permuting the
columns of $T$ (which correspond to the hyperplanes of the
space) under $M$ is in some sense equivalent to permuting the rows of $T$
(corresponding to the points of the space) under $N$. The reason that this occurs is the concept of ``duality''; in a finite projective space the
points and hyperplanes  play the same role.\\

Consider a finite projective space $\pi = \PG(n,q)$ with incidence matrix $T$.
With each hyperplane in $\pi$ we associate a tile in the first billiard,
and to each point in $\pi$ we associate a tile in the second billiard.
If we choose permutations $M^{(\mu)}$ in $\mathbf{P\Gamma L}(n + 1,q)$, then the commutation relation (\ref{commutation})
will ensure that there exist permutations $N^{(\mu)}$ verifying
\begin{equation}
TM^{(\mu)}=N^{(\mu)}T.
\end{equation}
Since these commutation relations imply
transplantability, they also imply isospectrality of the billiards
constructed from the graphs corresponding to $M^{(\mu)}$ and $N^{(\mu)}$.\\

\textsc{Constraints}.\quad
If the base tile has $r$ sides, we need to choose $r$
elements $M^{(\mu)}$, $1\leq\mu\leq r$, in $\mathbf{P\Gamma L}(n + 1,q)$ in such a way that (at least) the following remarks are taken into account.

\begin{itemize}
\item
Since $M^{(\mu)}$ represents the reflection of a tile with respect to one of
its sides, it has to be an involution.
\item
In order that the billiards be connected, no point should be left out by
the matrices $M^{(\mu)}$ | in other words, the graph associated to the
matrices $M^{(\mu)}$ should be connected.
\item
If we want the base tile to be of  ``any'' shape, there should be no closed circuit in the graph (in other words, it should be a finite tree).
\end{itemize}

Assume one is looking for a pair of isospectral billiards with
$d=(q^3-1)/(q-1)$ copies of a base tile having the
shape of an $r$-gon, $r\geq 3$.
We need to search for $r$ involutions such that the associated graph is
connected and does not admit a closed circuit. Such a graph connects $d$ vertices and hence
 requires $d-1$ edges.
For involutions with $s$ fixed
points, there are $(d-s)/2$ independent transpositions in its cycle decomposition, and
each transposition is represented by an edge in the graph. As a consequence,
 $q$, $r$ and $s$ have to satisfy the following condition:
\begin{equation} 
r(q^2+q+1-s)/2=q^2+q.
\end{equation}
More generally, we define ``projective isospectral data'' as  triples $(\mathbf{P},\{\theta^{(i)}\},r)$, where $\mathbf{P}$ is a finite projective space of dimension at least $2$, and
$\{\theta^{(i)}\}$ a set of $r$ nontrivial involutions of $\mathbf{P}$, satisfying the following equation
\begin{equation}
\label{eq1}
 r(\vert \mathbf{P}\vert - \mbox{Fix}(\theta)) = 2(\vert \mathbf{P}\vert - 1),
\end{equation}
for some natural number $r \geq 3$.
Here Fix$(\theta)$ = Fix$(\theta^{(i)})$ is a constant number of fixed points of $\mathbf{P}$ under each $\theta^{(i)}$, and $\vert \mathbf{P}\vert$ is the number of points of $\mathbf{P}$.\\

One can now generate all possible pairs of isospectral billiards
whose transplantation matrix is the incidence matrix of a $\PG(2,q)$,
with  $r$ and $q$ restricted by the previous analysis.

Using the classification of involutions for dimension $2$,  we examine the various cases.\\

\textsc{Let $q$ be even and not a square}.\quad
Then
any involution is an elation and therefore has $q + 1$
fixed points. Therefore, $q$ and $r$ are constrained by the relation
\begin{equation} 
r q^2/2=q^2+q.
\end{equation}
The only integer solution with $r\geq 3$ and $q\geq 2$ is $(r=3, q=2)$.
These isospectral billiards  correspond to the Fano plane $\PG(2,2)$ and will be made
of $d=7$ copies of a base triangle.\\

\textsc{Let $q$ be odd and not a square}.\quad
Then
any involution is a homology and therefore has $q+2$
fixed points. Therefore, $q$ and $r$ are constrained by the relation
\begin{equation} 
r (q^2-1)/2=q^2+q.
\end{equation}
The only integer solution with $r\geq 3$ and $q\geq 2$ is $(r=3, q=3)$.
These isospectral billiards correspond to $\PG(2,3)$ and will be made
of $d=13$ copies of a base triangle.\\

\textsc{Let $q=p^2$ be a square}.\quad
Then
any involution fixes all points in a Baer subplane $\PG(2,p)$
and therefore has $p^2+p+1$ fixed points. Therefore, $p$ and $r$ are
constrained by the relation
\begin{equation} 
r (p^4-p)/2=p^4+p^2.
\end{equation}
There is no integer solution with $r\geq 3$ and $q\geq 2$.\\

\textsc{Closed circuits}.\quad
One could also look for isospectral billiards {\em with closed circuits}: this will require
the base tile to have a shape such that the closed circuit  does not make the
copies of the tiles come on top of each other when unfolded.
If we allow just one closed circuit
in the graph describing the isospectral billiards, then there are $d$
edges in the graph instead of $d-1$ and the equation for $p$ and $r$
becomes
\begin{equation}
r (p^4-p)/2=p^4+p^2+1,
\end{equation}
which has the only integer solution
$(r=3, p=2)$. These isospectral billiards correspond to $\PG(2,4)$
and will be made of $d=21$ copies of a base triangle.\\

To summarize, we have the following:
\begin{itemize}
\item
The Fano plane $\PG(2, 2)$ provides three pairs (made of seven tiles).
\item
$\PG(2, 3)$ provides nine pairs (made of $13$ tiles).
\item
$\PG(2, 4)$ provides one pair (made of $21$ tiles).
\end{itemize}
It turns out that the pairs obtained in such a way are exactly those obtained by
\textcite{BusConDoySem} and \textcite{OkaShu}.\\

Now consider the space $\PG(3,2)$, which contains 15 points.
The collineation group of
$\PG(3,2)$ is the group
\begin{equation}
\PGL(4,2) \cong \mathbf{P\Gamma L}(4,2) \cong \mathbf{GL}(4,2).
\end{equation}
Generating all possible graphs from the $316$
involutions, one obtains four pairs of
isospectral billiards with $15$ triangular tiles, which completes
the list of all pairs found by \textcite{BusConDoySem} and \textcite{OkaShu}. This list can be found in Appendix \ref{gallery}.\\

For projective spaces of dimension $2$, we thus have the following result \cite{Giraud}.
Let $\mathbf{P} = \PG(2,q)$ be the two-dimensional projective space over the finite field $\mathbb{F}_q$,  and suppose there exists projective isospectral data
$(\mathbf{P},\{\theta^{(i)}\},r)$. If $q$ is not a square, then  $(r,q) \in \{(3,2),(3,3)\}$. If $q$ is a square, then there are no integer solutions of Eq.~$(\ref{eq1})$.

So the method introduced by \textcite{Giraud} explicitly gives the transplantation matrix $T$ for all these
pairs | each one is the incidence matrix of some finite projective space, and
the transplantation matrix provides the mapping between
eigenfunctions of both billiards. The inverse mapping is given by
\begin{equation} 
T^{-1}=(1/q^{n-1})(T^{T}-(\lambda/k)\JJ).
\end{equation}

\subsubsection{Generalized isospectral data}
\textcite{KT} obtained the next generalization for any dimension $n \geq 2$.
It turns out that all possible candidates $\PG(n,q)$ other than the ones
already obtained are ruled out by the following results.

\bt[\textcite{KT}]
 Let $\mathbf{P} = \PG(n,q)$ be the $n$-dimensional projective space over the finite field $\mathbb{F}_q$,  and suppose there exists projective isospectral data
$(\mathbf{P},\{\theta^{(i)}\},r)$.    Then $q$ cannot be a square. If $q$ is not a square, then $(r,n,q) \in \{(3,2,2),(3,n,3)\}$, where in the the case $(r,n,q) = (3,n,3)$
each $\theta^{(i)}$ fixes pointwise a hyperplane, and also a point not in that hyperplane. However, this class of solutions only generates planar isospectral pairs if $n = 2$.
\et

Call a triple $(\mathbf{P},\{\theta^{(i)}\},r)$, where $\mathbf{P}$ is a finite projective space of dimension at least $2$,
and $\{\theta^{(i)}\}$ a set of $r$ nontrivial involutory automorphisms of $\mathbf{P}$, satisfying
\begin{equation}
\label{eq12}
 r(\vert \mathbf{P}\vert) - \sum_{j=1}^r\mbox{Fix}(\theta^{(j)}) = 2(\vert \mathbf{P}\vert - 1),
\end{equation}
for some natural number $r \geq 3$, ``generalized projective isospectral
data''.

These data were completely classified in \cite{KT2}.

 \bt[\textcite{KT2}]
 \label{GID}
Let $\mathbf{P} = \PG(l,q)$ be the $l$-dimensional projective space over the finite field $\mathbb{F}_q$,  $l \geq 2$, and suppose there exists generalized projective isospectral data
$(\mathbf{P},\{\theta^{(i)}\},r)$ which yields isospectral billiards. Then either $l = 2$, the $\theta^{(i)}$ fix the same number of points of $\mathbf{P}$, and the solutions are as previously described,
or $l = 3$, $r = 3$ and $q = 2$, and again the examples are as before.
\et

\subsubsection{The operator group}
The same kind of results can be formulated at a more abstract level.
Suppose $D$ is a Euclidean TI-domain on $d$ base triangles, and  let $M^{(\mu)}$, $\mu \in \{1,2,3\}$, be the corresponding permutation $d\times d$-matrices.  Define again involutions $\theta^{(\mu)}$ on a set $X$ of $d$ letters $\Delta_1,\Delta_2,\ldots,\Delta_d$ (corresponding to the base triangles) as follows: $\theta^{(\mu)}(\Delta_i) = \Delta_j$ if $M^{(\mu)}_{ij} = 1$ and $i \ne j$. In the other cases, $\Delta_i$ is mapped onto
itself. Then clearly, $\langle \theta^{(\mu)} \parallel \mu \in \{1,2,3\}\rangle$ is a transitive permutation group
on $X$, which we call the {\em operator group} of $D$.\\

 Suppose that  $(D_1,D_2)$ is a pair of non-congruent planar isospectral domains constructed from unfolding an $r$-gon, $r \geq 3$, $d < \infty$ times. Since $D_i$ are constructed
 by unfolding an $r$-gon, we can associate $r$ involutions $\theta^{(j)}_i$ to $D_i$, $j = 1,2,\ldots,r$ and $i = 1,2$.  Define the operator groups
 \begin{equation} 
G_i = \langle \theta^{(j)}_i\rangle. 
\end{equation}
Now suppose that
\begin{equation} 
G_1  \cong \PSL(n,q) \cong G_2, 
\end{equation}
with $q$ a prime power and $n \geq 2$ a natural number.
The natural geometry on which $\PSL(n,q)$ acts (faithfully) is the $(n -
1)$-dimensional projective space $\PG(n - 1,q)$ over the finite field
$\mathbb{F}_q$. It should be mentioned that $\PSL(n,q)$ acts transitively on
the points of $\PG(n - 1,q)$. So we can see the involutions $\theta^{(j)}_i$
for fixed $i \in \{1,2\}$ as automorphisms of $\PG(n - 1,q)$ that generate
$\PSL(n,q)$. 

This implies (by nontrivial means) that for fixed $i \in \{1,2\}$ the triple
\begin{equation} 
(\PG(n - 1,q),\{\theta^{(j)}_i\},r) 
\end{equation}
yields generalized projective isospectral data.
Theorem \ref{GID} implies that $(n,q)$ is contained in
$\{(3,2),(3,3),(4,2),(3,4)\}$ if $n \geq 3$.  

Now suppose that $n = 2$.  We have to solve the equation
\begin{equation} 
r\vert \PG(1,q)\vert - \sum_{j = 1}^{r}\mathrm{Fix}(\theta^{(j)}_i) =
2(\vert \PG(1,q)\vert - 1),              
\end{equation}
for fixed $i \in \{1,2\}$,
where Fix$(\theta^{(j)}_i)$ is the number of fixed points in $\PG(1,q)$ of $\theta^{(j)}_i$. Since $\vert \PG(1,q)\vert = q + 1$ and
since a nontrivial element of $\PSL(2,q)$ fixes at most $2$ points of $\PG(1,q)$, an easy calculation leads to a contradiction if $q \geq 3$.

Now let $q = 2$. Then $\PSL(2,2)$ contains precisely $3$ involutions, and they each fix precisely one point of $\PG(1,2)$.
A numerical contradiction follows.
\eop \\

Thus, the only possible examples of isospectral billiards that can be
constructed from the third family of finite simple groups (see
\cite{ConCur}) are those obtained in  \cite{BusConDoySem,
  OkaShu, Gir05}. They are listed in Appendix \ref{gallery}.

%\newpage
%%%%%%%%%%%%%%%%%%%%%%%%%%%%%%%%%%%%%%%%%%%%%%%%%%%%%%%%%%%%%%%%%%%%%%%%%%%%%%%%%%%%5
\section{Semiclassical Investigation of Isospectral Billiards}
%%%%%%%%%%%%%%%%%%%%%%%%%%%%%%%%%%%%%%%%%%%%%%%%%%%%%%%%%%%%%%%%%%%%%%%%%%%%%%%%%%%%5
\label{properties}
The existence of isospectral pairs proves that the knowledge of the
infinite set of eigenenergies of a billiard does not suffice to uniquely determine
the shape of its boundary. 
%Weyl's formula
%(\ref{dvpweyl}) provides us with the area and the perimeter of the
%billiard, as well as with some information on the properties of its
%curvature or its scattering corners.
A natural question arises: if the set of eigenvalues is not
sufficient to distinguish between two isospectral billiards, then which
additional quantity would suffice to  uniquely specify which is
which? A parallel issue is to identify what kind of geometric information on the system one can extract from the spectrum. This type of inverse problem occurs in many fields of physics, from lasing cavities to stellar oscillations.

It is well known that
classical mechanics can be seen as a limit of quantum mechanics when
Planck's constant, seen as a parameter, goes to zero. It is therefore
natural that, for small enough values of this parameter, classical
characteristics of quantum systems begin to emerge. If one considers
an electron in a box, one can construct a certain linear combination
of stationary 
wave functions that describes its probability density distribution
at each point of the box. At the classical limit, this probability
distribution gets mainly concentrated on classically authorized 
trajectories. The quantum system thus somehow
``knows'' about classical trajectories of the underlying classical
system. As shown in this section, the semiclassical approach provides
a constructive way 
to retrieve geometric information on the system.

More formally, the time-dependent propagator of the Schr\"odinger equation 
can be expressed as a Feynman path integral, which is a sum over all 
continuous paths going from the initial to the final point.
Using a stationary phase approximation, \textcite{VVl} 
obtained a formula expressing the propagator (or, more precisely, its discretized 
version) in the semiclassical limit as a sum over all classical trajectories of the 
system. \textcite{BalBlo74} showed that the density of states
can be written as a sum over closed trajectories of the classical system.
Using a stationary phase approximation technique, the semiclassical Green function 
can be similarly expressed as a sum over all classical trajectories. This led to the
Gutzwiller trace formula for chaotic systems (see \cite{Gut89} and references therein)
or the Berry-Tabor trace formula for integrable systems
\cite{BerTab76}. These trace formulae relate the quantum spectrum to classical 
features of the system. While the leading terms of the mean spectral density provide geometric information about global quantities of the system, such as the area or perimeter, the trace formulae contain information about classical trajectories.
Corrections to these trace formulae account for the presence of other classical trajectories, such as diffractive orbits.

As mentioned in section \ref{paperfoldingproof}, the transplantation proof of isospectrality shows that pairs displaying any kind of classical dynamics can be constructed, from (pseudo-)integrability to chaos. One might ask whether the spectrum of a billiard uniquely determines its length spectrum.
As we will see, the transplantation method provides an answer to this
question. However, in the pseudo-integrable case where diffractive
contributions to the trace formula can be handled, it turns out that
transplantation properties of diffractive orbits are different from those of 
periodic orbits.

In this section we first introduce some tools relevant to semiclassical quantization and then review in more detail
various classical and quantum properties of isospectral pairs that
have been studied in the literature, either for generic isospectral
billiards, or for particular examples such as the celebrated example of Fig.~\ref{celebrated}.

\subsection{Mean density of eigenvalues}
\label{meandensity}
The problem of calculating the eigenvalue distribution for a given
domain $B$ (sometimes called Weyl's problem) is dealt with starting
from the density of energy levels
\begin{equation}
\label{densityofstates}
d(E)=\sum_n\delta(E-E_n),
\end{equation}
where $\delta$ is the Dirac delta function and 
the sum runs over all eigenvalues of the system. The counting
function is the integrated version of the eigenvalue distribution:
\begin{equation}
\cN(E)=\sum_n\Theta(E-E_n),
\end{equation}
where $\Theta$ is the Heaviside step function. Statistical functions of the energy
can be studied by proper smoothing of the delta functions in
\eqref{densityofstates}. The mean of a function $f$ of the energy is
defined by its convolution with a test-function $\xi$:
\begin{equation}
\bar{f}(E)=\int_{-\infty}^{\infty}f(e) \xi(E-e)de.
\end{equation}
The test-function $\xi$ is taken to be centered at 0, normalized to 1
and have an important weight only around the origin, with a width
$\Delta E$ large compared to the mean level spacing but small compared
to $E$.  

Isospectral billiards share by definition the same counting function $\cN(E)$.
Let us study the mean behavior of $\cN(E)$. Suppose the
Hamiltonian of an $N$-dimensional system is of the form 
\begin{equation}
H(q,p)=p^2/2m+V(q).
\end{equation}
The ``Thomas-Fermi
approximation'' consists in making the assumption that each quantum
state is associated with a volume $(2\pi\hbar)^N$ in phase
space. The mean value of $\cN(E)$ is given by
\begin{eqnarray}
\cNo(E)&\simeq&\int\frac{d^N p d^N q}{(2\pi\hbar)^N}\Theta\left(E-H(q,p)\right)\\
&\simeq&\frac{1}{\Gamma(\frac{N}{2}+1)}\left(\frac{m}{2\pi\hbar^2}\right)^{N/2}
\int_{V(q)<E}[E-V(q)]^{N/2}dq\nonumber
\end{eqnarray}
after integration over $p$. In the case where we
describe the movement in an $N$-dimensional domain of volume $\cV$ we get
\begin{equation}
\label{zeroweyl}
\cNo(E)\simeq\frac{\cV}{\Gamma(N/2+1)}
\left(\frac{m}{2\pi\hbar^2}\right)^{N/2}E^{N/2},
\end{equation}
which is the first term in a series expansion of $\cNo(E)$,
called the Weyl expansion. In particular two isospectral $N$-dimensional
domains necessarily have the same volume.

For two-dimensional billiards and under our conventions on units, this first
term of Weyl expansion reads
\begin{equation}
\label{weylexp}
\cNo(E)\simeq\frac{\cA}{4\pi}E,
\end{equation}
where $\cA$ is the area of the billiard. This means that a necessary
condition for isospectrality is that the billiards have the same area.
The asymptotic expansion of the Laplace transform of the density of states
\cite{SteWae} allows us to derive the following terms in the Weyl expansion 
\cite{BalHil76}. The expansion is given by
\begin{equation}
\label{dvpweyl}
\cNo(E)\simeq \frac{\cA}{4\pi} E \mp \frac{\cL}{4\pi} \sqrt{E}+\cK,
\end{equation}
where $\cA$ and $\cL$ are the area and the perimeter of
the billiard, respectively. The sign before $\cL$ is (--) for Dirichlet boundary 
conditions and (+) for Neumann boundary conditions. 
The constant $\cK$ depends on the  geometry of the boundary.
For boundaries with smooth arcs of length $\gamma_i$ and corners of 
angle $0<\alpha_j\leq 2\pi$ it reads
\begin{equation}
\label{constantK}
\cK=\frac{1}{24}\left(\frac{\pi}{\alpha_j}-\frac{\alpha_j}{\pi}\right)
+\sum_i\int_{\gamma_i}\frac{\kappa(l)}{2\pi}dl,
\end{equation}
where $\kappa(l)$ is the curvature measured along the arc.

The Weyl expansion \eqref{dvpweyl} shows that if two billiards have the
same spectrum, then they necessarily have the same area and the same
perimeter. Furthermore, a certain combination of the properties of
their angles and curvatures must be the same. In the case of polygonal
isospectral billiards, such as those given in the examples
in Appendix \ref{gallery}, the fact that $\cK$ must be the
same entails that a certain relation between the
angles $\alpha_i$ of the first billiard and  the
angles $\alpha_i'$ of the first billiard must hold, namely
\begin{equation}
\sum_{\textrm{first billiard }}\left(\frac{\pi}{\alpha_i}-\frac{\alpha_i}{\pi}\right)=
\sum_{\textrm{second billiard }}\left(\frac{\pi}{\alpha_i'}-\frac{\alpha_i'}{\pi}\right).
\end{equation}

\subsection{Periodic orbits}
\label{periodicorbits}
The previous section gives necessary relations that must hold
between two isospectral billiards, in particular the fact that they
must have the same area and perimeter. These relations were based on
the fact that the mean density of quantum eigenvalues (or the mean counting
function) could be related to classical features of the billiards. In
fact deeper relations exist between the quantum properties of a
billiard and its classical features. These relations are expressed
through ``trace formulas'', which express the density of energy levels
as a sum over classical trajectories, in the semiclassical approximation.
Semiclassical methods are based on the fact that the classical limit
of quantum mechanics is obtained for $\hbar\to 0$ in the path integral
expressing the propagator. The expansion of this path integral in powers of 
$\hbar$ allows us to calculate the sequence of quantum corrections to
classical theory. The semiclassical approximation keeps in this
expansion only the lowest-order term in $\hbar$. Corrections to this
approximation correspond to taking into account higher-order terms.
In this section we recall the main steps leading to a trace formula
for billiards, and apply it to isospectrality.

\subsubsection{Green function}
The propagator of the system is defined as the conditional probability
amplitude  $K(q_f, t_f; q_i, t_i)$ for the particle to be at point
$q_f$ at time $t_f$, if it was at point $q_i$ at time
$t_i$. The propagator is the only solution of the Schr\"odinger equation
that satisfies the condition
\begin{equation}
\lim_{t_f\to t_i} K(q_f, t_f; q_i, t_i)=\delta(q_f-q_i).
\end{equation}
One can then show that the
propagator can be written as a Feynman integral
\begin{equation}
\label{feynmanpath}
 K(q_f, t_f; q_i, t_i)=\int\cD q(t) e^{\frac{i}{\hbar}
\int dt L(\dot{q}, q, t)},
\end{equation}
where the sum runs over all possible trajectories going from $(q_i, t_i)$
to $(q_f, t_f)$ and $L$ is the Lagrangian. The notation (\ref{feynmanpath}) has to be understood as
the limit as $n$ goes to infinity of a discrete sum over all $n$ step paths 
going from $(q_i, t_i)$ to $(q_f, t_f)$: the integral
 (\ref{feynmanpath}) runs over all continuous, but not necessarily
derivable, paths. One immediately sees that the classical limit of
quantum mechanics corresponds to letting the constant 
$\hbar$ go to $0$: the main contributions to the probability $K$
then correspond to stationary points of the action 
$\int dt L(\dot{q}, q, t)$  \cite{FeyHib}.

The advanced Green function is the Fourier transform of the propagator, which is defined by
\begin{equation}
\label{defgreen}
G(q_f,q_i;E)=\frac{1}{i\hbar}\int_{0}^{\infty}dt\  K(q_f, t; q_i, 0)\ e^{iEt/\hbar}.
\end{equation}
It is a solution of
\begin{equation}
\label{caracgreen}
(-H+E)G(q_f, q_i; E)=\delta(q_f-q_i).
\end{equation}
The action along a trajectory can be defined as the integral of the momentum
\begin{equation}
\label{sint}
S(q_f, q_i; E)=\int_{q_i}^{q_f} p\ dq,
\end{equation}
and the Green function as
\begin{equation}
\label{greenchemins}
G(q_f,q_i;E)=\frac{1}{i\hbar}\int\cD q(t) e^{\frac{i}{\hbar}
S(q_f, q_i; E)},
\end{equation}
where the path integral now runs over all continuous 
paths going from $q_i$ to $q_f$ at a given energy $E$.

In many cases Eq.~\eqref{caracgreen} allows us to calculate the Green
function. In the case of free motion in Euclidean space, the
Hamiltonian reduces to the Laplacian (up to a sign), and the Green
function is solution of
\begin{equation}
(\Delta_{q_f}+E)G(q_f,q_i; E)=\delta(q_f-q_i),
\end{equation}
where the $q_f$ index recalls that the derivatives of the Laplacian
are applied on variable $q_f$. In two dimensions, the Green function reads
\begin{equation}
G(q_f,q_i; E)=\frac{1}{4i}H_{0}^{(1)}(k|q_f-q_i|)
\end{equation}
with $k=\sqrt{E}$ and $H_{0}^{(1)}$ the Hankel function of the first
kind.

\subsubsection{Semiclassical Green function}
\label{scgreen}
The expression $(\ref{greenchemins})$ for the Green function 
$G(q_f,q_i; E)$ is a sum over all continuous paths joining $q_i$
to $q_f$ at energy $E$. The semiclassical approximation
consists in keeping only the lowest-order term in the $\hbar$ expansion.
This term is given by the stationary phase approximation. The only paths
contributing to integral $(\ref{greenchemins})$ are paths for which the
action $S$ reaches a stationary value, that is,
 paths that correspond to classical 
trajectories. The semiclassical Green function can thus be expressed
as a sum, over all classical trajectories. Each term in the sum is an exponential
whose phase is given by the classical action integrated along the
trajectory. The prefactor is obtained by the stationary-phase
approximation around the classical trajectory.

Choosing a coordinate system $(q_{\parallel}, q_{\perp})$ such that
$q_{\parallel}$ is the coordinate along the trajectory
and $q_{\perp}$ is the coordinates perpendicular to the trajectory, one
obtains the semiclassical Green function as a sum over all classical
trajectories \cite{Gut89}:
\begin{eqnarray}
\label{greengutzwiller}
G^{\scl}(q_f,q_i; E)&=&\sum_{\textrm{cl}}\frac{2\pi}{(2i\pi\hbar)^{(N+1)/2}}\\
&\times&\left[\frac{1}{\dot{q}_{i_{\parallel}}\dot{q}_{f_{\parallel}}}
\det\left(-\frac{\partial^2 S}{\partial q_{f_{\perp}}\partial q}_{i_{\perp}}\right)
\right]^{1/2}\nonumber\\
&\times&\exp\left(\frac{i}{\hbar}S(q_f, q_i; E)-i\mu
\frac{\pi}{2}\right),\nonumber
\end{eqnarray}
where $N$ is the space dimension. The phase $\mu$ is called
the Maslov index of the trajectory. In two dimensions for
hard wall reflections, each reflection of the classical orbit 
yields a contribution 
$\mu=2$ for Dirichlet boundary conditions and $\mu=0$ for Neumann or
periodic boundary conditions.

\subsubsection{Semiclassical density of eigenvalues}
\label{formuledetraces}
We defined the Green function of a quantum system by
Eq.~\eqref{defgreen}. It will be more useful to express the Green function
as a sum over eigenvalues and eigenfunctions of the Hamiltonian. It can be verified that formally
\begin{equation}
\label{greensomme}
G(q_f,q_i; E)=\sum_{n}\frac{\overline{\Psi}_{n}(q_i)\Psi_{n}(q_f)}{E-E_n},
\end{equation}
where $\overline{\Psi}$ denotes the complex conjugate of $\Psi$,
is a solution of Eq.~(\ref{caracgreen}). 
In order to give a mathematically correct meaning to this expression,
we use the advanced Green function
\begin{equation}
G_{+}(q_f,q_i; E)=G(q_f,q_i; E+i\epsilon).
\end{equation}
The words ``Green function'' will always implicitly refer to the
limit of the advanced Green function for $\epsilon\to 0$. 
The density of energy levels
$(\ref{densityofstates})$ can be related to the Green function by
\begin{equation}
\label{formuledetrace}
d(E)=-\frac{1}{\pi}\int\im G(q, q;E)\ dq.
\end{equation}
To prove this, we use the fact that for $\epsilon\to 0$, 
\begin{equation}
\lim_{\epsilon\to 0}\frac{1}{x+i\epsilon}=\mathrm{P}\frac{1}{x}-i\pi\delta(x)
\end{equation}
(P denotes the principal value and $\delta$ is the Dirac delta function), 
and that, since $H$ is Hermitian, its eigenvectors verify
$\int\overline{\Psi}_m\Psi_n=\delta_{mn}$. 
The Green function $G(q', q;E)$ diverges for $q'\to q$ but not its
imaginary part. The expression $\im G(q, q;E)$ has to be understood as the
imaginary part of $G(q',q)$ taken at the limit $q'\to q$.
Thanks to this relation, the density of states can be expressed as the
trace of the Green function. Equation \eqref{formuledetrace} is the
starting point of trace formulae. Note that, if the density of states
(\ref{densityofstates}) is regularized as a sum of Lorentzians
\begin{equation}
d_{\epsilon}(E)=\frac{\epsilon}{\pi}\sum_{n}\frac{1}{(E-E_n)^2+\epsilon^2},
\end{equation}
one gets
\begin{equation}
\label{formuledetraceeps}
d_{\epsilon}(E)=-\frac{1}{\pi}\int\im G(q, q;E+i\epsilon)\ dq.
\end{equation}
Equation (\ref{formuledetrace}) must therefore be understood as the
limit, as $\epsilon\to 0$, of each member of Eq.~(\ref{formuledetraceeps}).
However, the density of states is usually calculated from the Green
function by first evaluating the integral for $q=q'$
(the ``trace'' of the Green function), and then taking the imaginary
part. This can be made rigorous, by multiplying the Green function by some
factor making the integral convergent in the limit  $q=q'$ \cite{BalBlo74}.\\

The semiclassical density of states is then obtained by use of
Eq.~\eqref{greengutzwiller} with $q_i$=$q_f$.
The density of states in the semiclassical approximation is then the sum
of a ``smooth part'' and an oscillating term that is a superposition
of plane waves,
\begin{equation}
d^{\scl}(E)=\bar{d}(E)+d^{\mathrm{osc}}(E).
\end{equation}
The term $\bar{d}$ is obtained from the first term \eqref{weylexp} of
the Weyl expansion. It gives a mean density of states equal to 
\begin{equation}
\label{bard}
\bar{d}=\frac{\cA}{4\pi}.
\end{equation}
The oscillating term reads
\begin{equation}
\label{dtracegutzwiller}
d^{\mathrm{osc}}(E)\simeq\frac{i}{(2i\pi\hbar)^{3/2}}\sum_{\textrm{pp,n}}
\frac{T_p}{|\det(M_p^{n}-\mathbb{I})|^{\frac{1}{2}}}
e^{i n\left(\frac{S_p}{\hbar}-\nu_p \frac{\pi}{2}\right)}+\cc
\end{equation}
The Gutzwiller trace formula (\ref{dtracegutzwiller}) is a sum over all primitive 
periodic orbits (pp), repeated $n$ times. Each primitive periodic orbit
has a certain action $S_p$, period $T_p$, monodromy matrix $M_p$, and
Maslov index $\nu_p$ (taking into account additional phases owing to 
integration). The identity matrix is denoted by $\mathbb{I}$, 
and c.c. denotes the complex conjugated.

In the case of integrable and pseudo-integrable systems (such as the isospectral pair of Fig.~\ref{celebrated}), 
periodic orbits are no longer isolated but appear within families
of parallel trajectories having the same length (``cylinders of periodic orbits''). 
The Gutzwiller trace formula no longer applies. Pseudo-integrable
billiards are both non-integrable and non-chaotic,
and their classical characteristics are intermediate between those of
integrable and those of chaotic billiards. Classical
trajectories appear within families of parallel trajectories of same
length, but nevertheless the equations of motion are not exactly
solvable because of the presence of diffraction corners.
\textcite{BerTab76} derived a trace formula for
multidimensional integrable systems that can be adapted to polygonal billiards.
In the case of a two-dimensional polygonal billiard, the trace formula becomes
\begin{equation}
\label{dtraceberry}
d^{\mathrm{osc}}(E)\simeq\sum_{\textrm{pp}}\frac{\cA_p}{2\pi}
\sum_{n=1}^{\infty}\frac{e^{i k n l_{pp}-3i\pi/4-i n \nu_{pp} \pi/2}}
{\sqrt{8\pi k n l_{pp}}}+\cc,
\end{equation}
where $\cA_p$ is the area occupied by the cylinder of periodic orbits
labeled by $p$. Equation \eqref{dtraceberry} gives us
a strong relationship between periodic orbits of billiards having the
same spectrum. The trace formulas must be the same, and one might
think that the equality of the sums over periodic orbits can be
achieved only if the periodic orbits are identical in the two billiards.\\

\begin{figure}[ht]
\begin{center}
\includegraphics[width=0.88\linewidth]{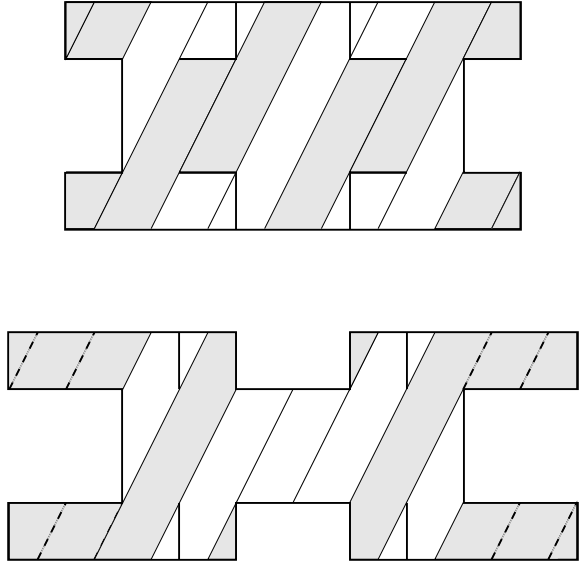}
\end{center}
\caption{Periodic and diffractive orbits in the unfolded pair of Fig.~\ref{squarebaseunfolded}.
\label{squarebaseunfolded_podo}}
\end{figure}

It turns out that this is true.
It can be proved fairly easily that two transplantable isospectral 
domains have the same length spectrum (i.e. both domains have periodic
orbits of the same length) \cite{OkaShu}.  The proof is given in
section \ref{okashuisolength}. Here we illustrate this fact on a
simple example. Consider the billiards of
Fig.~\ref{squarebaseunfolded}. It is possible to encode
any trajectory drawn on the billiard (provided it does not pass through 
vertices) by symbolic dynamics. Consider a trajectory $\cT_{ij}$ drawn on the
first billiard, going from tile $i$ to tile $j$.
Recall that the way the building blocks are glued together (or, equivalently, the
coloring of the associated graph) can be described by matrices 
$M^{(\mu)}, N^{(\mu)}$, $1\leq\mu\leq 3$, as introduced in section \ref{pedestrian}.
With the trajectory $\cT_{ij}$ one can associate a ``word'' $(a_1, a_2, \ldots, a_n)$ 
describing the sequence of edges crossed by the trajectory on its
way. To this trajectory we then associate the matrix $M=\prod M^{(a_i)}$.
Note that there exists a trajectory between tiles $i$ and $j$ if and only if
$M_{ij}=1$. We then define  $N=\prod N^{(a_i)}$.
The transplantation between the two billiards can be described by some
matrix $T$ such that $T M^{(\mu)}=N^{(\mu)} T$ for $1\leq\mu\leq
3$. These commutation relations imply that  $T M=N T$ also holds. 
In particular, if $k$ is a tile of the second billiard such that
$T_{ki}=1$ and $k'$ a tile of the second billiard such that
$T_{k'j}=1$, then $(TM)_{kj}=1=(NT)_{kj}$, which implies that
$N_{kk'}=1$. This is exactly equivalent to saying that the trajectory
$\cT_{ij}$ can be drawn on the second billiard between tiles $k$ and $k'$.
Fig.~\ref{squarebaseunfolded_podo} shows two pencils of 
periodic orbits on each billiard. One can check that these two pencils 
appear with the same length and the same width in both billiards.

\subsection{Diffractive orbits}
\label{diffraction}
The semiclassical trace formula \eqref{dtracegutzwiller}, which is
expressed in terms of classical periodic orbits,
is only a leading-order approximation for small values of $\hbar$.
Higher-order corrections to this formula take into account 
contributions from diffractive orbits: creeping trajectories,
trajectories between scattering points \cite{VatWirRos94, Kel62, PavSch95}, and
orbits almost tangent to a concave section of the boundary \cite{PriSchSmi97}.
In the case of polygonal billiards, the semiclassical trace
formula \eqref{dtraceberry} has to be corrected to take
into account scattering trajectories, that is, classical 
trajectories going from one scattering point to another, or combinations 
thereof.

As  in the case of
periodic orbits, one might believe that the equality of
densities $d(E)$ for isospectral billiards must translate to an
equality of diffractive orbits. Surprisingly, this is not the
case, as we now show. Again, we concentrate on the simple
example of polygonal isospectral billiards. 

In the case of polygonal billiards, \textcite{HanTha}
were able to derive an exact expansion for the Green function, 
as a sum over all scattering trajectories. The exact Green function
between a point $a$ and a point $b$ reads
\begin{eqnarray}
\label{greenstovicek}
G(a, b)&=&\sum_{n=0}^{\infty}\frac{1}{(2\pi)^n}\sum_{\genfrac{}{}{0pt}{}{n\ \textrm{vertex}}{\textrm{paths}}}
\frac{1}{2i}\int_{-\infty}^{\infty}ds_1 ds_2 ... ds_n\nonumber\\
&\times& H_0^{(1)}\left[k R(s_1, s_2, ..., s_n)\right]\nonumber\\
&\times&\prod_{k=1}^{n}\frac{2\pi}{(\gamma_k M_k+\theta_k+i s_k)^2-\pi^2},
\end{eqnarray}
where
\begin{eqnarray}
&R^2(s_1, s_2, ..., s_n)=\hspace{4cm}\\
&\left(r_0+r_1 e^{s_1}+r_2 e^{s_1+s_2}+\cdots+r_n e^{s_1+s_2+\cdots+s_n}\right)\nonumber\\
&\times\left(r_0+r_1 e^{-s_1}+r_2 e^{-s_1-s_2}+\cdots+r_n e^{-s_1-s_2-\cdots-s_n}\right).\nonumber
\end{eqnarray}
The Green function appears as a sum over paths made of $n+1$ straight
lines of length $r_i$, $0\leq i\leq n$. The first line goes from point
$a$ to a diffracting corner, then there are $n$ scattering
trajectories going from one diffracting corner to another, and finally a
trajectory going from one diffracting corner to point $b$. The diffraction angles
are $M_k \gamma_k+\theta_k$, $1\leq k\leq n$, with $\gamma_k$
the measure of the angle at the diffracting corner and $M_k$ the number
of times the path winds around the diffracting corner (thus, 
$0\leq \theta_k<\gamma_k$). 

\textcite{Gir04} showed, using the expansion
\eqref{greenstovicek} of the Green function, that isospectral domains can be
distinguished by the fact that in general the lengths of their 
diffractive orbits differ. This can be illustrated  in the
case of the billiard with rectangular base tile unfolded to a 
translation surface (Fig.~\ref{squarebaseunfolded}). If the sides
of the base tiles are incommensurate, then there cannot be diffractive
orbits of the same length as a given diffractive
orbit but in the same direction in the plane. For instance, for the dashed
 diffractive orbit drawn in the second billiard of 
Fig.~\ref{squarebaseunfolded_podo}, orbits starting from a diffractive
corner of the first billiard in the same direction never reach another
diffractive corner. This means that the dashed orbit has no partner in
the first billiard.

\begin{figure}[ht]
\begin{center}
\includegraphics[width=0.88\linewidth]{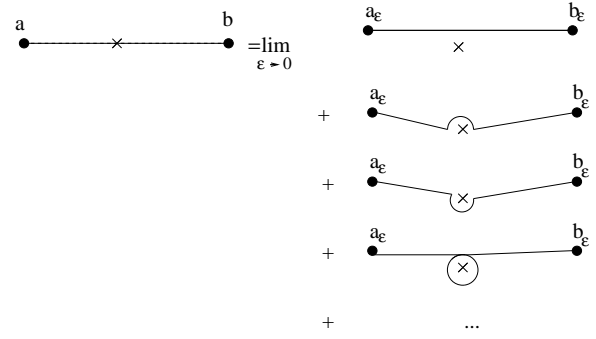}
\end{center}
\caption{A contribution to the Green function in the case of forward diffraction.
If the orbit goes through a vertex, the term in Eq.~\eqref{greenstovicek} should 
be interpreted as the limit for $\epsilon\to 0$ of an infinite number
of trajectories. If there is no vertex only the straight path contribution
remains, the other (winding) terms add up to zero.\label{abscat}}
\end{figure}

\begin{figure}[ht]
\begin{center}
\includegraphics[width=0.98\linewidth]{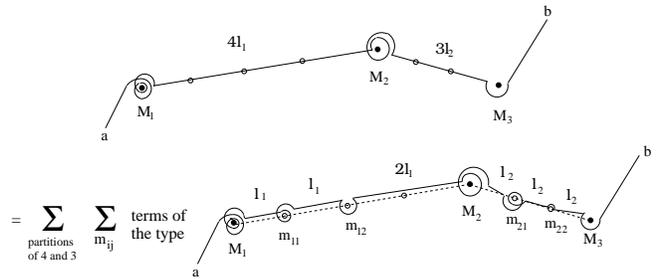}
\end{center}
\caption{A contribution to the Green function in the case of forward diffraction.
 The filled circles are scattering vertices; the empty ones are 
non-scattering vertices.\label{pathinfini}}
\end{figure}

The connection between the energy spectrum and the length spectrum through the
trace formula indicates, however, that these discrepancies between
diffractive orbits must be compensated in a certain way.
This compensation can be understood by analyzing the formula of Hannay and Thain 
\eqref{greenstovicek}. In fact each contribution to the Green function
in Eq.~\eqref{greenstovicek} has to be understood as an infinite sum over
all windings around vertices (see Fig.~\ref{abscat}). Here by vertices we
mean the four corners and the two points 
at the middle of the horizontal sides of each of the seven rectangular 
tiles in Fig.~\ref{squarebaseunfolded}.
If there is a diffracting corner (as is the case for instance at the bottom right 
corner of tile 7 in the second billiard of Fig.~\ref{squarebaseunfolded})
then there is a non-zero contribution, while if there is no scatterer
(e.g.~ at the bottom left corner of tile 7 in the second billiard)
the series of diffractive terms adds up to zero,
\begin{equation}
\sum_{M_k=-\infty}^{\infty}\frac{2\pi}{(2\pi M_k+\pi+i s_k)^2-\pi^2}=0.
\end{equation}
As a consequence, a diffractive contribution to the Green function,
 going from a point $a$ to a point $b$ through possibly several vertices, 
has to be understood as a sum of trajectories
winding around both scattering and non-scattering vertices (see 
Fig.~\ref{pathinfini}). Now each of these new ``fictitious'' trajectories
avoids vertices (since they wind around). The reasoning we used in
 section \ref{periodicorbits} applies: With any trajectory one
 can associate a matrix $M$ describing the edges crossed by the
 trajectory. The matrix $N$ corresponding to $M$ in the other billiard
 is such that  $T M=N T$, and a partner of the diffractive orbit can be
 found between tiles $i$ and $j$ such that $N_{ij}=1$.

Thus, even though diffractive orbit lengths might differ, each
``expanded'' diffractive orbit indeed has a partner of the same length.

\subsection{Green function}
This relation between diffractive orbits translates to a relation
between Green functions of the two domains \cite{Gir04}.
The matrices $M, N$ introduced in section \ref{periodicorbits}
verify the property
\begin{equation}
\label{identite}
\sum_{i', j'}T_{i i'}T_{jj'}M_{i'j'}=1+2 N_{ij},
\end{equation}
which can be proved using the commutation relation \eqref{commute} 
and the fact that $(T^2)_{ij}=1+2\delta_{ij}$.
Thus, in the expansion \eqref{greenstovicek} of the Green function 
between a point in tile $i$ and a point in tile $j$ in the first (second) billiard, 
each trajectory appears with a weight $M_{ij}$ ($N_{ij}$). 
But according to Eq.~\eqref{identite} we have
\begin{equation}
\label{mbma}
N_{ij}=\frac{1}{2}\sum_{i', j'}T_{i i'}T_{jj'}M_{i'j'}-\frac{1}{2}.
\end{equation}
Therefore from Eq.~\eqref{greenstovicek} and identity \eqref{mbma} one can
infer a relation between Green functions, namely,
\begin{equation}
\label{relationG}
G^{(B)}(a,i;b,j)=\frac{1}{2}\sum_{i',j'}T_{ii'}T_{jj'}G^{(A)}(a,i';b,j')
- \frac{1}{2}G^{(t)}(a;b),
\end{equation}
where $G^{(t)}(a;b)$ is the Green function on the base tile.
This relation between Green functions like the relations between
periodic orbits or diffractive orbits are all consequences of the
transplantation property which is the fundamental feature of all known
examples of isospectral billiards.

\subsection{Scattering poles of the exterior Neumann problem}
\label{fredholm}
In section \ref{diffraction} we considered the particular case of
polygonal isospectral billiards, for which it is possible to express
the exact Green function as an infinite expansion given by
Eq.~\eqref{greenstovicek}.  
In a more general setting, it is also possible to express the 
Green function of the billiard with Dirichlet boundary conditions as an infinite
sum taking into account all possible reflections on obstacles.
\textcite{BalBlo74} gave a general method, called 
``multiple reflection expansion'', which gives the Green function  
in terms of the free Green function $G_0$. Applied to a two-dimensional
billiard, this expansion is given by
\begin{eqnarray}
\label{dvpbalblo}
G(q, q';E)&=&G_0(q, q';E)\\
&-&2\int_{\partial B}ds\  
G_0(q,s;E)\partial_s G_0(s,q';E)\nonumber\\
&+&(-2)^2\int_{\partial B}ds\ ds'\
\partial_{s}G_0(q,s;E\nonumber)\\
&\times&
\partial_{s'}G_0(s,s';E)G_0(s',q';E)+\cdots,\nonumber
\end{eqnarray}
where $s$ and $s'$ are points along the boundary, and $\partial_x$ denotes the 
derivative along an outward vector normal to the boundary at point $x$.
The first term $G_0(q, q';E)$ on the right-hand side of Eq.~(\ref{dvpbalblo}) corresponds
to direct (free) propagation from $q$ to $q'$, the first integral corresponds to
trajectories from $q$ to $q'$ with one reflection on the boundary at
point $s$, and so on. We introduce the kernel $K_E(q,q')=-2\partial_{q'} G_0(q,q'; E)$, 
which is a continuous infinite-dimensional operator defined on $\partial
B\times\partial B$ ($\partial B$ is the boundary of the billiard).
One can express Eq.~\eqref{dvpbalblo} as 
\begin{eqnarray}
\label{dvpfredholm}
G(q, q';E)=G_0(q, q';E)\hspace{5cm}\\
-2\sum_{n=0}^{\infty}\int_{\partial B}ds\ ds'\
G_0(q,s;E)K_E^n(s,s')\partial_{s'}G_0(s',q';E).\nonumber
\end{eqnarray}
Formal performance of the sum over $n$ yields the infinite-dimensional operator 
$(\mathbb{I}-K_E)^{-1}$, where $\mathbb{I}$ is the identity operator. Fredholm theory \cite{Smi} 
showed that for
sufficiently ``nice'' billiards the operator $(\mathbb{I}-K_E)^{-1}$ is well-defined
and can be expressed as
\begin{equation}
(\mathbb{I}-K_E)^{-1}=\frac{N_E}{D(E)},
\end{equation}
where $D(E)$ is the Fredholm determinant $\det(\mathbb{I}-K_E)$, and 
$N_E$ is an infinite-dimensional operator defined on $\partial
B\times\partial B$.
The Fredholm determinant admits an expansion
\begin{equation}
D(E)=\sum_{n=0}^{\infty}D_n(E),
\end{equation}
with $D_0(E)=1$, and for $n\geq 1$
\begin{equation}
D_n(E)=\frac{(-1)^n}{n!}\int_{\partial B}dq_1\ldots\int_{\partial
  B}dq_n K\left(\bfq, \bfq\right)
\end{equation}
($\bfq$ is the vector $(q_1,q_2,\ldots,q_n)$). We have introduced the determinant
\begin{equation}
K\left(\bfq, \bfq'\right)=
\left|\begin{array}{cccc}K_E(q_1,q_1')&K_E(q_1,q_2')&\cdots&K_E(q_1,q_n')\\
K_E(q_2,q_1')&K_E(q_2,q_2')&\cdots&K_E(q_2,q_n')\\
\cdots&\cdots&\cdots&\cdots\\K_E(q_n,q_1')&K_E(q_n,q_2')&\cdots&K_E(q_n,q_n')
\end{array}\right|.
\end{equation}
The operator $N_E$ is defined on $\partial B\times\partial B$ by its expansion
$N_E=\sum_{n=0}^{\infty}N_n$ with
\begin{equation}
N_n=\sum_{k=0}^{n}D_k(E)K_E^{n-k}.
\end{equation}
The Fredholm determinant $D(E)$ appearing in the expression of the
Green function has the property that it has zeros at
eigenvalues of the system \cite{GeoPra}. A natural question is whether
isospectral billiards share the same Fredholm determinant.

It has been shown by \textcite{TasHarShu}
that, for billiards with $C^2$ boundary, $D(E)$ can be decomposed 
into an interior and an exterior
contribution, namely $D(E)=D(0)d_{\textrm{int}}(E)d_{\textrm{ext}}(E)$. The exterior
contribution $d_{\textrm{ext}}(E)$ is related to the scattering of a wave on 
an obstacle having the shape of the billiard with
Neumann boundary conditions, i.e. the zeros of its analytic continuation
are resonances of the exterior scattering problem. The interior
contribution reads
\begin{equation}
d_{\textrm{int}}(E)=e^{i\frac{\cA E}{4}}\left(\frac{\cL^2
  E}{4}\right)^{-\frac{\cA E}{4\pi}}e^{-\frac{\cA\gamma E}{2\pi}}\prod_{n=1}^{\infty}\left(1-\frac{E}{E_n}\right)e^{E/E_n},
\end{equation}
where $\cA$ and $\cL$ are the area and the perimeter of the billiard, respectively,
and $\gamma$ is a constant depending on the geometry of the
billiard. The zeros of $d_{\textrm{int}}(E)$ are thus the eigenenergies of
the interior Dirichlet problem. 

Obviously, isospectral billiards share the same interior part
$d_{\textrm{int}}(E)$. But the exterior part depends
on the shape of the billiard. In particular solutions
of the exterior Neumann scattering problem may differ between two
isospectral billiards. Therefore a conclusion of  \cite{TasHarShu} is
that isospectral pairs might be distinguished by measuring the sound
scattered by them.\\

To check this property, numerical investigations were performed by 
\textcite{OkaShuTasHar}. In fact, Fredholm theory applies only for billiards
with a smooth boundary, which is not the case for any of the known examples
of isospectral pairs.
For billiards with a piecewise smooth boundary, it is however possible to approximate
the Fredholm determinant $D(E)$ by a discretized version $D^{m}(E)$, depending on the number $m$ of
points taken on the boundary of the billiard, which converges to $D(E)$ for large
$m$. This convergence fails for boundaries with corners: All $D^{m}(E)$
tend to 0. Nevertheless, for domains with corners \cite{OkaShuTasHar05b} showed that 
it is possible to define a regularized version of $D^{m}(E)$
that converges to $D(E)/D(0)=d_{\textrm{int}}(E)d_{\textrm{ext}}(E)$.
Using this regularized version, \textcite{OkaShuTasHar} computed numerically
zeros of the regularized Fredholm determinant for various pairs of isospectral billiards. 
It was observed that zeros of the determinant close to the real axis coincide, as they should
since they are eigenvalues of the interior problem.
On the other hand, complex zeros (remote from the real axis), which correspond to 
resonances of the exterior Neumann problem, are shown to differ. 
To quantify this discrepancy between the resonances of the two billiards, the 
resonance counting number
\begin{equation}
N_{\delta}(r)=\left\{z\in\mathbb{C}; |z|<r, -\frac{\pi}{2}<\arg(z)<-\delta\right\}
\end{equation}
was studied by \textcite{OkaShuTasHar}. 
The best fit $N_{\delta}(r)=C_{\delta,R} r^2$, computed over the 
range $r\in [0, R]$, yields noticeably different values of $C_{\delta,R}$ for each billiard.
This clearly shows that isospectral pairs can indeed be distinguished by 
resonances of scattering waves.

%INVERSE PROBLEM?????????????????KOEN: zelditch (cf OkaShuTasHar05a):
%for certain class of planar domains the shape can be determined by its spectrum

\subsection{Eigenfunctions}

\subsubsection{Triangular states}
\label{triangular}
In general, analytical solutions to the Helmholtz equation
$(\Delta+E)\Psi=0$ with Dirichlet boundary conditions cannot be
found. However, it is possible to construct particular solutions of
this equation provided solutions are known on elementary
subdomains. This is, for instance, the case if the subdomains have the
shape of a half-square (billiards of Fig.~\ref{celebrated}), or a
rectangle (billiards of Fig.~\ref{squarebase}).

\begin{figure}[ht]
\begin{center}
\includegraphics[width=0.88\linewidth]{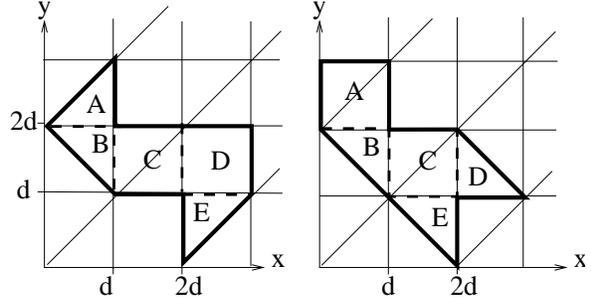}
\end{center}
\caption{Isospectral billiards divided into smaller regions.\label{modematching}}
\end{figure}

We take the example of the two billiards in Fig.~\ref{celebrated}.
Each billiard is made of seven triangular 
(half-square) tiles. 
Eigenfunctions for a $d\times d$--square with Dirichlet boundary conditions
are of the form
\begin{equation}
\label{evs}
s_{m,n}(x,y)=\frac{4}{d^2}\sin\left(\frac{m\pi x}{d}\right)\sin\left(\frac{n\pi y}{d}\right),
\end{equation}
with eigenvalues $E_{m,n}=\pi^2(m^2+n^2)/d^2$, $m,n\geq 1$.
Eigenfunctions for the elementary triangles with Dirichlet boundary conditions
are obtained from \eqref{evs} by antisymmetrization with respect to
the diagonal,
\begin{eqnarray}
\label{evt}
t_{m,n}(x,y)&=&\frac{4}{d^2}\left[\sin\left(\frac{m\pi x}{d}\right)
\sin\left(\frac{n\pi y}{d}\right)\right.\nonumber\\
&-&\left.\sin\left(\frac{m\pi
  y}{d}\right)\sin\left(\frac{n\pi x}{d}\right)\right],
\end{eqnarray}
and the corresponding eigenenergies are given by $\pi^2(m^2+n^2)/d^2$,
$m>n$. For the sake of definiteness, we consider the two isospectral
pairs on a Cartesian reference frame, following \cite{WuSprMar}, as in Fig.~\ref{modematching}.
The functions
$t_{m,n}$ turn out to be elementary solutions of the Helmholtz
equation for both isospectral billiards of Fig.~\ref{modematching}.
Indeed $t_{m,n}$ vanishes on all lines $x=k d$, $y=k d$, $y=x+ 2 k d$ and
$y=-x+2 k d$, $k\in\mathbb{Z}$, which are precisely the lines on which the boundaries
of both billiards lie (in the convention of Fig.~\ref{modematching}).
The particular solutions $t_{mn}$ are called ``triangular states''. An
example of the lowest-energy triangular state is given in Appendix
\ref{gallery}.

The labels  of the lowest-energy triangular states among the eigenvalues 
$E_1\leq E_2\leq\ldots$ of the billiards 
have been calculated by \textcite{GotMcM}. The results are 
displayed in Table \ref{triangularmodes}. 
\begin{table}
\begin{center}
\begin{tabular}{|c|c|c||c|c|c|}
m & n & Eigenvalue & m & n & Eigenvalue\\
1 & 2 & $E_9$ & 0 & 1 & $E_5$  \\
1 & 3 & $E_{21}$ & 1 & 1 & $E_9$  \\
2 & 3 & $E_{27}$ & 0 & 2 & $E_{15}$ \\
1 & 4 & $E_{38}$ & 1 & 2 & $E_{20}$ \\
2 & 4 & $E_{44}$ & 2 & 2 & $E_{29}$ \\
\end{tabular}
\caption{First triangular modes $t_{m,n}$. Left: Dirichlet boundary
  conditions. Right: Neumann boundary conditions.\label{triangularmodes}}
\end{center}
\end{table}
Each integer pair $(m,n)$, $m>n$, defines a 
triangular state $t_{m,n}$. Obviously, the fact that an integer can be
represented in more than one way as a sum of two squares leads to
degeneracies for triangular states, and hence for the isospectral pairs of Fig.~\ref{celebrated}.

Note that for Neumann boundary conditions, it can be easily checked that the functions
\begin{eqnarray}
u_{m,n}(x,y)&=&\frac{4}{d^2}\left[\cos\left(\frac{m\pi x}{d}\right)
\cos\left(\frac{n\pi y}{d}\right)\right.\nonumber\\
&+&\left.
\cos\left(\frac{m\pi y}{d}\right)\cos\left(\frac{n\pi x}{d}\right)\right]
\end{eqnarray}
for $0\leq m\leq n$, $(m,n)\neq (0,0)$, have a normal derivative that
vanishes on all lines $x=k d$, $y=k d$, $y=x+2kd$, and
$y=-x+2k d$, $k\in\mathbb{Z}$. Therefore $u_{m,n}$ are solutions of Helmholtz
equations for the billiards of Fig.~\ref{modematching} with Neumann boundary 
conditions. Their label among the eigenstates of the billiards is given
in Table \ref{triangularmodes} \cite{GotMcM}.

\subsubsection{Mode-matching method}
\label{modematchingsection}
\begin{figure}[ht]
\begin{center}
\includegraphics[width=0.98\linewidth]{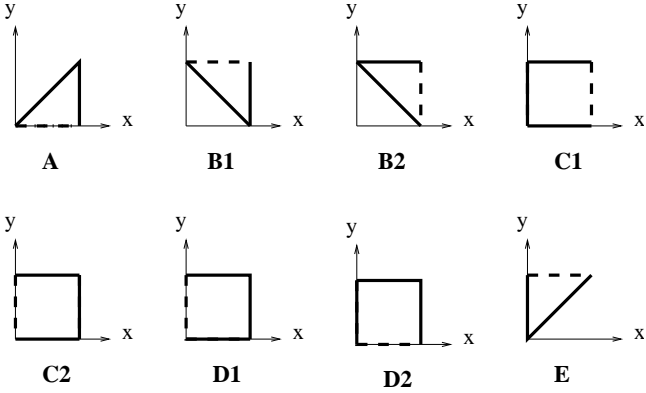}
\end{center}
\caption{Elementary regions building the isospectral pairs.\label{briqelem}}
\end{figure}

The knowledge of these particular triangular states is the starting point
for the so-called ``mode-matching method''.
It consists in dividing the billiards into
subdomains for which solutions of the Helmholtz equation are known
analytically. Consider for example the left billiard of Fig.~\ref{modematching}. It is made
of five elementary domains, three triangles $A, B, E$ and two squares $C, D$.
For each subdomain,
analytical solutions for the Dirichlet problem are given by (translations of) functions 
\eqref{evs} or \eqref{evt}. 
We define the function $\psi_n(x,y)=\sin(a_n x)\sin(b_n y)/\sin(b_n d)$, where
$d$ is the length of the side of the elementary square, 
and we have set $a_n=n\pi/d$ and $b_n=\sqrt{E-a_n^2}$.
If given boundary conditions are
imposed on the boundaries of these subdomains, as in Fig.~\ref{briqelem}, 
solutions can be written explicitly for these elementary subdomains as
superpositions of functions obtained from translations or reflections of $\psi_n$.
In particular, one can construct functions taking the value 0 on the plain boundary and 
$\sin(a_n x)$ ($\sin(a_n y)$)  on the dashed horizontal (vertical) boundary for each of the
domains shown in Fig.~\ref{briqelem}. Such functions are given by
\begin{eqnarray}
\Phi^{(A)}_n(x,y)&=&\psi_n(x, d-y)-\psi_n(y, d-x)\\
\nonumber\Phi^{(B1)}_n(x,y)&=&\psi_n(x, y)-\psi_n(d-y, d-x)\\
\nonumber\Phi^{(B2)}_n(x,y)&=&\psi_n(y,x)-\psi_n(d-x, d-y)\\
\nonumber\Phi^{(C1)}_n(x,y)&=&\psi_n(y,x)\\
\nonumber\Phi^{(C2)}_n(x,y)&=&\psi_n(y,d-x)\\
\nonumber\Phi^{(D1)}_n(x,y)&=&\psi_n(y,d-x)\\
\nonumber\Phi^{(D2)}_n(x,y)&=&\psi_n(x,d-y)\\
\nonumber\Phi^{(E)}_n(x,y)&=&\psi_n(x, y)-\psi_n(y, x).
\end{eqnarray}
The mode-matching method consists in looking for a solution $\Psi$ of the Helmholtz 
equation as a superposition of such functions, with
amplitudes chosen such that $\Psi$ and its partial derivatives be continuous at
each boundary between subdomains.
At the boundary between elementary subdomains, the eigenfunction $\Psi$
can be expanded on the functions $\varphi_n(x)=\sin (a_n x)$ as
\begin{eqnarray}
\Psi_{AB}(x,y)&=&\sum_n A_n\varphi_n(x)\\
\nonumber\Psi_{BC}(x,y)&=&\sum_n B_n\varphi_n(y-d)\\
\nonumber\Psi_{CD}(x,y)&=&\sum_n C_n\varphi_n(y-d)\\
\nonumber\Psi_{DE}(x,y)&=&\sum_n D_n\varphi_n(x-2d),
\end{eqnarray}
where the sum goes from 1 to some truncation  number $N$.
The eigenfunction $\Psi$ is entirely determined by knowledge of the vector
${\bf V}=(A_1,\ldots, A_N, B_1,\ldots, B_N, C_1,\ldots, C_N, D_1,\ldots, D_N)$. 
Therefore $\Psi$ can be written as
\begin{eqnarray}
\label{eigenfunctionABCDE}
\Psi_{A}(x,y)&=&\sum_n A_n\Phi^{(A)}_n(x,y-2d)\nonumber\\
\nonumber\Psi_{B}(x,y)&=&\sum_n A_n\Phi^{(B1)}_n(x,y-d)+\sum_n B_n\Phi^{(B2)}_n(x,y-d)\\
\nonumber\Psi_{C}(x,y)&=&\sum_n B_n\Phi^{(C1)}_n(x-d,y-d)\\
&+&\nonumber\sum_n C_n\Phi^{(C2)}_n(x-d,y-d)\\
\nonumber\Psi_{D}(x,y)&=&\sum_n C_n\Phi^{(D1)}_n(x-2d,y-d)\\
&+&\nonumber\sum_n D_n\Phi^{(D2)}_n(x-2d,y-d)\\
\Psi_{E}(x,y)&=&\sum_n D_n\Phi^{(E)}_n(x-2d,y),
\end{eqnarray}
where $\Psi_{X}$ is the restriction of the function $\Psi$ to the elementary
domain $X=A, B, C, D$, or $E$. The function $\Psi$ is indeed 
an eigenfunction of the billiard if its normal derivatives at the boundaries
between domains are continuous. This latter condition can be written as a
system of linear equations that
can be cast under the form $M {\bf V}=0$, where $M$ is a $4N\times 4N$-matrix given by
\begin{equation}
\label{matrixM}
M=\left(\begin{array}{cccc}
U-2W&PWP-PV/2&0&0\\
PWP-PV/2&U-W&-V/2&0\\
0&-V/2&U&W\\
0&0&W&U-PWP
\end{array}\right)
\end{equation}
with $U_{mn}=(b_n\cot b_n d)\delta_{mn}$, $V_{mn}=(b_n/\sin b_n d)\delta_{mn}$,
 $W_{mn}=a_m a_n/(E-a_m^2-a_n^2)$, and $P_{mn}=(-1)^n\delta_{mn}$. 
The matrix $M$ depends on $E$ through $b_n$. Eigenvalues of 
the billiard correspond either to values of $E$ where $\det M=0$ or to
 $V=0$. In the case $V=0$, the wave function vanishes on the boundaries 
between the domains, and the eigenfunction is a triangular state.
If $\det M=0$, Eqs.~\eqref{eigenfunctionABCDE} give the corresponding eigenfunction.\\

Interestingly, the mode-matching method provides an alternative proof
to isospectrality \cite{WuSprMar}. The matrix $M'$
corresponding to $M$ for the right billiard of Fig.~\ref{modematching} is
the  $4N\times 4N$-matrix given by
\begin{equation}
M'=\left(\begin{array}{cccc}
U-W&PWP-PV/2&0&0\\
PWP-PV/2&U-W&PV/2&W\\
0&PV/2&U-W&PWP\\
0&W&PWP&U-W
\end{array}\right).
\end{equation}
It can be easily checked that $M$ and $M'$ are related by
\begin{equation}
\label{mtmt}
M=^{t}TM'T,
\end{equation}
with
\begin{equation}
T=\frac{1}{\sqrt{2}}\left(\begin{array}{cccc}
0&1&0&P\\1&0&P&0\\0&-1&0&P\\-1&0&P&0\end{array}\right).
\end{equation}
If $\Psi$ is a solution of the Helmholtz equation for the first billiard, 
it can be written under the form \eqref{eigenfunctionABCDE} with constants specified by some
vector ${\bf V}$ verifying $M{\bf V}=0$. Let $\Psi'$ be the function defined
on the second billiard by some constants given by the vector 
${\bf V'}=T{\bf V}$. Because of Eq.~\eqref{mtmt} the vector ${\bf V'}$ verifies $M'{\bf V'}=0$, 
and therefore $\Psi'$ is a solution of the Helmholtz equation for the second billiard.
Since the relation between $\Psi$ and  $\Psi'$ is linear, the eigenenergy is the same
for both functions, and thus the billiards are isospectral.

\subsection{Eigenvalue statistics}
As explained in section \ref{pedestrian}, the shape of the elementary
building block of a pair of isospectral billiards can be varied at will
provided some conditions are satisfied.
Thus, examples of chaotic pairs, or pseudo-integrable
pairs, or even pairs with a fractal boundary can be produced.
However, the most popular examples of isospectral billiards, e.g., those
of  Fig.~\ref{celebrated}, are constructed
with a triangular-shaped base tile. The resulting billiards are thus
polygonal billiards. Billiards with a polygonal boundary can display
a whole range of classical behaviors from integrability to chaos.
Isospectral billiards made of tiles whose angles are 
rational multiples of $\pi$ are pseudo-integrable billiards
\cite{BerRic81}. The properties of these billiards were mentioned
in section \ref{transplantationproof}.

In the field of quantum chaos, many works have been concerned with
a characterization of the statistical properties of spectra of billiards. The question of the spectral properties displayed by polygonal isospectral billiards attracted some interest in the literature.
Eigenvalue statistics for the pair of Fig.~\ref{celebrated} have been studied
numerically by \textcite{WuSprMar}, based on the first 598 energy levels.
The short-range correlations of the spectrum were shown to lie
between the random matrix statistics of the Gaussian orthogonal ensemble (GOE) and Poisson statistics (see \cite{Por65} for a review on the seminal papers, and \cite{GuhMulWei98} for a recent review on random matrix theory). 
On removal of the 78 triangular states, it was observed that the nearest-neighbor level spacing distribution function $P(s)$, which characterizes the 
distribution of the spacings between nearest-neighbor energy
levels, agrees with the nearest-neighbor distribution for GOE matrices.
The spectral rigidity $\bar{\Delta}_3(L)$ (see, e.g., \cite{Meh90}
for a rigorous definition) measures the deviation of the integrated density of 
states $N(E)$ (the number of eigenvalues smaller than $E$) from a straight 
line, on an interval $[E-L/2,E+L/2]$. Computation of $\bar{\Delta}_3$ showed that it is also of GOE type for 
these billiards.
\textcite{AurBacSte97} calculated the functions $E(k,L)$, which give the
probability to find $k$ energy levels in an random interval of length $L$ \cite{AurSte90},
for isospectral billiards shaped as in Fig.~\ref{celebrated}, again showing a behavior
that is intermediate between the chaotic and integrable cases.

\subsection{Nodal domains}
\label{nodal}
Nodal lines for two-dimensional billiards are one-dimensional curves
on which eigenfunctions vanish. Nodal domains are connected regions of
the billiard where an eigenfunction has a constant sign. A theorem by 
\textcite{CouHil} states that the $n$th eigenfunction $\Psi_n$ has at most $n$
nodal domains. The number $\nu_n$ of nodal domains in  $\Psi_n$ can be further
estimated \cite{Ple56}. We define a rescaled nodal-domain number
$\xi_n=\nu_n/n\in[0,1]$. If $j_1$ is the first zero of the Bessel function
$J_0$, then $\limsup_{n\to\infty}\xi_n\leq (2/j_1)^2$. 
The limit distribution of $\xi_n$ is defined by
\begin{equation}
P(\xi)=\lim_{E\to\infty}\frac{1}{N_{I_g(E)}}\sum_{E_n\in I_g (E)}\delta\left(\xi-\xi_n\right),
\end{equation}
where $I_g (E)$ is the interval $[E, E+gE]$ for some fixed $g>0$, and
$N_I$ is the number of eigenvalues in the interval $I$. It has been shown 
by \textcite{BluGnuSmi02} that this distribution has universal features.

For some instances of isospectral pairs, such as flat tori in $\mathbb{R}^n$
with $n\geq 4$ \cite{GnuSmiSon} (see also \cite{LevParPol}),
it was conjectured that two isospectral domains produce a different
number of nodal domains (domains separated by nodal lines where $\Psi=0$). Heuristic arguments as well as
numerical investigations were collected by \textcite{GnuSmiSon} to support this conjecture.
A recent solution of this conjecture can be found in \cite{BruKlaPuh}.

\subsection{Isospectrality versus isolength spectrality}
\label{mushroom}

We now consider a related
important question. Since transplantation is a mapping between the two
billiards, the classical properties should map onto one another as
well. Here we investigate the mapping between periodic orbits.

\subsubsection{Okada and Shudo's result on isolength spectrality}
\label{okashuisolength}
Let $D$ be a planar domain obtained by unfolding $N$ times the same triangular building block $B$ with sides $1, 2, 3$. Then the length spectrum is the set of lengths of closed trajectories (periodic orbits) of $D$. Any periodic orbit on $D$ can be regarded as a ``lift'' of a closed trajectory on $B$, because its projection is always a periodic orbit on $B$.
(The converse is, of course, not necessarily true.) One observes that the number of closed lifts of a given closed
trajectory on $B$ is counted as
\begin{equation}
n^D(\gamma) = \mbox{Tr}(M^{(\gamma_m)}M^{(\gamma_{m-1})}\ldots M^{(\gamma_1)}),         
\end{equation}
 where $\gamma = \prod \gamma_i$ ($\gamma_i \in \{1,2,3\}$) denotes the sequence representing the order in which a given closed trajectory on $B$ hits the boundary segments. (The $M^{(\gamma_j)}$'s are adjacency matrices.) Note that such a sequence is not uniquely determined by a given closed orbit | the number of closed lifts, however, is.
So the length spectrum of $D$ is determined by the length spectrum of $B$
and by $n^D(\gamma)$. Hence, if one considers two domains $D$ and $D'$ that
are constructed by unfolding the same building block as above, it is
sufficient to prove that $n^D(\gamma) = n^{D'}(\gamma)$ for all possible
sequences $\gamma$ in order to deduce ``isolength spectrality''. 

 The following is now obvious.

\bt[\textcite{OkaShu}]
Let $D$ and $D'$ be two unfolded domains obtained by $N$ times successive reflections of the same building block. If $D$ and $D'$ are transplantable, 
then $n^D(\gamma) = n^{D'}(\gamma)$ for any sequence $\gamma$, so $D$ and $D'$ are isolength spectral.\eop \\
\et

Let $S$ be a finite set, say, $S = \{a_1,\ldots,a_k\}$ with $k \in \mathbb{N}_0$. 
The {\em free group} $\mathbf{F} = \mathbf{F}(S)$ generated by $S$ is defined as follows. Introduce a set $S^{-1} := \{a_1^{-1},\ldots,a_k^{-1}\}$ which consists of  the ``inverse symbols'' of $S$.
A {\em word} with alphabet $S$ (or $S \cup S^{-1}$) is just a finite sequence of elements of $S \cup S^{-1}$. A {\em reduced word} is a word in which 
any sequence consisting of an element of $S$ and its inverse is deleted. By definition, $\mathbf{F}$ consists of all reduced words with alphabet $S$, together with the empty word.
Group operation is just concatenating words, and reducing if necessary.

\bt[\textcite{OkaShu}]
Let $D$ and $D'$ be two unfolded domains obtained by $N$ times successive reflections of the same building block.
If $n^D(\gamma) = n^{D'}(\gamma)$ for any sequence $\gamma$, then $D$ and $D'$ are transplantable, so also
isospectral.
\et

{\em Proof}.\quad
Let $G$ and $G'$ | corresponding to $D$ and $D'$, respectively | be the groups generated by the adjacency matrices:
\begin{equation}  
G = \langle M^{(\mu)}\rangle, \ \ G' = \langle N^{(\mu)}\rangle;               
\end{equation}
then clearly $G$ and $G'$ are subgroups of the symmetric group $\mathbf{S}_N$ on $N$ letters.
Let $\mathbf{F}_3$ be the free group generated by symbols $a, b$ and $c$. Define the surjective homomorphism
\begin{equation}
\Phi_D: \mathbf{F}_3 \mapsto G: \gamma = \gamma_1\gamma_2\ldots\gamma_m \mapsto M^{(\gamma_m)}M^{(\gamma_{m - 1})}\ldots M^{(\gamma_1)}.
\end{equation}
Then 
\begin{equation}
G \cong \mathbf{F}_3/\mbox{ker}\Phi_D\ \ \mbox{and}\ \ G' \cong \mathbf{F}_3/\mbox{ker}\Phi_{D'},
\end{equation}
the latter notation being obvious.

Now assume that $n^D(\gamma) = n^{D'}(\gamma)$ for any sequence $\gamma$. Then
\begin{eqnarray}
\mbox{ker}\,\Phi_D&=&\{\gamma\parallel \Phi_D(\gamma) = \mathbb{I}\} = \{\gamma\parallel n^D(\gamma) = N\}\nonumber\\
&=& \{\gamma\parallel n^{D'}(\gamma) = N\} 
= \{\gamma\parallel \Phi_{D'}(\gamma) = \mathbb{I}\}\nonumber\\
&=& \mbox{ker}\,\Phi_{D'}. 
\end{eqnarray}
(Note that $\Phi_D(\gamma)$ is a $(0,1)$-matrix, so that $\Phi_D(\gamma) = \mathbb{I}$ if and only if $n^D(\gamma) = N$.)
So the map
\begin{equation}  
\Delta: G \mapsto G': \Phi_D(\gamma) \mapsto \Phi_{D'}(\gamma)           
\end{equation}
yields an isomorphism between $G$ and $G'$.

Let (identity maps)
\begin{equation} 
\rho^D: G \mapsto \mathbf{GL}(N,\mathbb{C}), \ \ \rho^{D'}: G' \mapsto
\mathbf{GL}(N,\mathbb{C})              
\end{equation}
be linear representations of $G$ and $G'$, respectively. Since the latter groups are isomorphic,
\begin{equation} 
\rho = \rho^{D'} \circ \Delta: \Phi_D(\gamma) \mapsto \Phi_{D'}(\gamma) \in
\mathbf{GL}(N,\mathbb{C})                
\end{equation}
is another linear representation of $G$. Since $n^D(\gamma)$ and $n^{D'}(\gamma)$ become (equal) characters of the representations $\rho^D$ and $\rho$ respectively, the representations are similar. So there exists an invertible matrix $T$ for which
\begin{equation}   
TM^{(\mu)} = N^{(\mu)}T           
\end{equation}
for any $\mu$. Thus $D$ and $D'$ are transplantable. \eop \\

\subsubsection{Penrose--Lifshits mushrooms}
Since transplantation implies isolength spectrality, one might wonder
if two billiards with the same length spectrum are, in general, necessarily isospectral.

M.~Lifshits, exploiting a  construction
attributed to R.~Penrose (see, e.g., \cite{Rau}), constructed a
class of pairs of $\mathbb{R}^2$-domains that, while not isometric, have
periodic geodesics of exactly the same lengths, including
multiplicities.
 When the boundaries are  ($C^\infty$) smooth,
 it follows that the two billiards have the same wave invariants,
in the sense that the traces of their wave groups,
$\,\cos(t\sqrt{\Delta})\,$,
 differ at most by a smooth function~\cite{Me}. Such billiards  provide  drums that 
sound different but are similar geometrically.

 In this section we describe a construction of smooth Penrose--Lifshits mushroom
 pairs that are not isospectral, following \cite{FulKuc}.
 The domains are smooth, so the spectral difference
 is not attributable to diffraction from corners.

 We  start from
 a half-ellipse $E$ with foci $F$ and $F'$ as shown in Fig.~\ref{fig10}:
\begin{figure}[ht]
\begin{center}
\includegraphics[width=0.8\linewidth]{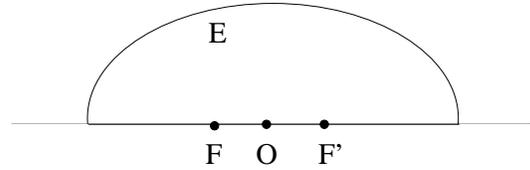}
\caption{Starting half-ellipse.\label{fig10}}
\end{center}
\end{figure}
The map
\begin{equation}
\xi \mapsto \xi',
\end{equation}
whether applied to regions, curves, or points,
indicates  reflection through the minor axis of
the ellipse.
 If  objects are interchanged by that reflection, we call them
\emph{dual}.
Now replace a line segment by a bounded smooth curve defined over the same interval, $B_1$, on the left and  $B_2$ on the
right, with $B_1' \ne B_2\,$,
 to form a smooth domain~$\Omega$ (Fig.~\ref{fig11}).
\begin{figure}[ht]
\begin{center}
\includegraphics[width=0.8\linewidth]{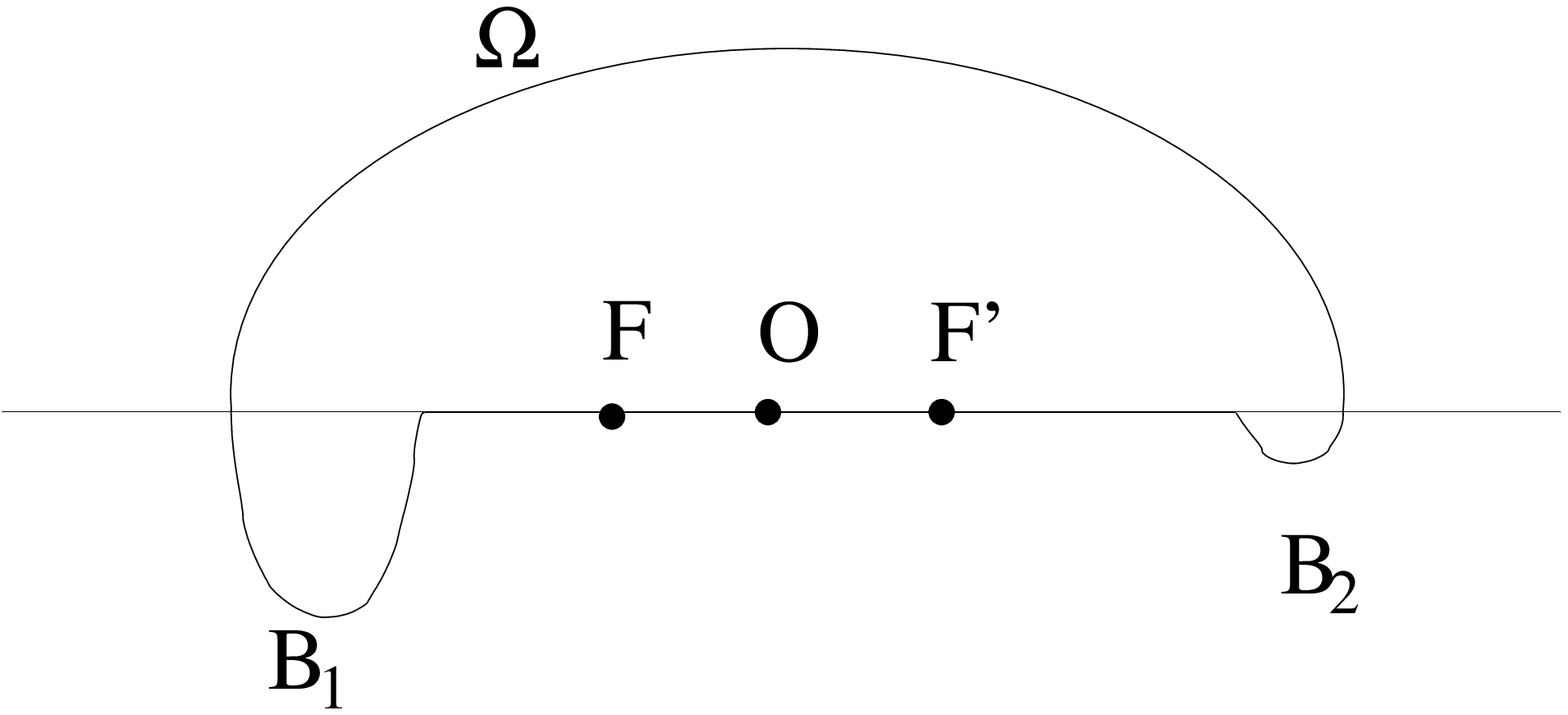}
\caption{Half-ellipse with two bumps.\label{fig11}}
\end{center}
\end{figure}
 Finally, carry out the same replacement operation (not self-dually)
 between the foci in two dual
 ways ($M$ and $M'$) to get two domains
$\Omega_1$ and $\Omega_2$ (Figs.~\ref{fig12} and \ref{fig13}).
We  call  domains $\Omega_1$ and $\Omega_2$ constructed in this manner
\emph{ Penrose--Lifshits mushroom pairs}.
\begin{figure}[ht]
\begin{center}
\includegraphics[width=0.8\linewidth]{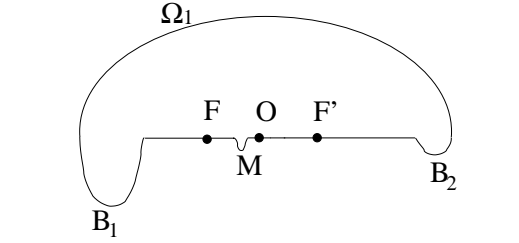}
\caption{Perturbed half-ellipse with two bumps.\label{fig12}}
\end{center}
\end{figure}
respectively.
\begin{figure}[ht]
\begin{center}
\includegraphics[width=0.8\linewidth]{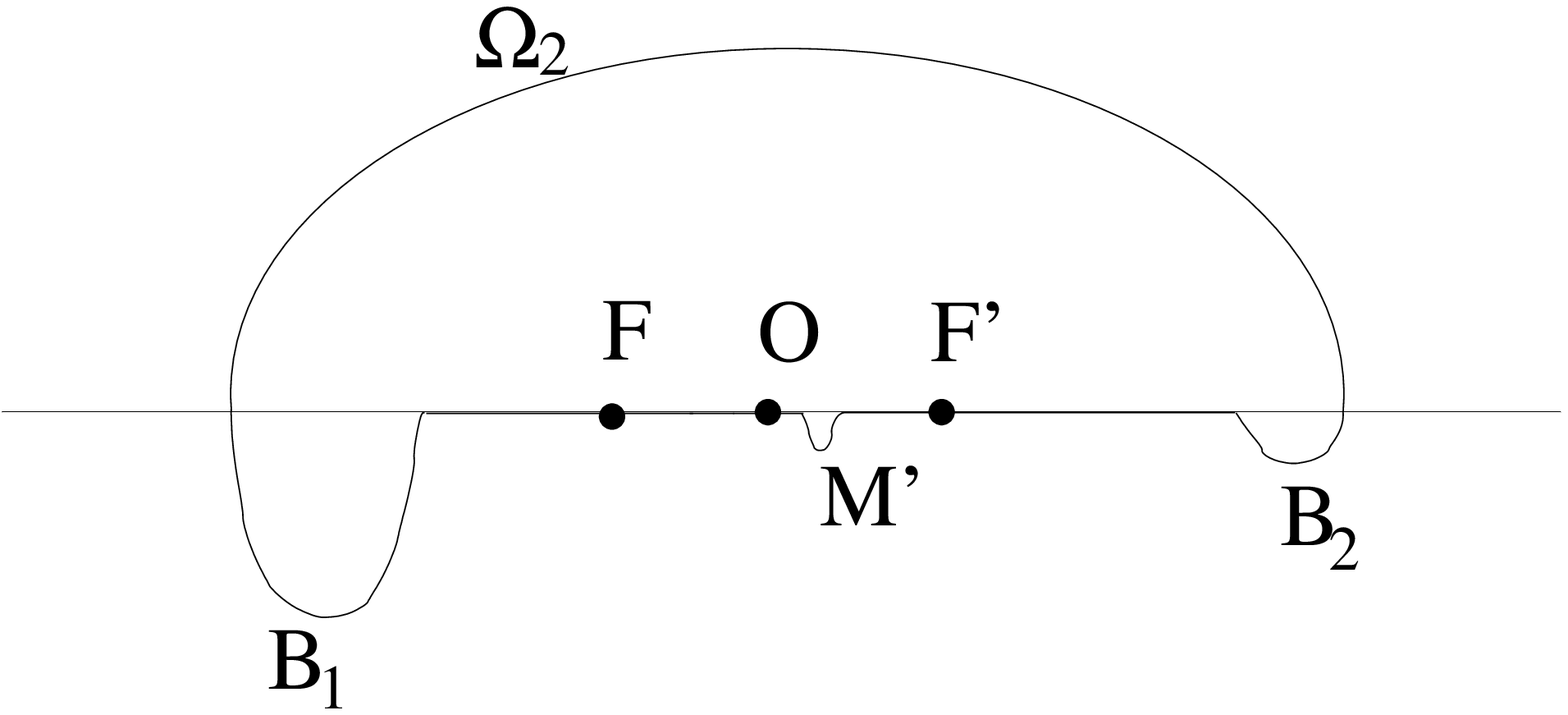}
\caption{Same as Fig.~\ref{fig12} but perturbed in a dual way.\label{fig13}}
\end{center}
\end{figure}

\begin{theorem}[\textcite{FulKuc}]
\label{T:non-isospectral}
If $B_1$ and $B_2$ are given and not dual,
%mirror-symmetric with respect to the vertical axis of the ellipse,
 then there exist dual bumps
$M$ and $\tilde M$ such that the resulting Penrose--Lifshits
mushrooms
  $\Omega_j$
have the same length spectra and wave invariants but are not
isospectral.
\end{theorem}

{\em Proof}.\quad
Following \cite{FulKuc} we first handle the length spectra
 \cite{Me,Zeld}. Geodesics in an ellipse (logically) fall into two disjoint categories
 \cite{KR,Rau,Berry}: those that intersect the major axis between the foci, and
those that do so at or beyond the foci. (The smoothness assumption guarantees that 
the major axis will not bifurcate  in $\Omega_j$ by diffraction.)

 A similar observation holds  for the domains
$\Omega_j$  just described before Theorem \ref{T:non-isospectral}:
Any geodesic originating in a curve $B_1$ or $B_2$ can never reach a
curve $M$ or $M'$, and vice versa.

  The geodesics that do not intersect the focal segment $FF'$ are  the same for
the two domains. Those for $\Omega_1$ that do intersect
this segment are identified one-to-one  with their duals in $\Omega_2$ by the
 reflection.  Hence the two billiards are length isospectral.
It now suffices to show that it is possible to choose an $M$ in such a
way that the two billiards are  nonisospectral. 
One considers the lowest eigenvalue of the
domains whose boundary is modified by a small perturbation. 
Using the  Rayleigh--Hadamard formula for change in the spectrum under domain perturbations (cf.
\cite{GS,Krein,Zeld} or \cite[section 15.1, Exercise 9]{Garab}), one
can prove that this eigenvalue is different for the two domains for a given choice of the
perturbation. Thus, one can construct non-isospectral billiards having
the same length spectrum, and the theorem is proved.
\eop\\

\subsection{Analytic domains}
\label{analyticdomains}
As all known counterexamples to the question ``Can one
hear the shape of a drum?'' are plane domains with corners, it might be possible
that analytic drumheads are spectrally determined.
The paper of \textcite{Zeld09} is part of a series (cf. \cite{Zeld04,Zeld04a})
devoted to the inverse
spectral problem for simply connected analytic Euclidean plane domains
$\Omega$, the
motivating problem being whether generic analytic Euclidean drumheads are
determined
by their spectra. 
The main results of \textcite{Zeld09} give the strongest evidence to date for
this conjecture by proving it for two classes of analytic drumheads:
those with an up/down symmetry, and those with dihedral symmetry.

{\bf Planar drumheads with symmetry}.\quad
We now state the results more precisely.  As before, by Lsp$(\Omega)$
we denote the length spectrum of $\Omega$,
that is, the set of lengths of closed trajectories of its billiard
flow. A bouncing ball orbit $\gamma$ is a two-leg
periodic trajectory that intersects $\partial\Omega$ orthogonally at both boundary
points. By rotating and translating
$\Omega$ we may assume
that $\gamma$ is vertical, with endpoints at $A = (0,L/2)$ and $B =
(0,-L/2)$. Zelditch's inverse results (\cite{Zeld09}) pertain to the following two
classes of drumheads, $\mathcal{D}_{1,L}$ and  $\mathcal{D}_{m,L}$, which are defined as follows:
(i) the class $\mathcal{D}_{1,L}$ of drumheads with one symmetry $\sigma$ and a
bouncing ball orbit of length $2L$ which is reversed by $\sigma$, and
(ii) the class $\mathcal{D}_{m,L}$ for $m \geq 2$ of drumheads admitting the
dihedral group $\mathbf{D}_{2m}$ (acting on $m$ letters) as symmetry group
and an invariant $m$-leg reflecting ray.

Note that the class $\mathcal{D}_{1,L}$ consists of simply connected
real-analytic plane domains $\Omega$ with
the property that there is an isometric involution $\sigma$ of $\Omega$
which ``reverses'' a nondegenerate
bouncing ball orbit (that is, $\sigma(\gamma)=\gamma^{-1}$, i.e.~the same orbit reversed)  of length $L_{\gamma} =
2L$. Other geometric properties can be found in
\cite{Zeld09}.\\

Let $\mathrm{Spec}_B$ denote the spectrum of the Laplacian $\Delta_B$ of
the domain $\Omega$ with boundary conditions $B$. The result of Zelditch is that
for Dirichlet (or Neumann) boundary conditions $B$,  the map
$\mathrm{Spec}_B: \mathcal{D}_{1,L} \mapsto \mathbb{R}^N_+$ is one-to-one.
As a corollary, one obtains the main result of \textcite{ISZ02} and \textcite{Zeld99, Zeld00}
that a simply connected analytic domain with the symmetries of an ellipse
and with
one axis of a prescribed length $L$ is spectrally determined within this
class. The above theorem admits a generalization to the special piecewise
analytic mirror symmetric domains with corners that are formed by reflecting the
graph of an analytic function, see \cite{Zeld09}.
For $m \geq 2$, that is, ``dihedrally symmetric domains'', 
\textcite{Zeld09} similarly proves that the map
$\mathrm{Spec}_B: \mathcal{D}_{m,L} \mapsto \mathbb{R}^N_+$ is one-to-one.\\

{\bf Higher dimensional drumheads with the symmetry of an ellipsoid}.\quad
More generally, \textcite{prep09} proved
that bounded analytic domains $\Omega \subset \mathbb{R}^n$ with $\pm$
mirror symmetries across all coordinate axes and with one axis height
fixed (and also satisfying some generic non-degeneracy conditions)
are spectrally determined among other such domains.  That is, you can hear
the shape of a real analytic drum in any number of dimensions {\em if you
know in advance that the mystery drums have the symmetries of an
ellipsoid}.  It is one of the first positive higher dimensional inverse
spectral results for Euclidean domains that is not restricted to balls.

%\newpage
%%%%%%%%%%%%%%%%%%%%%%%%%%%%%%%%%%%%%%%%%%%%%%%%%%%%%%%%%%%%%%%%%%%%%%%%
\section{Experimental and Numerical Investigations}
%%%%%%%%%%%%%%%%%%%%%%%%%%%%%%%%%%%%%%%%%%%%%%%%%%%%%%%%%%%%%%%%%%%%%%%%
\label{expandnum}

Although isospectrality is proved on mathematical grounds, the knowledge of exact
eigenvalues and eigenfunctions can not be obtained analytically for such
systems. Experimental as well as numerical simulations occurred very early
in the history of billiards. In 1909, in the Bulletin international de 
l'Acad\'emie des Sciences de Cracovie, Zaremba proposed a
way of ``numerically'' calculating solutions of the Dirichlet and
Neumann problem at a given point \cite{Zar09}.
To solve the eigenvalue problem for the Helmholtz equation, one standard
method is the boundary element method \cite{Rid79, BerWil84}. However,
this approach faces problems when the billiard has corners. Such
situations have been addressed e.g. by \textcite{Pis96} and \textcite{OkaShuTasHar05b}.
The usual numerical methods to compute eigenvalues and eigenfunctions in polygonal
billiards are based on the so-called ``method of particular solutions'' 
introduced by Fox, Henrici and Moler (FHM) in \cite{FoxHenMol}.
At a diffracting corner with angle $\pi\alpha$, a wave function $\Psi$ admits a
``corner'' decomposition into Bessel functions valid at a distance smaller than to
the nearest diffracting corner. In polar coordinates centered
around the corner $\pi\alpha$ this decomposition reads
\begin{equation}
\label{expansionbessel}
\Psi(r,\theta)=\sum_k a_k J_{k\alpha}(k r)\sin(k\alpha\theta),
\end{equation}
where $J_{\nu}$ are Bessel functions of the first kind, and
$k=\sqrt{E}$. The sine function in Eq.~\eqref{expansionbessel} ensures that
the function $\Psi(r,\theta)$ is zero on the boundary
edges connected to corner $\pi\alpha$. The idea of FHM is
to require that $\Psi$ also vanish on the rest of the boundary at a finite number of points,
and  to truncate the sum \eqref{expansionbessel}. This gives a
system of $m$ linear equations, which admits a non-zero
solution $\{a_k, 1\leq k\leq m\}$if and only if the  matrix corresponding
to this linear system is singular. The FHM method therefore consists in varying
the energy $E$ and tracking the singularities of the matrix $M$.
 
Unfortunately, for more than one diffracting corner it becomes virtually impossible
to track singularities, especially since in various circumstances the FHM method
fails to converge when the number of terms
included in  Eq.~\eqref{expansionbessel} is increased. Even for the
paradigmatic pair with half-square base shape (Fig.~\ref{celebrated}), 
which is one of
the simplest isospectral billiards, each pair has four diffractive
angles: two $3\pi/2$ and two $3\pi/4$ angles, and the FHM method fails to give eigenvalues with 
a good accuracy. This is why attention has been focused on
physical experiments. 

All known pairs of isospectral billiards are built on the same principle as
the ''historical'' pair $7_3$ of Fig.~\ref{celebrated}. As  explained in section
\ref{pedestrian} any initial building block possessing three sides along which to
unfold the block can be used to construct an isospectral pair. In particular, 
the properties of the resulting pair will depend strongly on the choice of the
initial building block. Physicists have mainly concentrated on the paradigmatic
example of Fig.~\ref{celebrated}. This allows us to make comparisons between 
the different approaches.

In this section we review both experimental and
numerical investigations which give insight into the behavior of
eigenvalues and eigenfunctions for isospectral billiards.

\subsection{Numerical investigations}
\subsubsection{Mode-matching method}
Numerical approaches to the study of isospectrality for the billiards 
of Fig.~\ref{celebrated} have followed the experiments of Sridhar and
Kudrolli that will be reviewed in the next subsection. Various approaches have been used in order to solve the
Helmholtz equation $(\Delta+E)\Psi=0$ with Dirichlet boundary conditions
inside the billiards.
The first numerical results were obtained by Wu, Sprung and Martorell
and reported in \cite{WuSprMar}. Using the mode-matching method described in section
\ref{modematchingsection}, they obtained eigenvalues of the billiard as the
values for which the determinant of the matrix $M$, given by Eq.~\eqref{matrixM},
vanishes. The results obtained by this method are displayed in column
2 of Table \ref{ev_iso}. As expected, 
both billiards yield the same values. The numerical results
were found to vary linearly in $1/N$.
\textcite{WuSprMar} compared their results
to results obtained by a finite-difference method consisting in 
 discretizing the Laplacian $\Delta$. This finite-difference method
gives the results displayed in column 1 of Table \ref{ev_iso} (the numerical 
results are again exactly the same for both billiards).
As a check of the validity of this approach, one can identify the eigenvalues of triangular
states. The lowest-energy triangular state are expected to have eigenenergies equal to  
$5\pi^2/d^2$ and  $10\pi^2/d^2$. As one can see in Table \ref{ev_iso}, these
eigenvalues correspond to the ninth and 21st mode, respectively, consistently with 
Table \ref{triangularmodes}.

\subsubsection{Expansion of eigenfunctions around the corners with the domain-decomposition method}
The main drawback of the mode-matching method of  \cite{WuSprMar} is the fact that one
has to know analytic solutions of the Helmholtz equation on subdomains of
the billiard. \textcite{Dri} used a numerical method based on an
algorithm by \textcite{DesTol}, particularly suited to
treating the case of polygonal billiards. The idea is again to
decompose the billiard into domains, each domain $\cD_i$ containing only one
diffracting angle $a_i$. On each domain the restriction of the
eigenfunction $\Psi$ is supposed to be some $\Psi_i$ that admits a
Bessel function expansion around corner $a_i$, according to Eq.~\eqref{expansionbessel}.
Truncation of this expansion to some finite order reduces the problem to that of
finding the coefficients of the expansion for the
$\Psi_i$. Mode-matching numerically leads to undesired
singularities. Instead, \textcite{DesTol} used an algorithm minimizing a
function that measures discrepancies between the $\Psi_i$ and between
their derivatives at the boundaries between subdomains. Improvement of this
algorithm allowed Driscoll to obtain the first 25 eigenvalues for both
billiards of Fig.~\ref{celebrated} with an accuracy of up to 12 digits.
\textcite{BetTre} used a modified method of particular solutions
using 140 expansion terms at each singular corner, 140 boundary points on
each side of the polygon, and 50 interior points to obtain following estimates
for the first three eigenvalues: $2.537943999798$, $3.65550971352$ and $5.17555935622$.

\subsection{Experimental realizations}

\subsubsection{Electromagnetic waves in metallic cavities}
Many experimental studies have been carried out on chaotic quantum billiards
to check the various properties conjectured analytically for
chaotic systems \cite{BohGiaSch84}.
One commonly used method is based on the correspondence
between the stationary Schr\"odinger equation and the Helmholtz equation for
electromagnetic waves in two dimensions (which is also the equation obeyed
by vibrating plates). The experiments are carried out by sending 
electromagnetic microwaves into a cylindrical copper cavity. The
height $h$ of the cavity is small, and the two other dimensions 
are shaped according to the desired billiards to investigate. 
For wavelengths $\lambda>2h$, i.e. frequencies below $\nu_0=c/2h$, all 
modes obey the two-dimensional wave equation $(\Delta+k^2)\Psi=0$. The $E_z$ component 
of the electric field plays the role of the quantum wave and vanishes 
on the boundary. Probes allow one to send an electromagnetic wave into the cavity
and to measure the transmission spectrum. In particular, 
eigenvalues correspond to resonances in the transmission
spectrum. Various choices of the probe locations ensure that no resonance
is missed.\\
If $\cA$ is the area of the cavity, the number of 
resonances below $\nu_0$ is approximately given by 
$\frac{\cA}{4\pi}(\pi/h)^2$. But the quality factor of the cavity is 
proportional to $h$; therefore one has to find a compromise between a high
quality factor and a large number of resonances.\\
Measurements of the intensity of the wave function (or here the electric
field) were achieved by the perturbation body method \cite{SriHogWil}: 
The resonance frequency of the cavity is shifted by the presence of a 
small metallic body inside the cavity. This shift is a function of the 
square of the electric field at the point of the metallic body.  

The first experimental investigation of isospectral billiards was realized
at Northeastern University, Boston, by \textcite{SriKud}. 
Sridhar and coworkers carried out various studies
on chaotic quantum billiards, such as the Sinai billiard (a square billiard with a circular
obstacle in the interior), and the Bunimovitch stadium-shaped billiard,
observing the scarring of eigenfunctions \cite{Sri} or localization phenomena 
\cite{SriHel} for such billiards. 
The experiments aimed at investigating isospectrality were realized on cavities
having the shape of the isospectral pair of Fig.~\ref{celebrated}.

Experimentally, each cavity has nine rectangular sides. The base
shape is an isosceles rectangular triangle (a half-square) whose smaller side is
$d=76$ mm ($3$ in.) long. The height of the cavity is $h=6.3$ mm 
($\simeq 0.25$ in.), so that microwaves at frequencies below $\nu_0=25$ GHz 
are actually two-dimensional. Measurements carried out to obtain 
the 54 lowest eigenvalues showed that, as expected, the eigenvalues 
of the two cavities are equal. Relative discrepancies of $0.01$ to 
$0.2\%$ between pairs of eigenvalues were found. These discrepancies and the
width of the resonances were assumed to be caused by imperfections 
due to the assembly of the pieces forming the cavity.
This experiment also allowed insight into the properties of 
eigenvalues of isospectral pairs. It was checked that the eigenvalues found
experimentally agree with the Weyl formula \eqref{weylexp}
 for the integrated density
of states:
\begin{equation}
\cNo(E)\simeq \frac{\cA}{4\pi}E-\frac{\cL}{4\pi}\sqrt{E}+\cK.
\end{equation}
For the choice $d=3$ in. one gets an area $\cA=31.5$ in.$^2$ and a perimeter 
$\cL=27$ in.; the constant $\cK$ is given
by Eq.~\eqref{constantK} and yields $\cK=5/12$. It was observed that at least for the 
lowest eigenvalues no degeneracy occurred.
By measuring the electric field inside the cavity, some of the lowest
eigenfunctions were obtained. The results for the ten first eigenvalues 
are displayed in Table \ref{ev_iso}. It is interesting to note that 
these pairs of eigenfunctions look quite different, although they
possess the same eigenvalues. It was checked that one eigenfunction could
be deduced from the other by transplantation.  
The particular case of the ninth mode, which is a triangular state, is
well reproduced. Indeed, as shown in Table \ref{ev_iso}, the measured
$9$-th eigenvalue is very close to its theoretical value $E=5\pi^2/d^2$.

Later \textcite{DhaMadUdaSri} applied a similar technique
to a chaotic isospectral billiard made of the billiard with half-square
base tile with scattering circular disks inside, showing
experimentally that isospectrality is indeed retained, provided scatterers
are added in a way consistent with the unfolding rules.

\subsubsection{Transverse vibrations in vacuum for liquid crystal smectic films}
Another experimental realization of Kac's membranes was achieved using liquid
crystal films in a smectic phase, spanning a shape of the form of the isospectral 
billiard \cite{EvePie}. 
First, the shapes were etched in circular stainless-steel
wafers of diameter 4cm and thickness 125 $\mu$m. 
The smectic film is then drawn on the shape, and after a 
few hours it reaches an equilibrium with uniform thickness $e$ of several 
hundred nanometers (corresponding to a few dozens monomolecular layers) over the 
whole surface. The whole experiment is set in vacuum. The film then 
obeys the wave equation
\begin{equation}
\gamma \Delta z=\rho e \frac{\partial^2 z}{\partial t^2},
\end{equation}
where $\gamma$ is the intrinsic tension of the film (in the experiments
$\gamma\sim 5.10^{-2}$N/m), and $\rho$ is the density, with a vertical 
displacement $z$ vanishing on the border. The film is excited by a voltage applied
by an electrode under the film, and the amplitude and phase of its oscillations 
are measured by sending in a laser beam and measuring its deviations with a photodiode. 
The signal detected is proportional to the height of the film at the position 
of the electrode. 
The frequency of the excitation is varied from a few Hertz to several kiloHertz, and
eigenfrequencies correspond to resonance peaks. Displacing the
electrode over the whole shape allows to reconstruct eigenmodes.  

The experiment was carried out on isospectral billiards with an isosceles triangular 
base shape: two angles $\beta=\gamma$ are equal, while the third one is varied
from $\alpha=67.5^{\circ}$ to $97.5^{\circ}$. The angle $\alpha=90^{\circ}$
corresponds to the example of Fig.~\ref{celebrated}.
The first 30 modes for both shapes were measured. The average relative
difference between two eigenvalues for a given mode is 0.3$\%$, which is
within the estimated experimental error of order 0.5$\%$. 
For the right angle triangle ($\alpha=90^{\circ}$)
the modes can be compared with other numerical or experimental results.
Data for the ten first eigenvalues of the $\alpha=90^{\circ}$ billiards were given
by \textcite{EvePie} and are displayed in Table \ref{ev_iso}. 
When the parameter $\alpha$ is varied, there is an avoided crossing between
eigenvalues of the eighth and ninth mode. Since the ninth mode is a triangular
mode (see section \ref{triangular}) and the eighth is not, the coupling between 
these two modes necessarily comes from experimental imperfections.

This experiment has also been tested on a billiard where the gluing scheme
of the  base triangles is modified. That is, tile $E$ in Fig.~\ref{modematching}
(left) is flipped around the line $x=5d/2$. This leads to a significantly different spectrum. 
In particular, the triangular modes are no longer eigenstates of such a billiard.
Again this is a check that the way the tiles are glued together, according to the
rules constructed from finite projective spaces or from Sunada triples, is
of primary importance for isospectrality.

\begin{center}
\begin{table*}[ht]
\begin{footnotesize}
\begin{tabular}{c|c|c|cc|cc|}
rank & Finite differences & Mode matching & \multicolumn{2}{c}{Electromagnetic waves} &  \multicolumn{2}{c}{Smectic films (relative values)}\\
1 & 1.028936 & 1.028535 & 1.02471 & 1.02481 & 1.000000 & 1.000000 \\
2 & 1.481865 & 1.481467 & 1.46899 & 1.47194 & 1.438000 & 1.430000 \\
3 & 2.098249 & 2.097467 & 2.08738 & 2.08831 & 2.040000 & 2.027000 \\
4 & 2.649715 & 2.649547 & 2.64079 & 2.63985 & 2.571000 & 2.548000 \\
5 & 2.938176 & 2.937434 & 2.93297 & 2.92949 & 2.854000 & 2.823000 \\
6 & 3.732689 & 3.732334 & 3.72695 & 3.71892 & 3.623000 & 3.570000 \\
7 & 4.295193 & 4.294728 & 4.28393 & 4.28388 & 4.184000 & 4.153000 \\
8 & 4.677665 & 4.677532 & 4.67021 & 4.66917 & 4.554000 & 4.507000 \\
9 & 5.000002 & 5.000000 & 4.98838 & 4.98531 & 4.861000 & 4.811000 \\
10 & 5.291475 & 5.290275 & 5.27908 & 5.27278 & 5.150000 & 5.095000 \\
11 & 5.801531 & 5.801138 & 5.78755 & 5.78371 &  &  \\
12 & 6.433894 & 6.432156 & 6.41357 & 6.43781 &  &  \\
13 & 6.866260 & 6.866226 & 6.84891 & 6.84718 &  &  \\
14 & 7.159802 & 7.159343 & 7.15242 & 7.16045 &  &  \\
15 & 7.694737 & 7.692417 & 7.67783 & 7.70604 &  &  \\
16 & 8.463655 & 8.463257 & 8.44285 & 8.45947 &  &  \\
17 & 8.613536 & 8.611169 & 8.57859 & 8.62220 &  &  \\
18 & 9.012405 & 9.010349 & 8.99495 & 8.97209 &  &  \\
19 & 9.609968 & 9.609791 & 9.60312 & 9.59562 &  &  \\
20 & 9.921131 & 9.921040 & 9.92583 & 9.93689 &  &  \\
21 & 10.000008 & 10.000000 & 10.00330 & 10.03932 &  &  \\
22 & 10.571020 & 10.569736 & 10.55227 & 10.55740 &  &  \\
23 & 11.066916 & 11.065727 & 11.09578 & 11.10035 &  &  \\
24 & 11.419551 & 11.418850 & 11.41874 & 11.40569 &  &  \\
25 & 11.984650 & 11.984080 & 11.99364 & 11.98033 &  &  
\end{tabular}
\end{footnotesize}
\caption{Comparison between the first eigenvalues $E_i$ of the isospectral pair
obtained by various methods, expressed in units of $\pi^2/d^2$. The ninth
mode corresponds to the triangular mode: its normalized eigenvalue is expected
to be equal to 5. (The conversion from frequencies to lengths is done
assuming vacuum in the cavity; \textcite{WuSprMar} gave the values for 
electromagnetic cavities with a factor of 1.0006 corresponding to the presence
of air in the cavity.)\label{ev_iso}}
\end{table*}
\end{center}

\subsubsection{Isospectral electronic nanostructures}
Recently \cite{Moon} an experiment was done involving electrons confined in
isospectral billiards, with the purpose of using transplantation to
reconstruct the quantum phase of measured wavefunctions. Each billiard
consisted of a wall of 90 CO molecules, constructed by
positioning the molecules with the tip of a scanning tunneling
microscope. The chosen billiards were built according to the pattern
of Fig.~\ref{celebrated}, but the base shape was chosen to be a triangle with angles
$(\pi/2, \pi/3,\pi/6)$. As in \cite{EvePie}, it was checked that billiards
violating the isospectral construction rule led to a different result. 

Amusingly,  \textcite{Moon} took Kac's question literally by converting
the average measured spectra into audio frequencies, checking that one could
indeed ''hear'' non-isospectrality.

%\newpage
%%%%%%%%%%%%%%%%%%%%%%%%%%%%%%%%%%%%%%%%%%%%%%%%%%%%%%%%%%%%%%%%%%%%%%%%%%%%%%%%%%%%5
\section{Sunada Theory}
%%%%%%%%%%%%%%%%%%%%%%%%%%%%%%%%%%%%%%%%%%%%%%%%%%%%%%%%%%%%%%%%%%%%%%%%%%%%%%%%%%%%5
\label{sunadasection}
The examples of isospectral billiards considered so far can be
proved to be isospectral by quite simple tools. However historically
they were constructed by a group-theoretical approach. The
mathematical theory of isospectrality rests on a theory by
Sunada. We first review the necessary basic notions of group theory.
Then, in section \ref{ST}, we introduce Sunada Theory.

\subsection{Permutations}

Following the usual conventions, we denote permutation action
exponentially (i.e. the image of an element $x$ by the permutation $g$ is $x^g$)
and let elements act on the right. We denote the identity element
of a group by $\id$\index{$\id$} or $\mathbf{1}$\index{$\mathbf{1}$}, if no special symbol has been
introduced for it before. A group $G$ without its identity $\id$
is denoted $G^\times$\index{$G^{\times}$}. The number of elements of a group $G$ is denoted by $\vert G\vert$.
 A \emph{permutation group} $(G,X)$ is a pair consisting of a group
$G$ and a set $X$ such that each element $g$ of $G$ defines a
permutation $g:X\rightarrow X$ of $X$, and the permutation defined
by the product $gh$, $g,h\in G$, is given by $gh:X\rightarrow X:x\mapsto
(x^g)^h$. An {\em involution} in a group is an element $g$ of order $2$, that is, 
such that $g^2 = \id$.

\subsection{Commutator notions}

The group-theoretic setting of Sunada theory requires introduction of some notions such as
the commutator of two groups and perfect groups.
The {\em conjugate}\index{conjugate} of $g$
by $h$ is $g^h=h^{-1}gh$\index{$g^h$}.
Let $H$ be a group. The \emph{commutator}\index{commutator} of two group elements
$g,h$ is equal to $[g,h]=g^{-1}h^{-1}gh$\index{$[g,h]$}.
The {\em commutator}\index{commutator} of two subsets $A$ and
$B$ of a group $G$ is the subgroup $[A,B]$\index{$[A,B]$} generated by all
elements $[a,b]$, with $a\in A$ and $b\in B$. The \emph{commutator
subgroup}\index{commutator!subgroup} of $G$ is $[G,G]$, also denoted by $G'$\index{$G'$}. Two
subgroups $A$ and $B$ \emph{centralize} each other if $[A,B]=\{\id\}$.
The subgroup $A$ \emph{normalizes} $B$ if $B^a=B$ for all $a\in
A$, which is equivalent to $[A,B]$ being a subgroup of $B$.
If $A$ and $B$ are two subgroups of the group $G$, then they are {\em conjugate(d)} if there is an 
element $g$ of $G$ such that $A^g = B$. The subgroup $A$ of $G$ is (a) {\em normal} (subgroup) in (of) $G$ if 
$A^g = A$ for all $g \in G$. In such a case, we write $A \unlhd G$. If $A \ne G$, we also write $A \lhd G$.\\ 

Inductively, we define the {\em $n$th central
derivative}\index{central@$n$th central derivative} $[G,G]_{[n]}$\index{$[G,G]_{[n]}$}
of a group $G$ as $[G,[G,G]_{[n-1]}]$, and the {\em $n$th normal
derivative}\index{normal@$n$th normal derivative} $[G,G]_{(n)}$\index{$[G,G]_{(n)}$} as
$[[G,G]_{(n - 1)},[G,G]_{(n - 1)}]$. For $n=0$, the zeroth central and
normal derivatives are by definition equal to $G$ itself. If, for
some natural number $n$, $[G,G]_{(n)}=\{\id\}$, and
$[G,G]_{(n-1)}\neq\{\id\}$, then we say that $G$ is
\emph{solvable}\index{solvable group} ({\em soluble}\index{soluble}) of length $n$. If
$[G,G]_{[n]}=\{\id\}$ and $[G,G]_{[n-1]}\neq\{\id\}$, then we say
that $G$ is \emph{nilpotent}\index{nilpotent group} of class $n$.
The \emph{center}\index{center!of a group} of a group is the set
of elements that commute with every other element, i.e.,
$Z(G)=\{z\in G\parallel [z,g]=\id, \forall g\in G\}$\index{$Z(G)$}. Clearly, if a
group $G$ is nilpotent of class $n$, then the $(n-1)$th central
derivative is a nontrivial subgroup of $Z(G)$.\\

A group $G$ is the {\em central product} of its subgroups $A$ and $B$ if $AB = G$, $A \cap B$ is contained
in the center of $G$, and $A$ and $B$ centralize each other. Sometimes we write $G = A \circ B$ in such a case.\\ 

A group $G$ is called \emph{perfect}\index{perfect group} if
$G=[G,G] = G'$.\\

Let $R$ be a finite group.
The {\em Frattini group} $\phi(R)$\index{Frattini group}\index{$\phi(R)$} of $R$ is the intersection of all proper maximal
subgroups, or is $R$ if $R$ has no such subgroups.

\subsection{Finite simple groups}

A group is {\em simple} if it does not contain nontrivial normal subgroups.

The finite simple groups are often regarded as the elementary particles in finite group theory.
Before we explain this more precisely, recall that a {\em composition series} of a group $G$ is a normal series
\begin{equation}
    1 = H_0 \lhd H_1 \lhd \cdots \lhd H_n = G,
\end{equation}
such that each $H_i$ is a maximal normal subgroup of $H_{i+1}$. Equivalently, a composition series is a normal series such that each factor group $H_{i+1}/H_i$ is simple.
The factor groups are called {\em composition factors}.

A normal series is a composition series if and only if it is of maximal length. That is, there are no additional subgroups that can be ``inserted'' into a composition series. The length $n$ of the series is called the {\em composition length}.

If a composition series exists for a group $G$, then any normal series of $G$ can be refined to a composition series. Furthermore, every finite group has a composition series.

A group may have more than one composition series. However, the Jordan-H\"{o}lder theorem states that  any two composition series of a given group are equivalent.

The classification of finite simple groups (see \cite{Sol} for a survey) states that
every finite simple group is cyclic, or alternating, or is contained in one of $16$ families of groups of Lie type (including the Tits group, which strictly speaking is not of Lie type), or one of $26$ sporadic groups.

\textcite{ConCur} provided a list of the finite simple
groups, see also  \cite[p. 490-491]{Gor80}. In this review, we encounter several aspects of certain simple
groups in the construction theory of counter examples to Kac's initial question.

\subsection{$p$-Groups and extra-special groups}
The present section will be useful for construction of examples in section \ref{ExP}.

For a prime number $p$, a {\em $p$-group} is a group of order $p^n$ for
some natural number $n \ne 0$. A {\em Sylow $p$-subgroup}\index{Sylow
$p$-subgroup} of a finite group $G$ is a $p$-subgroup of 
order $p^n$ such that $p^{n+1}$ does not divide $|G|$.\\

A $p$-group $P$  is {\em special}\index{special $p$-group} if either $[P,P] = Z(P) = \phi(P)$ is elementary Abelian or $P$ itself is. (A group is {\em elementary Abelian} if it is Abelian, and if there exists a prime $p$ such that each of its nonidentity elements has order $p$.)
Note that $P/[P,P]$ is elementary Abelian
in that case. So 
\begin{equation} 
P/[P,P] \cong V(n,p), 
\end{equation}
where $V(n,p)$\index{$V(n,p)$} is the $n$-dimensional vector space over $\mathbb{F}_p$ (here seen as its additive group),  and $\vert P\vert = p^n\vert [P,P]\vert$.

Hence we have the exact sequence
\begin{equation}   
\1 \mapsto [P,P] \mapsto P \mapsto V(n,p) \mapsto \1.            
\end{equation}

If furthermore $\vert Z(P)\vert = \vert [P,P]\vert = \vert\phi(P)\vert
= p$, $P$ is called  {\em extra-special}\index{extra-special}.\\

We now present a classification for extra-special groups that depends on the knowledge of the nonabelian $p$-groups of order $p^3$.

There are four nonabelian $p$-groups of order $p^3$ | see \cite{Gor80}.
 First we have $M = M(p)$:
\begin{eqnarray} 
M(p)\index{$M(p)$}  &=& \langle  x,y,z \parallel x^p = y^p = z^p = \1, \nonumber\\
&&[x,z] =[y,z] = \1, [x,y] = z  \rangle.                  
\end{eqnarray}
 (Note that this is the general Heisenberg group of order $p^3$ which we will encounter later on.)
 Next, define
 \begin{equation} 
M_3(p)\index{$M_3(p)$} = \langle x,y \parallel x^{p^2} = y^p = \1, x^y =
x^{p + 1}\rangle.            
\end{equation}
 Finally, we have the dihedral group $D$\index{$D$} of order $8$ and the generalized quaternion group $Q$\index{$Q$} of order $8$.\\

\medskip
\bt[\textcite{Gor80}]
An extra-special $p$-group $P$ is the central product of $r \geq 1$ nonabelian subgroups of order $p^3$. Moreover, we have the following.
\begin{itemize}
\item[{\rm (1)}]
If $p$ is odd, $P$ is isomorphic to $N^kM^{r - k}$, while if $p = 2$, $P$ is isomorphic to $D^kQ^{r - k}$ for some $k$. In either case, $\vert P\vert = p^{2r + 1}$.
\item[{\rm (2)}]
If $p$ is odd and $k \geq 1$, $N^kM^{r - k}$  is isomorphic to $NM^{r - 1}$, the groups $M^r$ and $NM^{r - 1}$ are not isomorphic and $M^r$ is of exponent $p$.
\item[{\rm (3)}]
If $p = 2$, then $D^kQ^{r - k}$ is isomorphic to $DQ^{r - 1}$ if $k$ is odd and to $Q^r$ if $k$ is even, and the groups $Q^r$ and $DQ^{r - 1}$ are not isomorphic.
\end{itemize}
(All the products considered are central products.)
\et

\subsection{Sunada Theory}
\label{ST}
We now turn to the main theorems of Komatsu and Sunada, which allowed
Gordon {\it et al.} to produce the first known example of isospectral billiards. Sunada's idea was to reduce the problem of finding isospectral
 manifolds to a group-theoretical problem, namely, constructing triplets of groups having a certain property. As the groups that appear in Sunada's
proof are Galois groups, we need some more definitions.\\

A field extension $\mathbb{L} /\mathbb{K}$ is called {\em algebraic} if every element of L is {\em algebraic} over $\mathbb{K}$, i.e., if every element of $\mathbb{L}$ is a root of some non-zero polynomial with coefficients in $\mathbb{K}$. (Field extensions which are not algebraic, i.e. which contain transcendental elements, are called {\em transcendental}.)

Let $\mathbb{K}$ be an algebraic number field of degree $n$.
Recall that a {\em number field} is a finite, algebraic field extension of $\mathbb{Q}$; its degree is the dimension over $\mathbb{Q}$ as a $\mathbb{Q}$-vector space. A standard example is $\mathbb{Q}(\sqrt{2})$.

%A {\em ring} is a set R equipped with two binary operations $+Ê: R \times R \mapsto R$ and  $\circÊ: R \times R \mapsto R$, called addition and multiplication. To qualify as a ring, the %set and the two operations, $(R, +, \circ)$, must satisfy the following requirements:
%\begin{itemize}
%	\item
%$R, +$ is an Abelian group;
%\item
%$R,\circ$ is a monoid;
%\item
%the distributive laws hold.
%\end{itemize}

%A subset $I$ of a ring $R$ is said to be a {\em right ideal} (in $R$) if:
%\begin{itemize}
%\item
%$I, +$ is a subgroup of the underlying additive group in $(R, +, \circ)$;
%\item
%for every $x \in I$ and $r \in R$, $x\circ r$ is in $I$. 
%\end{itemize}

%In other words, $I$ is a right $R$-submodule. {\em Left ideals} are defined in a similar way. If an ideal is two-sided, one often drops ``left'' and ``right''. \\
The {\em ring of integers} of an algebraic number field $\mathbb{K}$, often denoted by 
$O_{\mathbb{K}}$, is the ring of algebraic integers contained in $\mathbb{K}$. 
An {\em algebraic integer} is an element of $\mathbb{K}$ that is a root of some monic 
polynomial with coefficients in $\mathbb{Z}$.

 The {\em (Dedekind) zeta function}  $\zeta_{\mathbb{K}}(s)$
 (associated with $\mathbb{K}$), $s$ being a complex variable, is
 defined by
\begin{equation} 
\zeta_{\mathbb{K}}(s) =  \sum_I[N_Q^\mathbb{K}(I)]^{-s},            
\end{equation}
taken over all ideals $I$ of the ring of integers $O_{\mathbb{K}}$ of $\mathbb{K}$, $I \ne \{0\}$.
Note that $N_Q^\mathbb{K}(I)$ denotes the norm of $I$ (to $\mathbb{Q}$), equal to $\vert O_{\mathbb{K}}/I\vert$.\\

An ideal $P$ of a ring $R$ is a {\em prime ideal} if it is a proper ideal and  if for any two ideals $A$ and $B$ in $R$ such that $AB \subseteq P$, we have that $A \subseteq P$ or $B \subseteq P$.
 Let $p$ be a rational prime. Let $P_1,\ldots,P_g$ be the prime ideals of $O_{\mathbb{K}}$ lying above $p$. Then
\begin{equation} 
\langle p\rangle = \prod_{i = 1}^gP_i^{e_i},             
\end{equation}
where
\begin{equation}  
e_i = e_{\mathbb{K}}(P_i).       
\end{equation}
Here $e_{\mathbb{K}}(P_i)$ is the {\em ramification index} of $P_i$ over $\mathbb{K}$. 
If $e_i > 1$ for some $i \in \{1,\ldots,g\}$, then $p$ is said to be {\em ramified} in $\mathbb{K}$. If $e_i = 1$ for all $i$,
$p$ is {\em unramified} in $\mathbb{K}$.\\

The {\em conjugate elements} of an algebraic element $\alpha$, over a field $\mathbb{K}$, are the roots of the minimal polynomial of $\alpha$ over $\mathbb{K}$.
(For example, the cubic roots of 1 are $1, -1/2 + \sqrt{3}/2i, -1/2 - \sqrt{3}/2i$.
The latter two roots are conjugate elements in the field $\mathbb{K} = \mathbb{Q}[\sqrt{-3}]$.)

Let $\mathbb{K} = \mathbb{Q}(\theta)$ be as above, that is, an algebraic number field of degree $n$ ($\theta \in \mathbb{C}$). Suppose $\theta_1,\theta_2,\ldots,\theta_n$ are the conjugates of $\theta$ over $\mathbb{Q}$. If 
\begin{equation} 
\mathbb{Q}(\theta_1) = \cdots = \mathbb{Q}(\theta_n) = \mathbb{K}, 
\end{equation}
then $\mathbb{K}$ is a {\em Galois extension} of $\mathbb{Q}$.\\ 

Suppose that $\mathbb{E}$ is an extension of the field $\mathbb{F}$ (written as $\mathbb{E}/\mathbb{F}$). Consider the set of all automorphisms of $\mathbb{E}/\mathbb{F}$ (that is, isomorphisms $\alpha$ from $\mathbb{E}$ to itself such that $\alpha(x) = x$ for every $x \in \mathbb{F}$). This set of automorphisms with the operation of function composition forms a group, sometimes denoted by $\mathrm{Aut}(\mathbb{E}/\mathbb{F}$).
If $\mathbb{E}/\mathbb{F}$ is a Galois extension, then $\mathrm{Aut}(\mathbb{E}/\mathbb{F}$) is called the {\em Galois group} of (the extension) $\mathbb{E}$ over $\mathbb{F}$, and is usually denoted by $\mathrm{Gal}(\mathbb{E}/\mathbb{F}$).\\

A number-theoretic exercise which asks for non-isomorphic number fields $\K_1$ and $\K_2$ with the same
zeta function has the following answer:

\bt[\textcite{Ko}]
\label{galois}
Let $\K$ be a finite Galois extension of $\mathbb{Q}$ with Galois group $G = \mathrm{Gal}(\K/\mathbb{Q})$, and let $\K_1$ and
$\K_2$ be the subfields of $\K$ corresponding to subgroups $G_1$ and $G_2$ of $G$, respectively. Then the following conditions are equivalent:
\begin{itemize}
\item[{\rm (i)}]
Each conjugacy class of $G$ meets $G_1$ and $G_2$ in the same number of elements;
\item[{\rm (ii)}]
The same primes $p$ are ramified in $\K_1$ and $\K_2$ and for the unramified $p$ the decomposition of $p$ in $\K_1$ and $\K_2$ is the same;
\item[{\rm (iii)}]
The zeta functions of $\K_1$ and $\K_2$ are the same.
\end{itemize}
\et

In particular, if $G_1$ and $G_2$ are not conjugate in $G$, then $\K_1$ and $\K_2$ are not
isomorphic while having the same zeta function. It should be noted that several such triples $(G,G_1,G_2)$ are known | see the examples in this section.

Any group triple $(G,G_1,G_2)$ satisfying Theorem \ref{galois}(i) is said to satisfy ``Property (*)''.\\

Sunada's idea was to establish a counterpart of this theorem for Riemannian geometry. In that context, there  is
an analogue for the Dedekind zeta function. For $\M$ a Riemannian manifold, one defines
\begin{equation}  
\zeta_{\M}(s) = \sum_{i = 1}^{\infty}\lambda_i^{-s}, \ \ \mathfrak{Re}(s)\gg 0,      
\end{equation}
where
\begin{equation}  
0 < \lambda_1 \leq \lambda_2 \leq \cdots
\end{equation}
are the non-zero eigenvalues of the Laplacian for $\M$. 
The function $\zeta_{\M}$ has an analytic continuation to the whole plane, and it is well-known that $\zeta_{\M_1}(s) = \zeta_{\M_2}(s)$ if and only if $\M_1$ and $\M_2$ are isospectral.

The following theorem gives sufficient conditions for two manifolds to have the same zeta function. 

\bt[\textcite{Su}]
\label{sunada}
Let $\pi: \M \mapsto \M_0$ be a normal finite Riemannian covering with covering transformation group $G$, and let $\pi_1: \M_1 \mapsto \M_0$ and $\pi_2: \M_2 \mapsto \M_0$ be the coverings corresponding to the subgroups $H_1$ and $H_2$ of $G$, respectively. If the triplet $(G,H_1,H_2)$ satisfies Property (*), then the zeta functions $\zeta_{\M_1}(s)$ and $\zeta_{\M_2}(s)$ are identical.
\et

The proof of the latter theorem makes use of an interesting trace formula, which we present now.

If $A$ is a non-negative self-adjoint operator of a Hilbert space, one defines the {\em trace} of $A$ as an extended real number by the possibly divergent sum
$\sum_k\langle Ae_k,e_k\rangle$, where $\{e_j\}_j$ is an orthonormal base of the space.
It is of {\em trace class} if and only if $\mathrm{Tr}(A) < \infty$.

Let $V$ be a Hilbert space on which a finite group $G$ acts as unitary transformations and let $A: V \mapsto V$ be a self-adjoint operator of trace class such that $A$ commutes with the $G$-action. For a
subgroup $H$ of $G$, denote by $V^H$ the subspace of $H$-invariant vectors.\\

{\bf Trace Formula}.\quad
{\em The restriction of $A$ to the subspace $V^G$ is also of trace class, and}

\begin{equation}  \mbox{tr}(A \vert_{V^G}) = \sum_{[g] \in [G]}(\vert G_g\vert)^{-1}\mbox{tr}(gA),           \end{equation}

{\em where $[G] =\{ [g]\}$, $[g]$ is the conjugacy class of $g$ in $G$ and $G_g$ is the centralizer of
$g$ in $G$}.\\
\bigskip

If the triplet $(G,G_1,G_2)$ satisfies Property (*), then

\begin{equation}  \mbox{tr}(A \vert_{V^{G_1}}) =  \mbox{tr}(A \vert_{V^{G_2}}).          \end{equation}

Even if $G_1$ and $G_2$ are not conjugate, the manifolds $\M_1$ and $\M_2$ could possibly be isometric.

\bt[\textcite{Su}]
There exist finite coverings
$\pi_1: \M_1 \mapsto \M_0$ and $\pi_2: \M_2 \mapsto \M_0$ of Riemann surfaces with genus $\geq 2$ such that for a generic metric $g_0$ on $\M_0$, the surfaces $(\M_1,\pi_1^*g_0)$ and $(\M_2,\pi_2^*g_0)$ are isospectral, but not isometric.
\et

\bigskip
Sunada's theorem allows us to construct isospectral pairs provided we find triples $(G,G_1,G_2)$ satisfying Property (*) | ``Sunada triples''.

Now we give examples of such triples.

\subsection{Examples of Sunada triples}
\label{ExP}

{\bf Example 1 | see \textcite{Ge}}.\quad Let $G$ be the semidirect product $\mathbb{Z}/8\mathbb{Z}^{\times}\ltimes\mathbb{Z}/8\mathbb{Z}$, and define $G_1$ and $G_2$ by
\begin{eqnarray}
G_1 &=& \{(1,0),(3,0),(5,0),(7,0)\},\nonumber\\
G_2 &=& \{ (1,0),(3,4),(5,4),(7,0)\}.
\end{eqnarray}

\bigskip
{\bf Example 2 | see \textcite{Ga}}.\quad
Let $G = \mathbf{S}_6$ be the symmetric group on six letters $\{a, b, c, d, e, f  \}$.
Set
\begin{equation} G_1 = \{\1,(ab)(cd), (ac)(bd), (ad)(bc) \}         \end{equation}
and
\begin{equation} G_2 = \{\1, (ab)(cd), (ab)(ef), (cd)(ef) \}.            \end{equation}
\bigskip
{\bf Example 3 | see \textcite{Ko}}.\quad
Let $G_2$ and $G_2$ be two finite groups with the same order, and suppose that their exponents (equal to the least common multiples of the orders of their elements) both equal
the same odd prime $p$. Set $\vert G_1\vert = \vert G_2\vert = p^h$ for $h \in \mathbb{N}^{\times}$ and
embed $G_1$ and $G_2$ in the symmetric group $\mathbf{S}_{p^h}$ on $p^h$ letters by their left action on themselves. For a conjugacy class $[g]$ corresponding to the partition
\begin{equation} 
\vert \mathbf{S}_{p^h}\vert = p^h! = p + p + \cdots + p,           
\end{equation}
we have
\begin{equation}  
\vert ([g] \cap G_1)\vert =  {p^h} - 1 = \vert ([g] \cap G_2)\vert,         
\end{equation}
while $\vert ([g] \cap G_i)\vert = 0$ otherwise.

Concretely, let $G_1 = (\mathbb{Z}/p\mathbb{Z})^3$, and let $G_2$ be the group

\begin{eqnarray}
G_2 = \langle a,b \parallel a^p = b^p = [a,b]^p = \1,\hspace{2cm}\nonumber\\ 
a[a,b] = [a,b]a, b[a,b] = [a,b]b \rangle,  \hspace{1cm}             
\end{eqnarray}

that is, $G_2$ is the extra-special group of order $p^3$. Then $(\mathbf{S}_{p^3},G_1,G_2)$ verifies
Property (*).\\

One can in fact generalize Komatsu's example by
defining the following group.
The {\em general Heisenberg group}  $\mathbf{H}_n$ of dimension $2n + 1$  over $\mathbb{F}_q$, with $n$ a natural number,
is the group of square $(n + 2)\times(n + 2)$-matrices with entries in $\mathbb{F}_q$, of the following form (and with the usual matrix multiplication):
\begin{equation} \left(
 \begin{array}{ccc}
 1 & \alpha & c\\
 0 & \II_n & \beta^T\\
 0 & 0 & 1\\
 \end{array}
 \right),                                         
\end{equation}
 \noindent
 where $\alpha, \beta \in \mathbb{F}_q^n$, $c \in \mathbb{F}_q$, and with $\II_n$  the $n\times n$-unit matrix.
 Let $\alpha, \alpha',\beta, \beta' \in \mathbb{F}_q^n$ and $c,c' \in \mathbb{F}_q$; then
 \begin{eqnarray}       
\left(
 \begin{array}{ccc}
 1 & \alpha & c\\
 0 & \II_n & \beta^T\\
 0 & 0 & 1\\
 \end{array}
 \right)
 \times
  \left(
 \begin{array}{ccc}
 1 & \alpha' & c'\\
 0 & \II_n & {\beta'}^T\\
 0 & 0 & 1\\
 \end{array}
 \right)\hspace{2cm}\nonumber\\
 =\left(
 \begin{array}{ccc}
 1 & \alpha + \alpha'& c + c' + \langle \alpha,\beta' \rangle\\
 0 & \II_n & \beta + \beta'\\
 0 & 0 & 1\\
 \end{array}
 \right).\hspace{.5cm}                      
\end{eqnarray}
Here $\langle x,y\rangle$, with $x = (x_1,x_2,\ldots,x_n)$ and $y = (y_1,y_2,\ldots,y_n)$ elements of $\mathbb{F}_q^n$, denotes $x_1y_1 + x_2y_2 + \ldots + x_ny_n$.

The following properties hold for $\mathbf{H}_n$.

\begin{itemize}
\item[{\rm (i)}]
$\mathbf{H}_n$ has exponent $p$ if $q = p^h$ with $p$ an odd prime; it has exponent $4$ if $q$ is even.
\item[{\rm (ii)}]
The center of $\mathbf{H}_n$ is given by

\begin{equation}    \{(0,c,0) \parallel c \in \mathbb{F}_q\}.         \end{equation}
\item[{\rm (iii)}]
$\mathbf{H}_n$ is nilpotent of class $2$.
\end{itemize}

Then, as above, $(\mathbf{S}_{p^{2n + 1}},\mathbf{H}_n,(\mathbb{Z}/p\mathbb{Z})^{2n + 1})$  verifies Property (*).\\

Any finite group arises as the fundamental group of a compact smooth manifold of dimension $4$. For a triplet
$(G,G_1,G_2)$ of the type described in Example 3, we find a compact manifold $\M_0$ with fundamental group
$G$. Let $\M$ be the universal covering of $\M_0$. Then the quotients $\M_i = \M/G_i$ have non-isomorphic fundamental groups $G_i$, $i = 1,2$. By Theorem \ref{sunada} the manifolds $(\M_1,\pi_1^*g_0)$ and
$(\M_2,\pi_2^*g_0)$ are isospectral for any metric $g_0$ on $\M_0$, but not isometric.\\

%\newpage
%%%%%%%%%%%%%%%%%%%%%%%%%%%%%%%%%%%%%%%%%%%%%%%%%%%%%%%%%%%%%%%%%%%%%%%%%%%%%%%%%%%%%%%%%%%%%%%%%%%%%%%%%%
\section{Related Questions}
%%%%%%%%%%%%%%%%%%%%%%%%%%%%%%%%%%%%%%%%%%%%%%%%%%%%%%%%%%%%%%%%%%%%%%%%%%%%%%%%%%%%%%%%%%%%%%%%%%%%%%%%%%
\label{RQ}
The literature on isospectrality is large, and it is out of the question to
review the entire field. In the present paper we have concentrated on
the questions addressed by planar two-dimensional domains with
Dirichlet boundary conditions. To open the topic further, we now
mention some questions related to the main one discussed
in the present paper, some of which have been addressed in the literature, and some of which remain open problems.

\subsection{Boundary conditions}
\label{levitin}
\begin{figure}[ht]
\begin{center}
\includegraphics[width=0.7\linewidth]{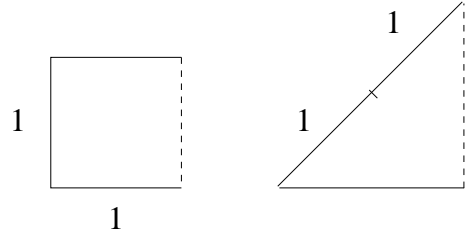}
\end{center}
\caption{Isospectral billiards with mixed Neumann-Dirichlet boundary
  conditions. Solid line, Dirichlet; dashed line, Neumann.\label{mixedbc}}
\end{figure}

So far we have mainly dealt with billiards with Dirichlet boundary
conditions. More recently attention has been concentrated on 
mixed Dirichlet-Neumann boundary conditions, that is, having 
either $\Psi = 0$ or $\partial_{\bf n}\psi=0$ on different intervals of the
boundary (${\bf n}$ being the normal to the boundary). This
is much simpler than the Dirichlet problem.
Simple instances of mixed-boundary condition isospectral pairs
are proposed in \cite{LevParPol} (see also \cite{JakLevNadPol}). 
Their simplest example is reproduced
in Fig.~\ref{mixedbc}. The eigenfunctions are given by
\begin{equation}
\sin\frac{\pi(m+1/2)x}{d}\sin\frac{\pi n y}{d},\ \ \ \ n\geq 1, m\geq
0,
\end{equation}
for the square of size $d$, and
\begin{eqnarray}
\sin\frac{\pi(m+1/2)x}{d\sqrt{2}}\sin\frac{\pi (n+1/2) y}{d\sqrt{2}}\hspace{3cm}\\
-\sin\frac{\pi(n+1/2)x}{d\sqrt{2}}\sin\frac{\pi (m+1/2) y}{d\sqrt{2}},\ \ \ \ m>n\geq 0,\nonumber
\end{eqnarray}
for the triangle of size $d\sqrt{2}$.

These examples can be generalized:  \textcite{LevParPol} gave a procedure
to construct similar pairs. The idea is to construct an elementary
domain, or ``construction block'', whose boundary is made of two line segments
$a$ and $b$ on the plane, with ends joined by two arbitrary curves. Imposing any mixed 
Neumann-Dirichlet boundary conditions on the construction block boundary, one obtains
a Neumann-Dirichlet isospectral pair by gluing the construction block together with its 
reflection with respect to either $a$ or $b$ (and imposing Neumann boundary conditions  to the segment
itself, Dirichlet to its image). This technique can be further generalized by gluing
together more copies of the construction block, yielding more complicated examples.
In particular this method shows that for mixed boundary
conditions it is possible to construct isospectral pairs such that one member is connected and the other is not; isospectral pairs such that one member is smooth and the other is not; isospectral 4-tuples; billiards whose spectrum remains invariant when Dirichlet and
  Neumann boundaries are swapped. These billiards were investigated by 
 \textcite{JakLevNadPol}. The simplest example is a billiard of semi-circular
shape: if the equation of the billiard on the complex plane is given by
$\{z\in\mathbb{C};0\leq \arg(z)\leq\pi;|r|\leq 1\}$, the Dirichlet
boundary conditions correspond to 
$\{z\in\mathbb{C};|r|=1, \pi/4\leq \arg(z)\leq 3\pi/4, \Re(z)<0\}$.
A necessary condition for this Dirichlet-Neumann isospectrality is that the Dirichlet
boundary has the same total length as the Neumann boundary. 

Such domains have been investigated numerically \cite{DriGot}
as well as analytically \cite{OkaShu}, and experimental setups have 
been proposed by \textcite{DriGot}.

All these examples have the property that the length difference between the
 Dirichlet boundary and the Neumann boundary are the same. This turns out
to be a necessary condition similar to those obtained from Weyl's law \eqref{dvpweyl}
applying to isospectral billiards
derived by \textcite{LevParPol} for mixed-boundary condition isospectral billiards. 
In particular, such 
isospectral pairs need to have the same area, the same length difference between the
 Dirichlet boundary and the Neumann boundary, and the same curvature-singularity 
properties, namely, the quantity
\begin{equation}
2\int_{\partial B}\kappa(s)ds+\sum_{DD}\frac{\pi^2-\beta^2}{\beta}
+\sum_{NN}\frac{\pi^2-\beta^2}{\beta}
-\frac{1}{2}\sum_{DN}\frac{\pi^2+2\beta^2}{\beta},
\end{equation}
where $\kappa$ is the curvature and $\beta$ represents the angles at the
Dirichlet-Dirichlet, Neumann-Neumann or Dirichlet-Neumann boundary
intersections, must be the same for both billiards.

Finally, we observe that, for some of the examples produced by \textcite{LevParPol},
it was shown that two isospectral domains produce a different
number of nodal domains (domains separated by nodal lines where $\Psi=0$; 
see section \ref{nodal}).

\subsection{Homophonic pairs}

{\em Homophonic pairs} in $\mathbb{R}^2$ are nonisometric compact domains that have a distinguished point such that the corresponding (normalized) Dirichlet eigenfunctions take equal values at that point.
This could be interpreted in the following way: If the corresponding drums are struck at these special points, then they sound the same in such a way that every frequency is excited to the same intensity for each.\\
An example of two billiards that are isospectral and homophonic 
\cite{BusConDoySem} is provided in Appendix
\ref{gallery} (example $21_1$ right). These billiards 
sound the same when struck at the interior points where six triangles meet.\\

\subsection{Spectral problems for Lie geometries}

There exists a vast literature on spectral problems for (finite) graphs
| see the excellent paper \cite{VDH}.
In this section we consider a spectral (''Kac type'') problem
for graphs that are associated with the most important incidence geometries.

We have seen in the previous sections that the construction of isospectral pairs is based on properties of finite projective spaces and their automorphism groups.
In this section we show that this construction is a special case of a wider
class of similar constructions based on so-called generalized polygons,
which are the natural generalization of projective planes.

One defines a finite axiomatic {\em projective plane $\Pi$} of {\em order
$n$}, where $n\in\mathbb{N}$, as a point-line incidence structure satisfying the following conditions: (i) each point is incident with $n + 1$ lines
 and each line is incident with  $n + 1$ points; 
(ii) any two distinct lines intersect in exactly one point and any two
 distinct points lie on exactly one line.
One also traditionally requires that $n$ be $\ge2$ to exclude the
uninteresting cases
of a single line and a point not on it ($n=-1$), a single line and one
point on it ($n=0$),
or the three vertices and three sides of a triangle ($n=1$). This is
equivalent to
requiring that $\Pi$ contains an ordinary quadrangle (four points with no
three on a line)
as subgeometry. It is easily seen
that a finite projective plane of order $n$ has $n^2+n+1$ points and
$n^2+n+1$ lines.

The obvious examples of finite projective planes are the projective planes
$\PG(2,q)$ over
finite fields $\mathbb{F}_q$ as defined in section \ref{transplantation}.  In this case the order $n
= |\mathbb{F}_q|$ is a prime power, and in fact no examples
of finite projective planes of non prime power order are known. A
classical theorem of Moufang
 states that a finite projective plane is isomorphic to some $\PG(2,q)$ if
and
only if a certain configurational property corresponding to the classical
theorem of Desargues
is satisfied.  Projective planes of this type are therefore often called
{\em Desarguesian}, and since these correspond to planes coordinatized
over finite fields, we also use this terminology for projective spaces of
dimension $n \geq 3$, as already mentioned. 
However, many finite projective planes are known which are not
Desarguesian; see \cite{HP}.\\

\begin{figure}
\centering
\includegraphics[width=6cm]{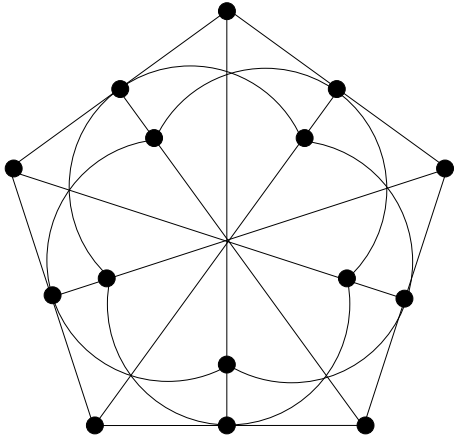}
\caption{The unique generalized quadrangle of order $2$: the symplectic
$\mathbf{W}(3,2)$.\label{fig20}}
\end{figure}

{\bf Generalized polygons.}
Let $n \geq 3$ be a natural number. A (thick) {\em generalized $n$-gon} or
(thick) {\em generalized polygon} (GP)
is a point-line geometry $\Gamma = (\mP,\mB,\I)$, where $\mP$ is the point
set, $\mB$ is the line set and $\I \subset (\mP \times \mB)\cup(\mB \times
\mP)$ is a symmetric incidence relation,
so that the following axioms are satisfied:
\begin{itemize}
\item[(i)]
$\Gamma$ contains no  $k$-gon (in the ordinary sense) for $2 \leq k < n$;
\item[(ii)]
Any two elements $x,y \in \mP \cup \mB$ are contained in some ordinary
$n$-gon  in
$\Gamma$;
\item[(iii)]
There exists an ordinary $(n + 1)$-gon  in $\Gamma$.
\end{itemize}
The {\em point graph} of a point-line geometry is the graph of which the
vertices are the points of the geometry, and for which two vertices
are joined by an edge if they are collinear in the geometry. Equivalently,
a generalized polygon could be defined as a point-line geometry
for which the point graph is bipartite of diameter $n$ and girth $2n$ 
(see, e.g., Fig.~\ref{fig20}).

The generalized $3$-gons are precisely
the aforementioned projective planes.
If (iii) is not satisfied for $\Gamma$, then $\Gamma$ is called {\em thin}.
Otherwise, it is called {\em thick}.
Each thick generalized $n$-gon, $n \geq 3$, $\Gamma$ has an {\em order}:
there are (not necessarily finite)
constants $s > 1$ and $t > 1$ so that each point is incident with $t + 1$
lines and each line is incident with $s + 1$ points. We then
say that $\Gamma$ has order $(s,t)$. Note that, for a point $x$ and a
line $L$, $x \I L$ means that $(x,L) \in \I$ (and so also $(L,x) \in
\I$).\\

{\bf Collinearity matrices and a spectral problem. } 
Suppose $\Gamma = (\mP,\mB,\I)$ is a finite  GP ($\Gamma$ has a finite
number of points and lines) of order $(s,t)$, set $\vert \mP\vert = v$,
and let  $\{x_1,x_2,\ldots,x_v\} = \mP$ be the point set.
Define the {\em collinearity matrix} $\A(\Gamma) = \A$ as the $v\times
v$-matrix $(a_{ij})$ for which $a_{ij} = 1$ if $x_i \sim x_j \ne x_i$ (the
latter notation meaning that $x_i$ and $x_j$ are different collinear
points),
and $0$ otherwise. So it is the adjacency matrix
of the point graph of $\Gamma$.
The {\em (point) spectrum} of $\Gamma$ is the spectrum of $\A$, and we
denote it by $\spec(\A)$.\\

The following quantum mechanical question is the Kac inverse problem for
the theory of GPs.
\bq
\label{KT}
Let $\Gamma$ and $\Gamma'$ be distinct  finite thick generalized polygons
with associated collinearity matrices $\A$ and $\A'$, respectively.
Does $\spec(\hA) = \spec(\hA')$ imply that $\Gamma \cong \Gamma'$?
\eq
Clearly, a similar problem can be posed for the line spectrum, but as
points and lines play essentially the same role in a GP, we only
consider the question in its above form.

Question \ref{KT} can be reduced to an important question in the theory of
GPs:
\bt[\textcite{KT5}]
\label{KT2}
 Let $\Gamma$ and $\Gamma'$ be distinct  finite thick generalized polygons
with associated collinearity matrices $\A$ and $\A'$, respectively.
 Then $\spec(\A) = \spec(\A')$ if and only if $\Gamma$ and $\Gamma'$ have
the same order.
\et
Details of the proof can be found in Appendix \ref{appB}.

\subsection{Further questions}
As mentioned, the literature on isospectrality is huge and
continuously growing. There is also a vast literature on isospectral graphs.  
In this section we state some fundamental open problems (which reflect our
personal interest) on billiards and graphs. 

Interesting problems in construction theory are numerous: We state only some of them. Perhaps the single most important open problem 
in Kac theory is the following: We have constructed pairs of isospectral billiards made
 of 7, 13, 15, or 21 tiles. Is it possible to go beyond that number? In mathematical words,
can one show that for all $N \in \mathbb{N}$ there exists an $N^* \geq N$ such that there are isospectral pairs on $N^*$ tiles? Equivalently, can one show that there are infinitely many pairs of involution graphs that yield isospectral pairs?

All examples constructed so far are polygonal examples. Even if different base tiles can be chosen, the unfolding rule imposes the presence of corners
in the boundary of the billiard. A natural question is thus whether one 
can construct isospectral $\mathbb{R}^2$-domains with smooth boundaries.

We have seen that point-line duality in finite projective spaces is at the root of billiard isospectrality and provides a transplantation property
between billiards. Since only one recipe is known for constructing
isospectral pairs, one may ask the following:
Is it possible to construct isospectral pairs which are not transplantable?
More generally, are the following statements achievable:
Derive criterions for pairs of involution graphs to yield isospectral plane domains; 
Construct isospectral pairs on $\infty$ tiles (perhaps by a free construction);
Find examples of (planar) isospectral pairs not coming from Sunada triples, 
or still arising from Sunada triples but not being transplantable.

\bigskip
On the group theoretical level, we pose the following question: 
Are the operator groups of (transplantable) isospectral pairs always two-transitive?
If so, the classification of finite simple groups could be used to classify such operator groups. 
In the same spirit, one could ask as to whether other finite simple groups can act as operator groups.
A related question is to develop a theory of isospectral ``domains'' on general buildings.
Note that the projective completion of $\mathbb{R}^2$ is a rank $2$-building
over $\mathbb{R}$ (see \cite{Tit:74} for an introduction on buildings).
The same questions could all be formulated for ``isospectral $n$-tuples'', $n > 2$.

%\newpage

\begin{acknowledgments}
K.~T. acknowledges the Fund for Scientific Research --- Flanders (Belgium) for financial support.
This paper  was partly written while K.~T.~was visiting the Discrete Mathematics group of the
Technical University of Eindhoven, The Netherlands, whose
hospitality he gratefully acknowledges.
The paper was finished while both authors were hosted by Institut Henri
Poincar\'{e} (IHP) at Paris, whose hospitality is gratefully acknowledged.
\end{acknowledgments}

%\newpage
\appendix
%%%%%%%%%%%%%%%%%%%%%%%%%%%%%%%%%%%%%%%%%%%%%%%%%%%%%%%%%%%%%%%%%%%%%%%%%%%%%%%%%%%%%%%%%%%%%%%%%%%%%%%%%%
\section{Gallery of examples}
%%%%%%%%%%%%%%%%%%%%%%%%%%%%%%%%%%%%%%%%%%%%%%%%%%%%%%%%%%%%%%%%%%%%%%%%%%%%%%%%%%%%%%%%%%%%%%%%%%%%%%%%%%
\label{gallery}
\subsection{Some modes}
\begin{figure}[ht]
\begin{center}
\includegraphics[width=0.675\linewidth]{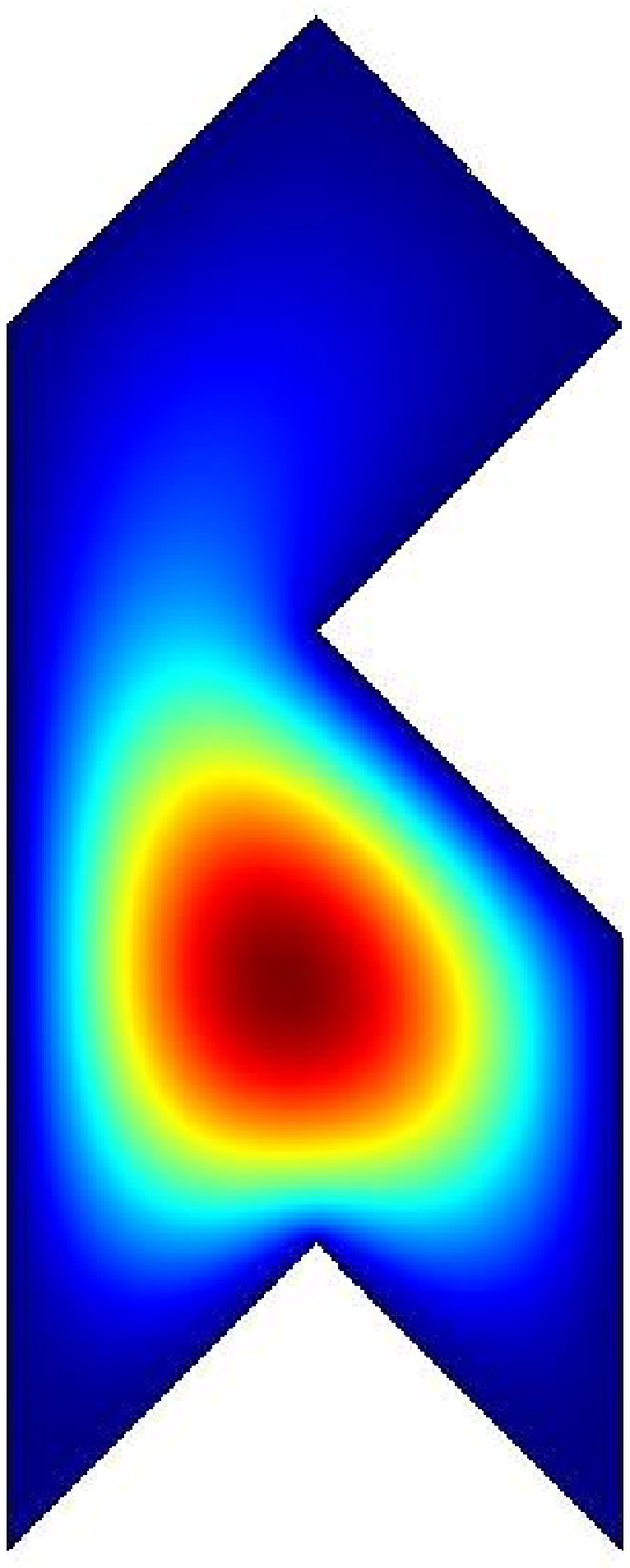}\hspace{-2.3cm}
\includegraphics[width=0.525\linewidth]{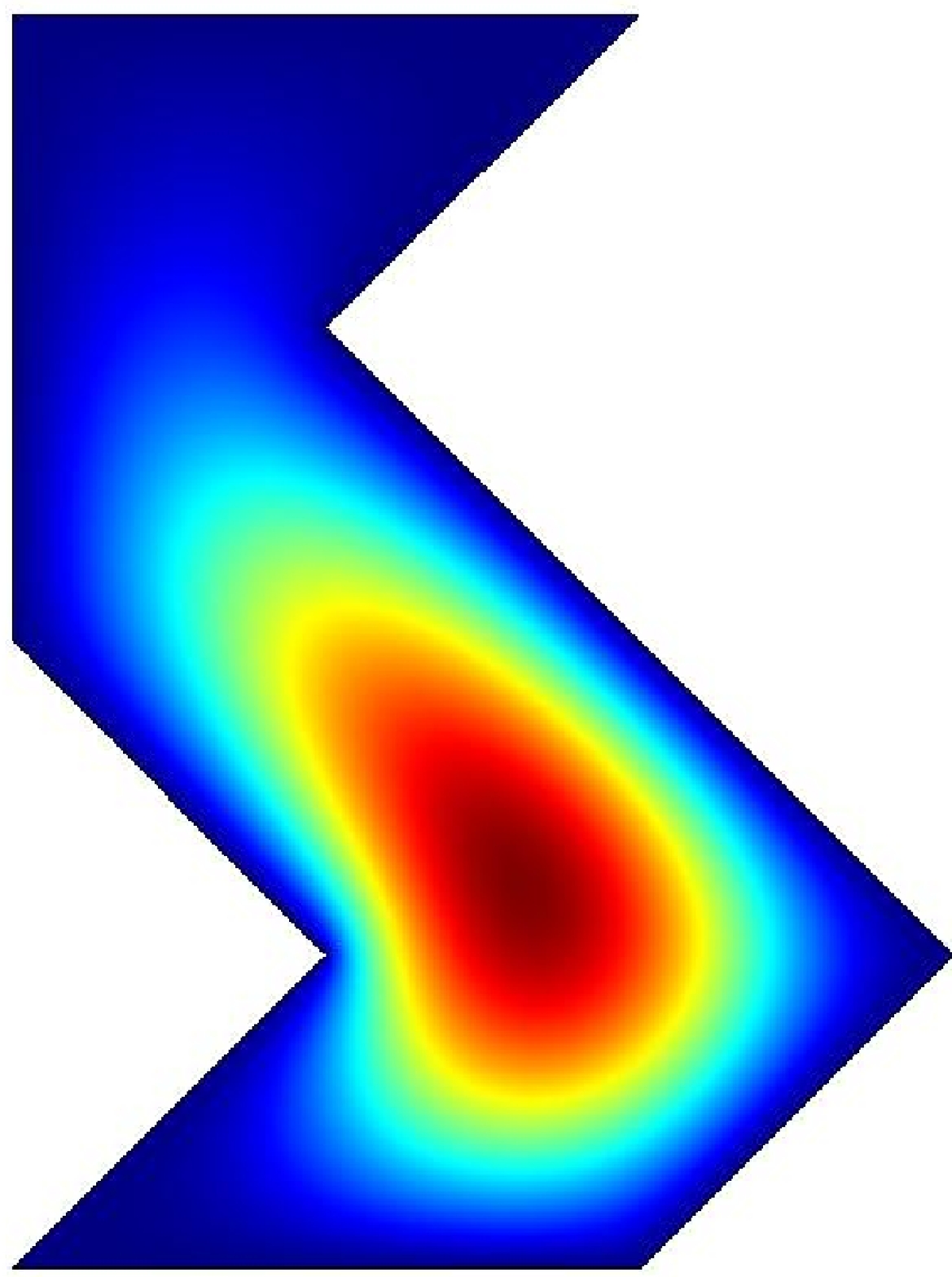}
\end{center}
\caption{(Color online) Fundamental mode.\label{fig21}}
\end{figure}

\begin{figure}[ht]
\begin{center}
\includegraphics[width=0.675\linewidth]{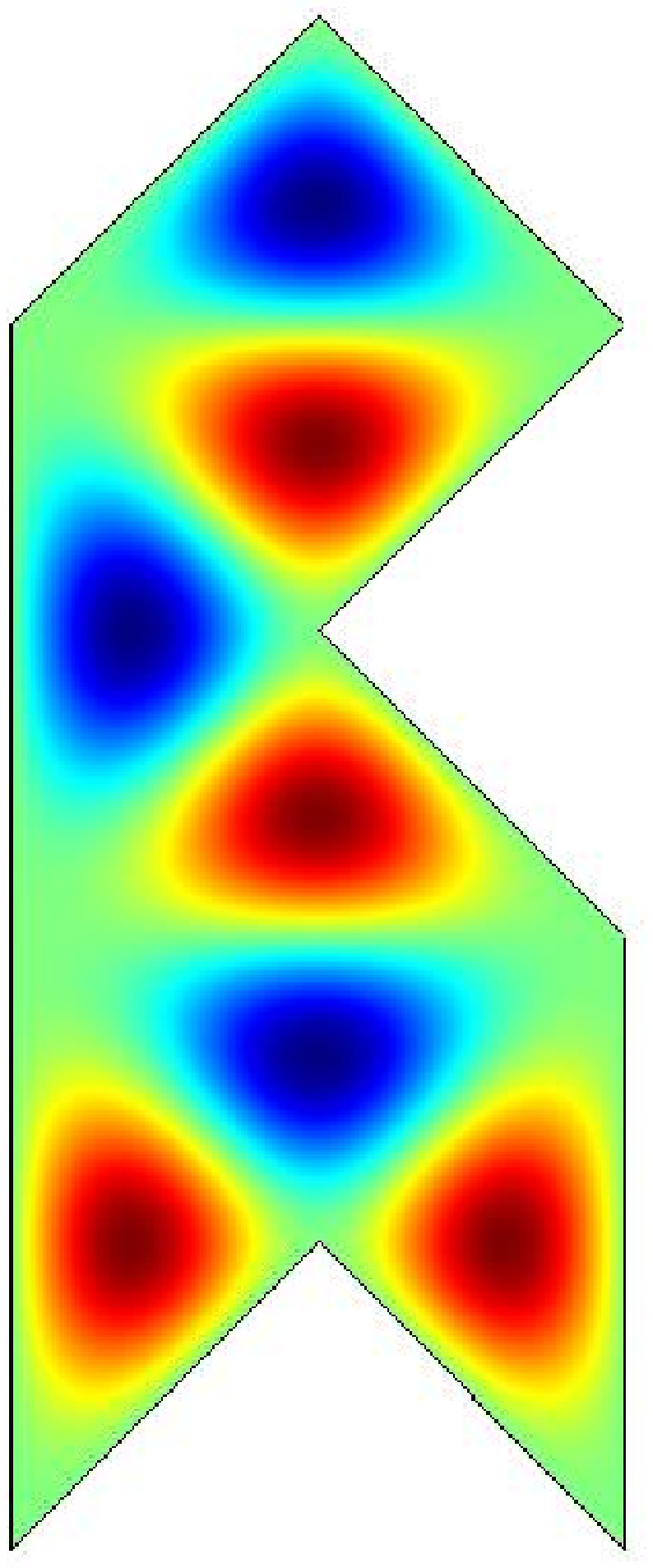}\hspace{-2.3cm}
\includegraphics[width=0.525\linewidth]{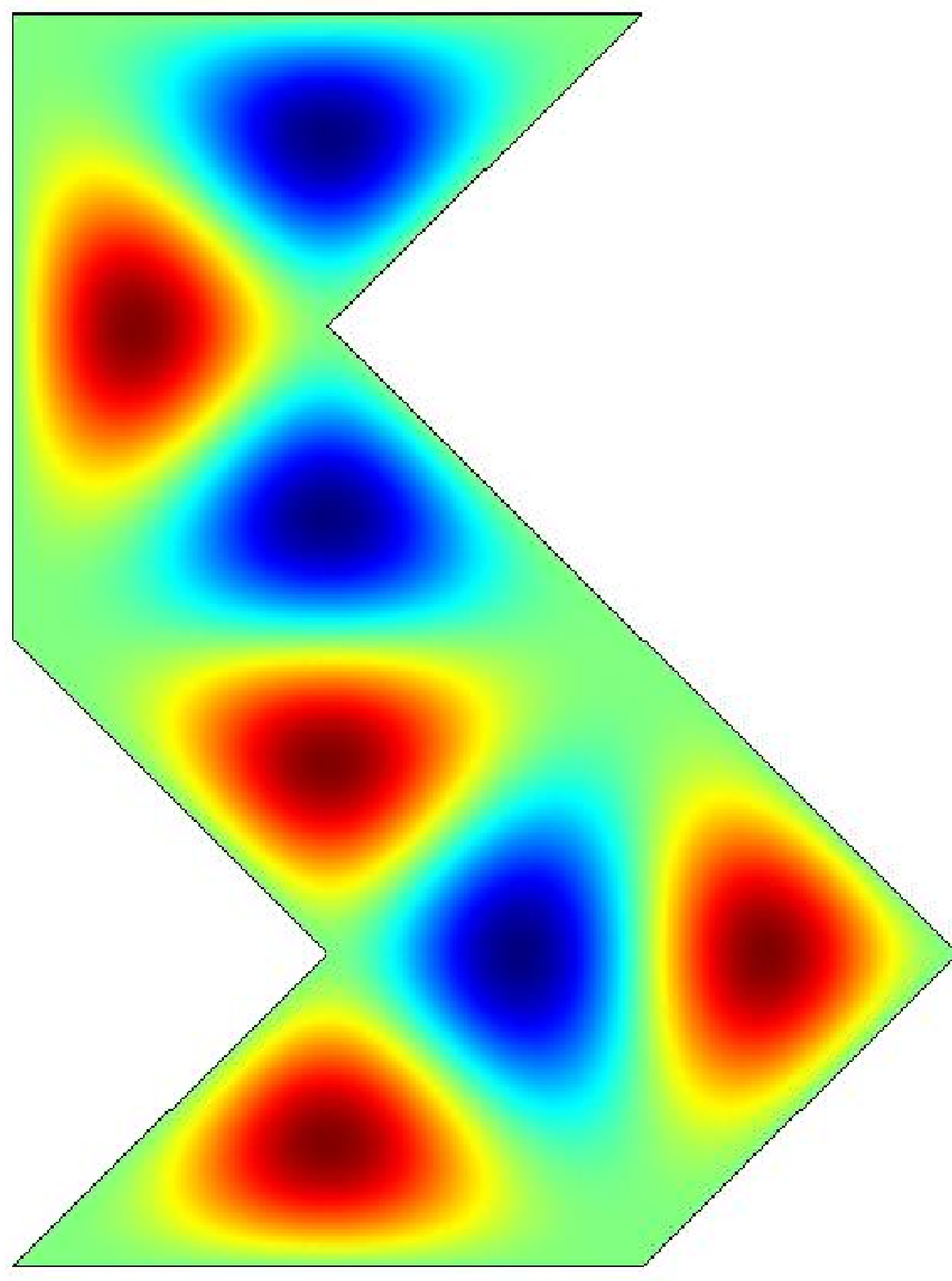}
\end{center}
\caption{(Color online) First triangular mode (ninth mode).\label{fig22}}
\end{figure}

\begin{figure}[ht]
\begin{center}
\includegraphics[width=0.675\linewidth]{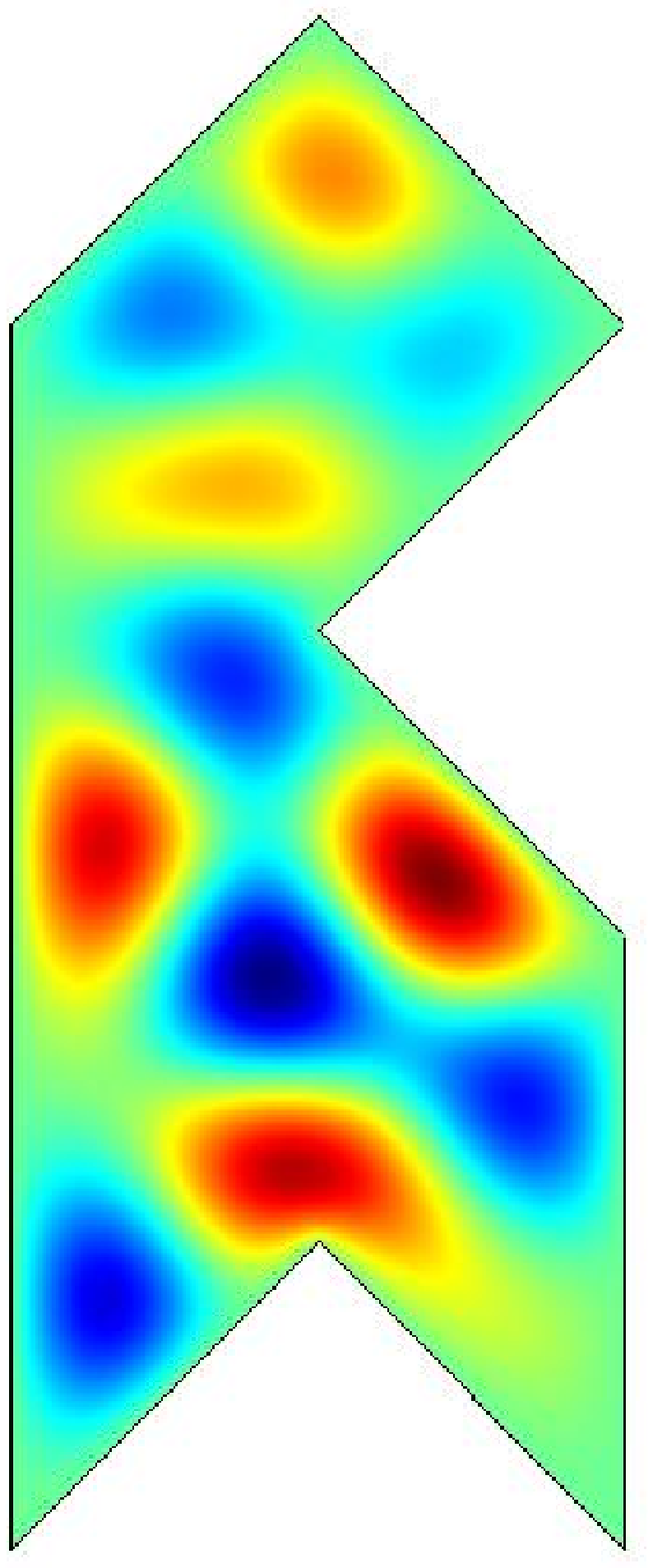}\hspace{-2.3cm}
\includegraphics[width=0.525\linewidth]{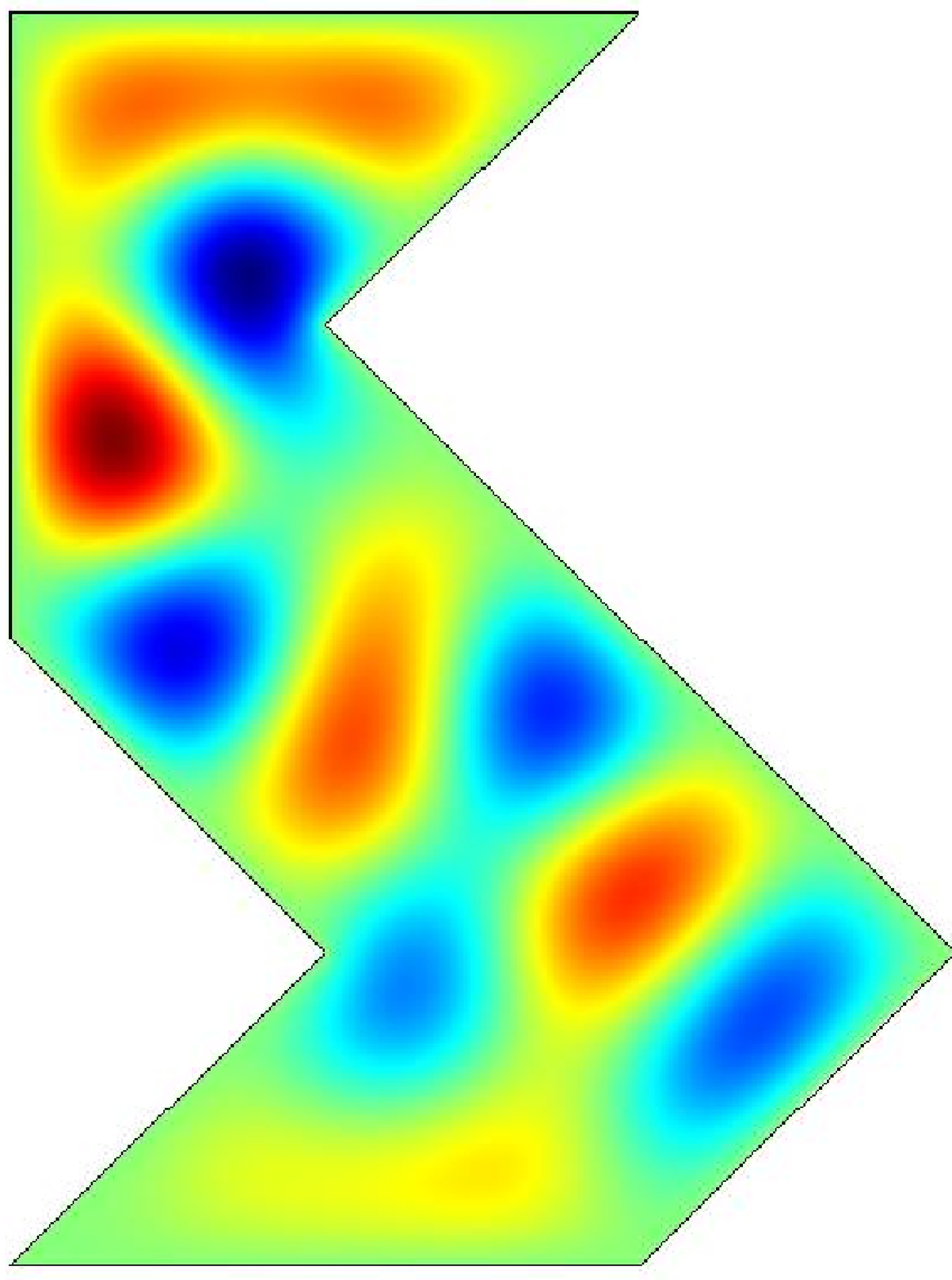}
\end{center}
\caption{(Color online) $15$th mode.\label{fig23}}
\end{figure}

Here we plot some eigenfunctions for the pair of billiards
of Fig.~\ref{celebrated}. Figure \ref{fig21} corresponds to the fundamental
mode, Fig.~\ref{fig22} to the first triangular mode, whose
nodal lines coincide with edges between the triangular
tiles. Figure \ref{fig23} corresponds to an excited state.

%\newpage
\subsection{The $17$ families of isospectral pairs and their mathematical construction}
\begin{figure}[h]
\begin{center}
\includegraphics[width=0.45\linewidth]{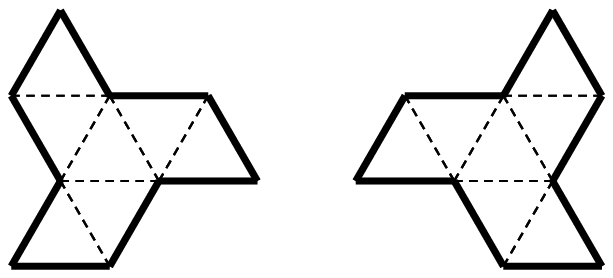}\hspace{0.08\linewidth}
\includegraphics[width=0.45\linewidth]{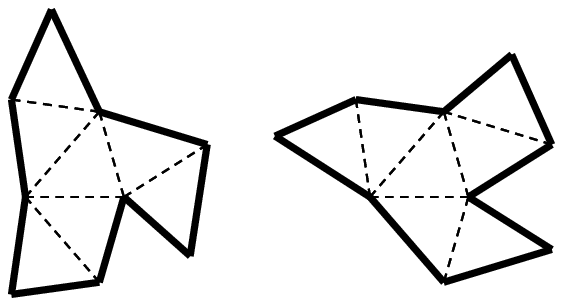}\caption{Pair $7_1$. Sunada triple $G=\PSL(3, 2)$, $G_i=\langle a_i, b_i, c_i\rangle$, $i=1,2$, with
$a_1=(0\ 1)(2\ 5)$, 
$b_1=(0\ 2)(3\ 4)$, 
$c_1=(0\ 4)(1\ 6)$, 
$a_2=(0\ 4)(2\ 3)$, 
$b_2=(0\ 1)(4\ 6)$, 
$c_2=(0\ 2)(1\ 5)$.
}
\label{7_1}
\end{center}
\end{figure}

\begin{figure}[h]
\begin{center}
\includegraphics[width=0.45\linewidth]{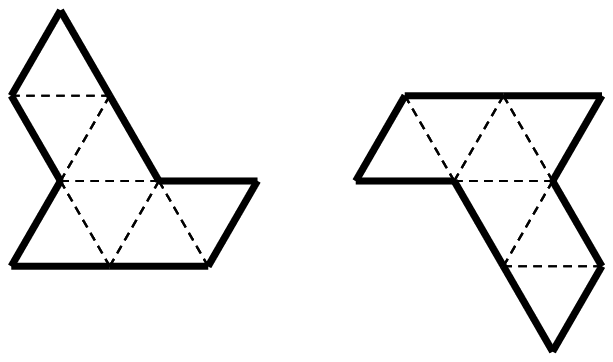}\hspace{0.08\linewidth}
\includegraphics[width=0.45\linewidth]{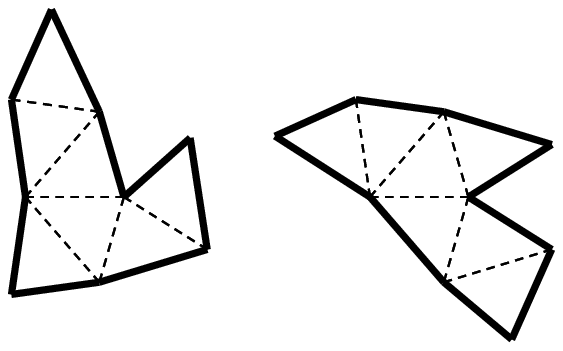}\caption{Pair $7_2$. Sunada triple $G=\PSL(3, 2)$, $G_i=\langle a_i, b_i, c_i\rangle$, $i=1,2$, with
$a_1=(0\ 1)(2\ 5)$, 
$b_1=(1\ 5)(3\ 4)$, 
$c_1=(0\ 4)(1\ 6)$, 
$a_2=(0\ 4)(2\ 3)$, 
$b_2=(0\ 6)(1\ 4)$, 
$c_2=(0\ 2)(1\ 5)$.
}
\label{7_2}
\end{center}
\end{figure}

\begin{figure}[h]
\begin{center}
\includegraphics[width=0.45\linewidth]{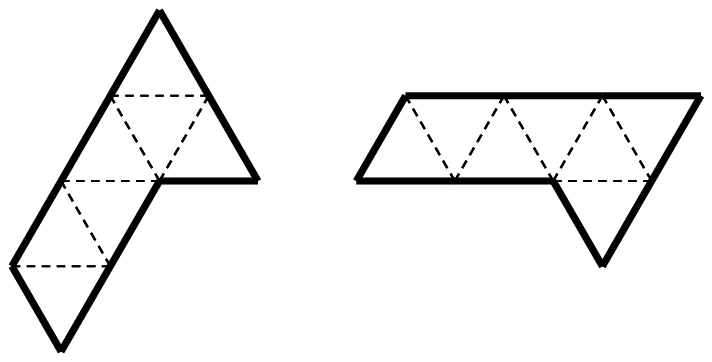}\hspace{0.08\linewidth}
\includegraphics[width=0.45\linewidth]{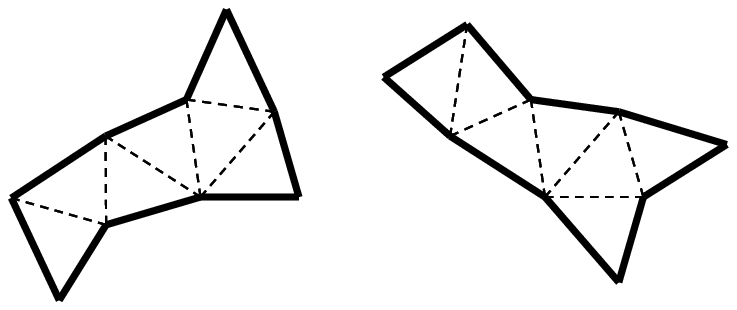}\caption{Pair $7_3$. Sunada triple $G=\PSL(3, 2)$, $G_i=\langle a_i, b_i, c_i\rangle$, $i=1,2$, with
$a_1=(2\ 5)(4\ 6)$, 
$b_1=(1\ 5)(3\ 4)$, 
$c_1=(0\ 4)(1\ 6)$, 
$a_2=(0\ 3)(2\ 4)$, 
$b_2=(0\ 6)(1\ 4)$, 
$c_2=(0\ 2)(1\ 5)$.
}
\label{7_3}
\end{center}
\end{figure}

\begin{figure}[h]
\begin{center}
\includegraphics[width=0.45\linewidth]{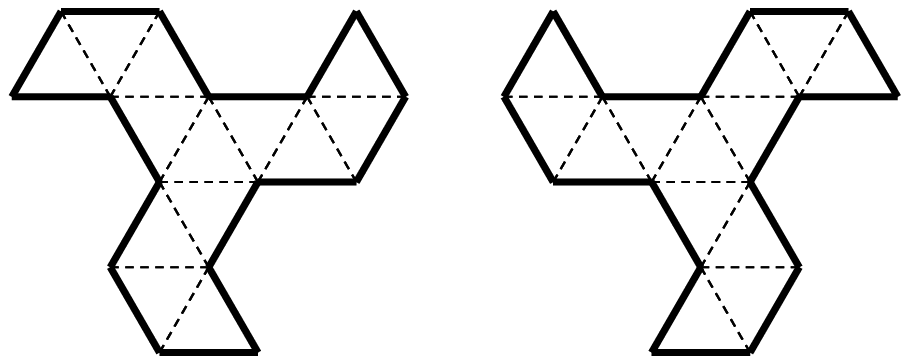}\hspace{0.08\linewidth}
\includegraphics[width=0.45\linewidth]{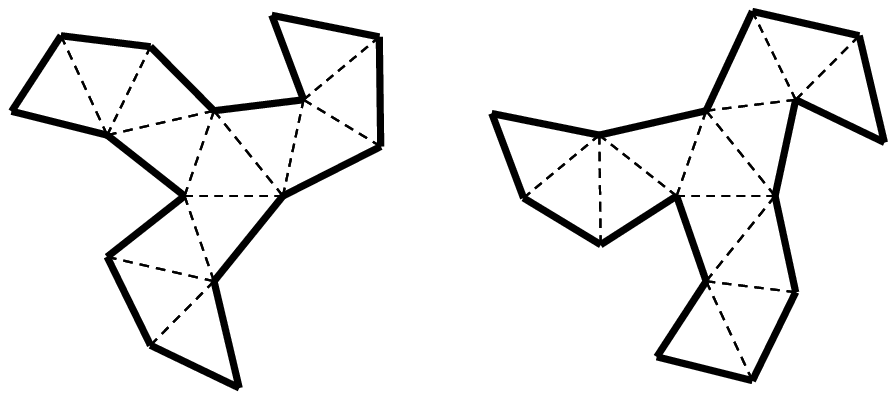}\caption{Pair $13_1$. Sunada triple $G=\PSL(3, 3)$, $G_i=\langle a_i, b_i, c_i\rangle$, $i=1,2$, with
$a_1=(0\ 12)(1\ 10)(3\ 5)(6\ 7)$, 
$b_1=(0\ 10)(2\ 9)(3\ 4)(5\ 8)$, 
$c_1=(0\ 4)(1\ 6)(2\ 11)(9\ 12)$, 
$a_2=(0\ 4)(2\ 3)(6\ 8)(9\ 10)$, 
$b_2=(0\ 12)(1\ 4)(5\ 11)(6\ 9)$, 
$c_2=(0\ 10)(1\ 5)(2\ 7)(3\ 12)$.
}
\label{13_1}
\end{center}
\end{figure}

\begin{figure}[h]
\begin{center}
\includegraphics[width=0.45\linewidth]{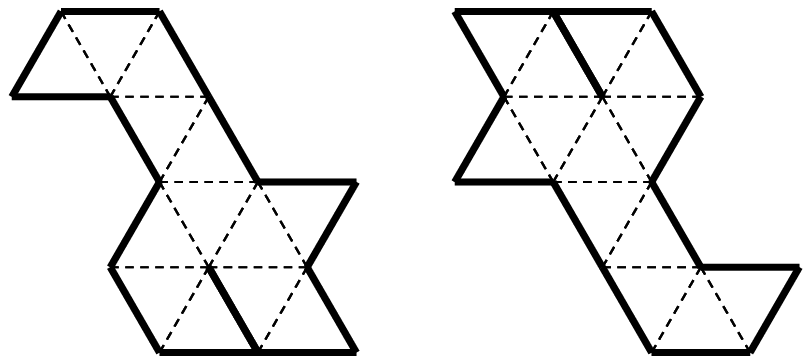}\hspace{0.08\linewidth}
\includegraphics[width=0.45\linewidth]{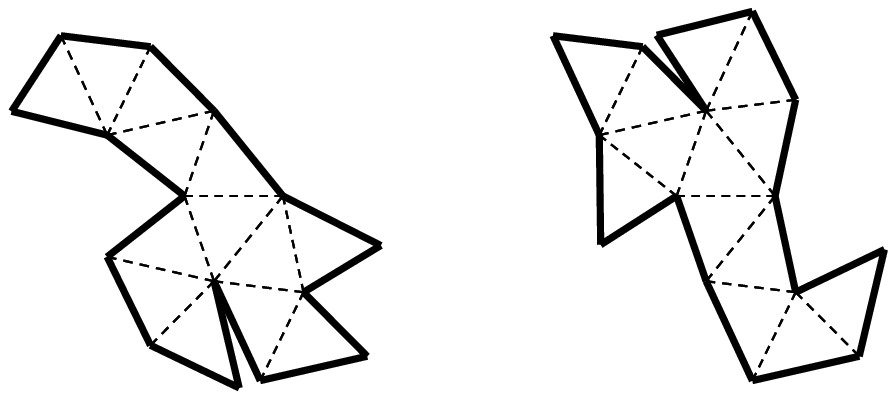}\caption{Pair $13_2$. Sunada triple $G=\PSL(3, 3)$, $G_i=\langle a_i, b_i, c_i\rangle$, $i=1,2$, with
$a_1=(0\ 12)(1\ 10)(3\ 5)(6\ 7)$, 
$b_1=(1\ 12)(2\ 9)(3\ 8)(4\ 5)$, 
$c_1=(0\ 4)(1\ 6)(2\ 11)(9\ 12)$, 
$a_2=(0\ 4)(2\ 3)(6\ 8)(9\ 10)$, 
$b_2=(0\ 1)(4\ 12)(5\ 11)(8\ 10)$, 
$c_2=(0\ 10)(1\ 5)(2\ 7)(3\ 12)$.
}
\label{13_2}
\end{center}
\end{figure}

\begin{figure}[h]
\begin{center}
\includegraphics[width=0.45\linewidth]{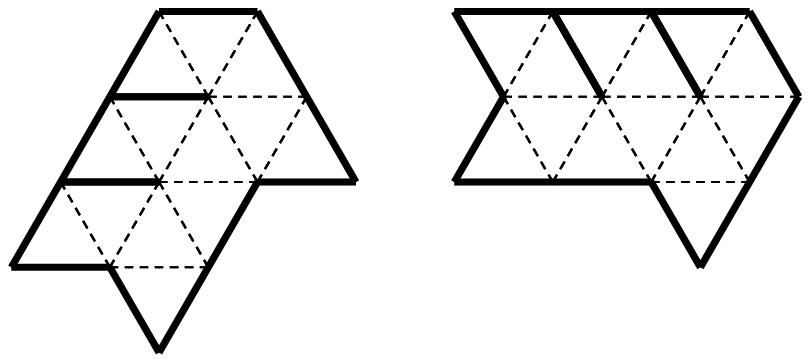}\hspace{0.08\linewidth}
\includegraphics[width=0.45\linewidth]{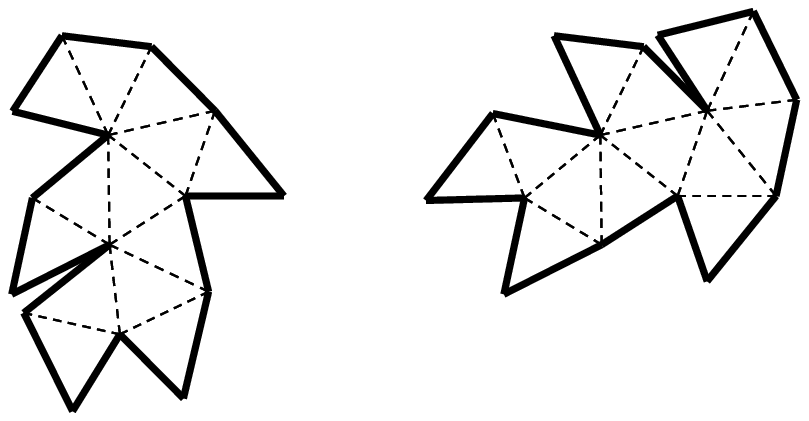}\caption{Pair $13_3$. Sunada triple $G=\PSL(3, 3)$, $G_i=\langle a_i, b_i, c_i\rangle$, $i=1,2$, with
$a_1=(1\ 7)(3\ 5)(4\ 9)(6\ 10)$, 
$b_1=(1\ 12)(2\ 9)(3\ 8)(4\ 5)$, 
$c_1=(0\ 4)(1\ 6)(2\ 11)(9\ 12)$, 
$a_2=(0\ 9)(4\ 10)(6\ 8)(7\ 12)$, 
$b_2=(0\ 1)(4\ 12)(5\ 11)(8\ 10)$, 
$c_2=(0\ 10)(1\ 5)(2\ 7)(3\ 12)$.
}
\label{13_3}
\end{center}
\end{figure}

\begin{figure}[h]
\begin{center}
\includegraphics[width=0.45\linewidth]{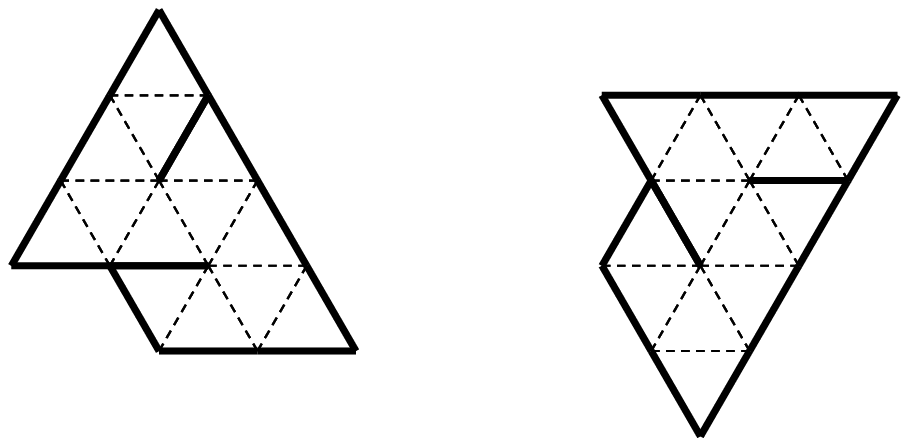}\hspace{0.08\linewidth}
\includegraphics[width=0.45\linewidth]{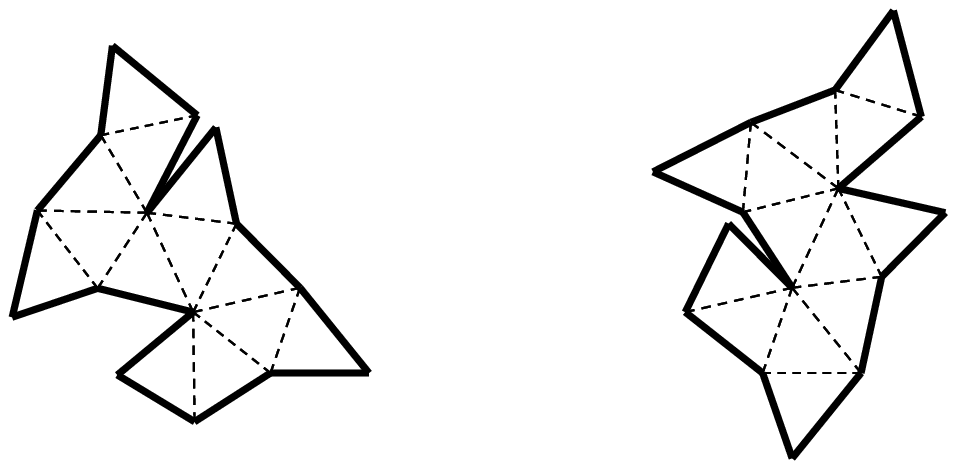}\caption{Pair $13_4$. Sunada triple $G=\PSL(3, 3)$, $G_i=\langle a_i, b_i, c_i\rangle$, $i=1,2$, with
$a_1=(1\ 7)(3\ 5)(4\ 9)(6\ 10)$, 
$b_1=(0\ 5)(1\ 2)(6\ 12)(9\ 11)$, 
$c_1=(0\ 4)(1\ 6)(2\ 11)(9\ 12)$, 
$a_2=(0\ 9)(4\ 10)(6\ 8)(7\ 12)$, 
$b_2=(0\ 11)(1\ 8)(2\ 7)(3\ 4)$, 
$c_2=(0\ 10)(1\ 5)(2\ 7)(3\ 12)$.
}
\label{13_4}
\end{center}
\end{figure}

\begin{figure}[h]
\begin{center}
\includegraphics[width=0.45\linewidth]{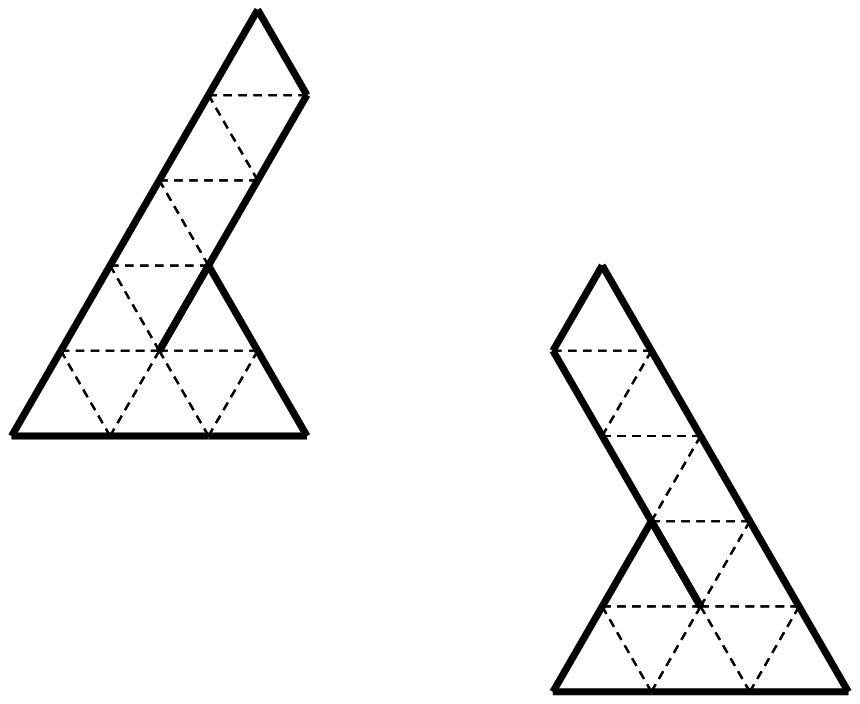}\hspace{0.08\linewidth}
\includegraphics[width=0.45\linewidth]{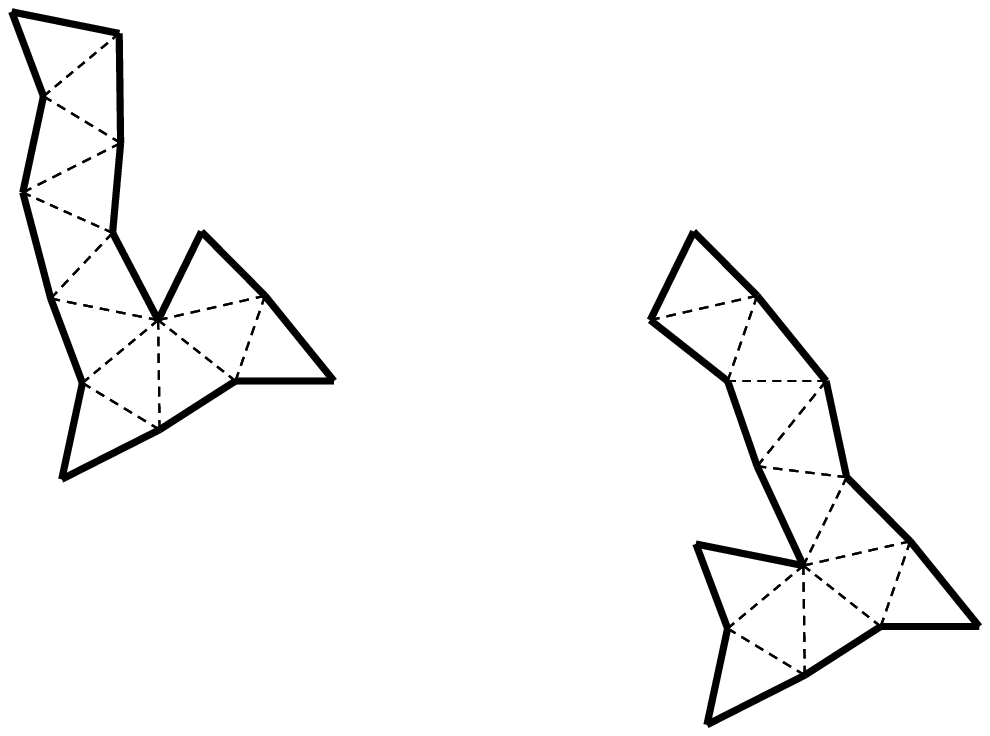}\caption{Pair $13_5$. Sunada triple $G=\PSL(3, 3)$, $G_i=\langle a_i, b_i, c_i\rangle$, $i=1,2$, with
$a_1=(1\ 7)(3\ 5)(4\ 9)(6\ 10)$, 
$b_1=(0\ 5)(1\ 2)(6\ 12)(9\ 11)$, 
$c_1=(0\ 4)(1\ 6)(2\ 11)(9\ 12)$, 
$a_2=(0\ 9)(4\ 10)(6\ 8)(7\ 12)$, 
$b_2=(0\ 11)(1\ 8)(2\ 7)(3\ 4)$, 
$c_2=(0\ 10)(1\ 5)(2\ 7)(3\ 12)$.
}
\label{13_5}
\end{center}
\end{figure}

\begin{figure}[h]
\begin{center}
\includegraphics[width=0.45\linewidth]{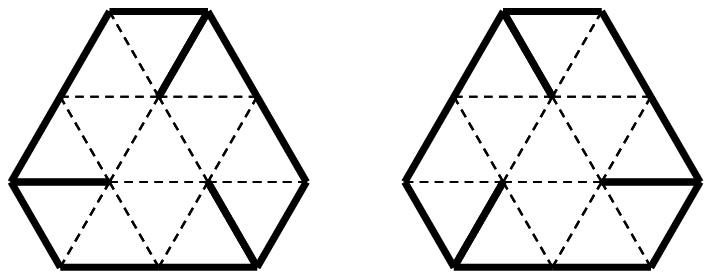}\hspace{0.08\linewidth}
\includegraphics[width=0.45\linewidth]{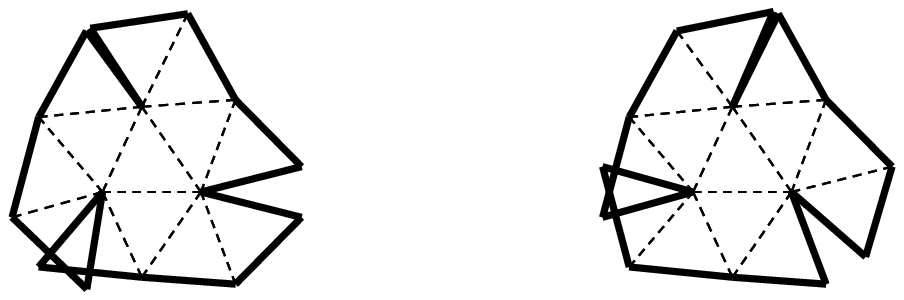}\caption{Pair $13_6$. Sunada triple $G=\PSL(3, 3)$, $G_i=\langle a_i, b_i, c_i\rangle$, $i=1,2$, with
$a_1=(0\ 2)(1\ 7)(3\ 6)(5\ 10)$, 
$b_1=(0\ 6)(2\ 4)(3\ 8)(5\ 9)$, 
$c_1=(0\ 5)(1\ 2)(6\ 12)(9\ 11)$, 
$a_2=(0\ 7)(3\ 11)(6\ 8)(9\ 12)$, 
$b_2=(0\ 8)(1\ 10)(5\ 11)(7\ 9)$, 
$c_2=(0\ 11)(1\ 8)(2\ 7)(3\ 4)$.
}
\label{13_6}
\end{center}
\end{figure}

\begin{figure}[h]
\begin{center}
\includegraphics[width=0.45\linewidth]{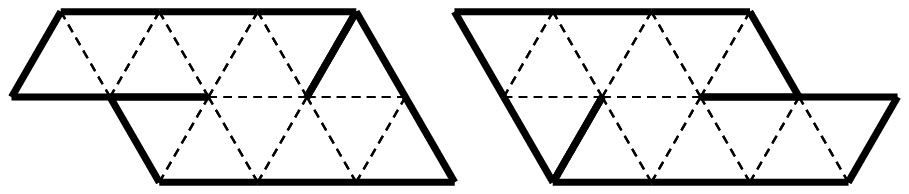}\hspace{0.08\linewidth}
\includegraphics[width=0.45\linewidth]{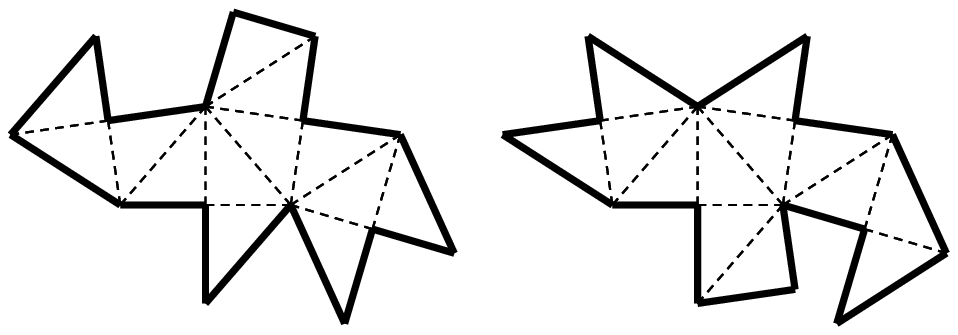}\caption{Pair $13_7$. Sunada triple $G=\PSL(3, 3)$, $G_i=\langle a_i, b_i, c_i\rangle$, $i=1,2$, with
$a_1=(0\ 2)(1\ 7)(3\ 6)(5\ 10)$, 
$b_1=(0\ 4)(2\ 3)(6\ 8)(9\ 10)$, 
$c_1=(0\ 5)(1\ 2)(6\ 12)(9\ 11)$, 
$a_2=(0\ 7)(3\ 11)(6\ 8)(9\ 12)$, 
$b_2=(0\ 12)(1\ 10)(3\ 5)(6\ 7)$, 
$c_2=(0\ 11)(1\ 8)(2\ 7)(3\ 4)$.
}
\label{13_7}
\end{center}
\end{figure}

\begin{figure}[h]
\begin{center}
\includegraphics[width=0.45\linewidth]{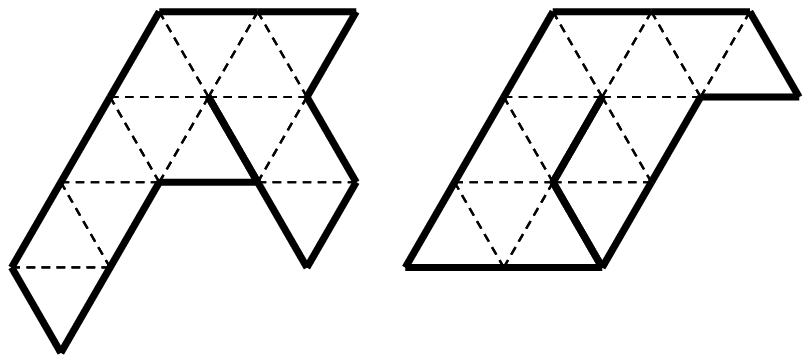}\hspace{0.08\linewidth}
\includegraphics[width=0.45\linewidth]{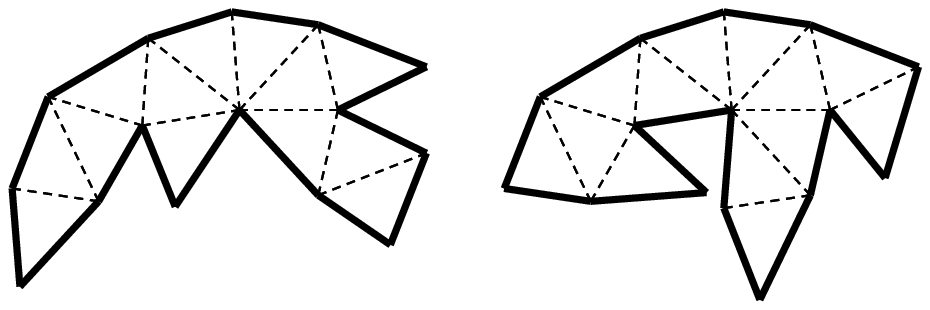}\caption{Pair $13_8$. Sunada triple $G=\PSL(3, 3)$, $G_i=\langle a_i, b_i, c_i\rangle$, $i=1,2$, with
$a_1=(0\ 10)(1\ 5)(2\ 7)(3\ 12)$, 
$b_1=(0\ 4)(2\ 3)(6\ 8)(9\ 10)$, 
$c_1=(0\ 5)(1\ 2)(6\ 12)(9\ 11)$, 
$a_2=(0\ 4)(1\ 6)(2\ 11)(9\ 12)$, 
$b_2=(0\ 12)(1\ 10)(3\ 5)(6\ 7)$, 
$c_2=(0\ 11)(1\ 8)(2\ 7)(3\ 4)$.
}
\label{13_8}
\end{center}
\end{figure}

\begin{figure}[h]
\begin{center}
\includegraphics[width=0.45\linewidth]{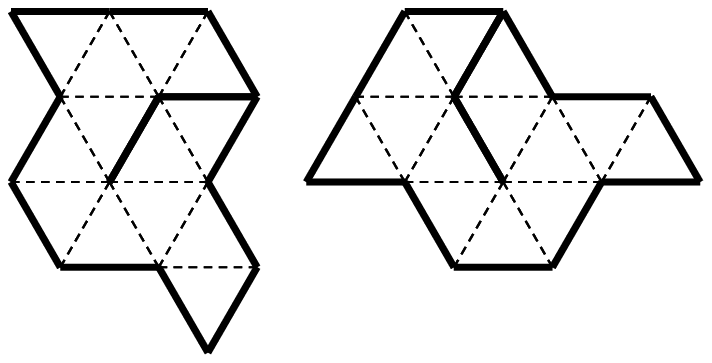}\hspace{0.08\linewidth}
\includegraphics[width=0.45\linewidth]{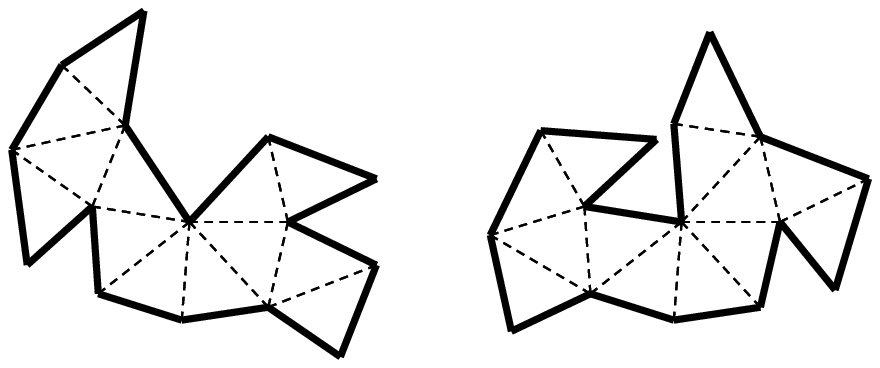}\caption{Pair $13_9$. Sunada triple $G=\PSL(3, 3)$, $G_i=\langle a_i, b_i, c_i\rangle$, $i=1,2$, with
$a_1=(0\ 10)(1\ 5)(2\ 7)(3\ 12)$, 
$b_1=(1\ 10)(3\ 6)(5\ 7)(9\ 11)$, 
$c_1=(0\ 5)(1\ 2)(6\ 12)(9\ 11)$, 
$a_2=(0\ 4)(1\ 6)(2\ 11)(9\ 12)$, 
$b_2=(0\ 3)(2\ 4)(6\ 8)(7\ 11)$, 
$c_2=(0\ 11)(1\ 8)(2\ 7)(3\ 4)$.
}
\label{13_9}
\end{center}
\end{figure}

\begin{figure}[h]
\begin{center}
\includegraphics[width=0.45\linewidth]{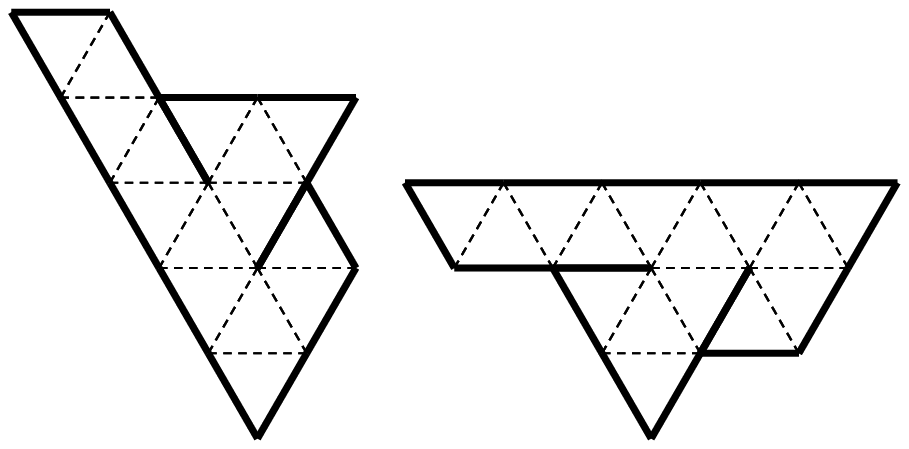}\hspace{0.08\linewidth}
\includegraphics[width=0.45\linewidth]{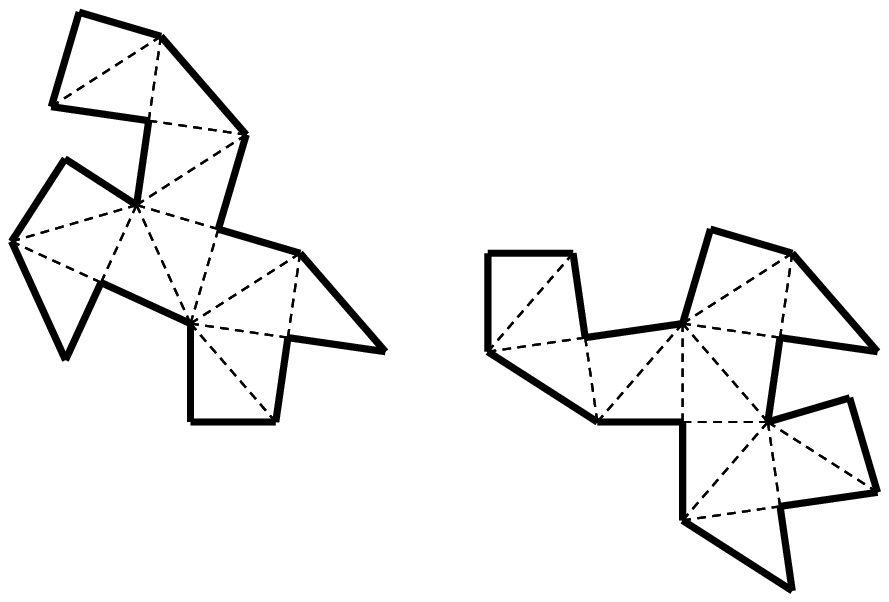}\caption{Pair $15_1$. Sunada triple $G=\PSL(4, 2)$, $G_i=\langle a_i, b_i, c_i\rangle$, $i=1,2$, with
$a_1=(0\ 14)(1\ 12)(2\ 6)(4\ 5)(7\ 11)(9\ 10)$, 
$b_1=(1\ 13)(2\ 7)(4\ 6)(8\ 9)$, 
$c_1=(1\ 14)(2\ 12)(3\ 4)(8\ 11)$, 
$a_2=(0\ 11)(1\ 5)(3\ 4)(6\ 10)(8\ 9)(13\ 14)$, 
$b_2=(0\ 10)(1\ 2)(6\ 9)(12\ 14)$, 
$c_2=(0\ 5)(2\ 4)(6\ 7)(11\ 14)$.
}
\label{15_1}
\end{center}
\end{figure}

\begin{figure}[h]
\begin{center}
\includegraphics[width=0.45\linewidth]{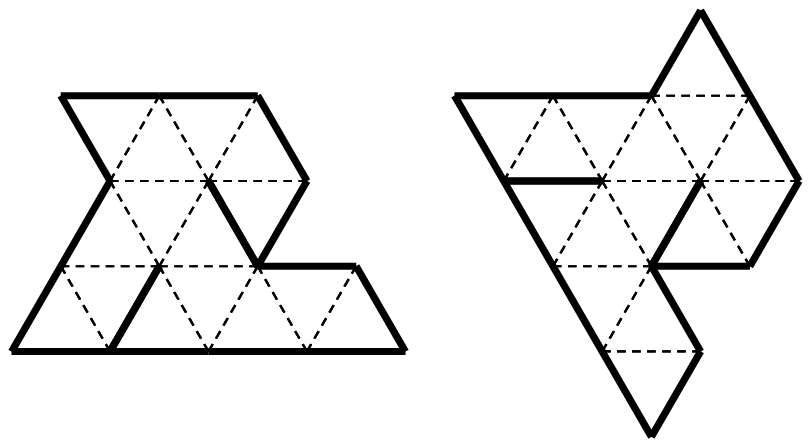}\hspace{0.08\linewidth}
\includegraphics[width=0.45\linewidth]{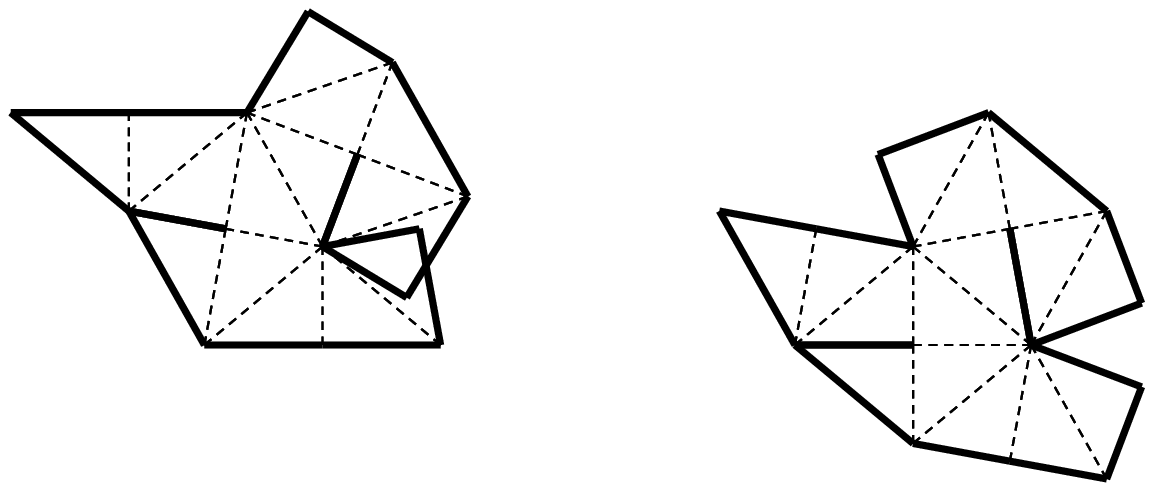}\caption{Pair $15_2$. Sunada triple $G=\PSL(4, 2)$, $G_i=\langle a_i, b_i, c_i\rangle$, $i=1,2$, with
$a_1=(0\ 14)(1\ 12)(2\ 6)(4\ 5)(7\ 11)(9\ 10)$, 
$b_1=(1\ 13)(2\ 7)(4\ 6)(8\ 9)$, 
$c_1=(0\ 12)(1\ 6)(3\ 5)(7\ 8)$, 
$a_2=(0\ 11)(1\ 5)(3\ 4)(6\ 10)(8\ 9)(13\ 14)$, 
$b_2=(0\ 10)(1\ 2)(6\ 9)(12\ 14)$, 
$c_2=(0\ 13)(1\ 11)(2\ 3)(7\ 10)$.
}
\label{15_2}
\end{center}
\end{figure}

\begin{figure}[h]
\begin{center}
\includegraphics[width=0.45\linewidth]{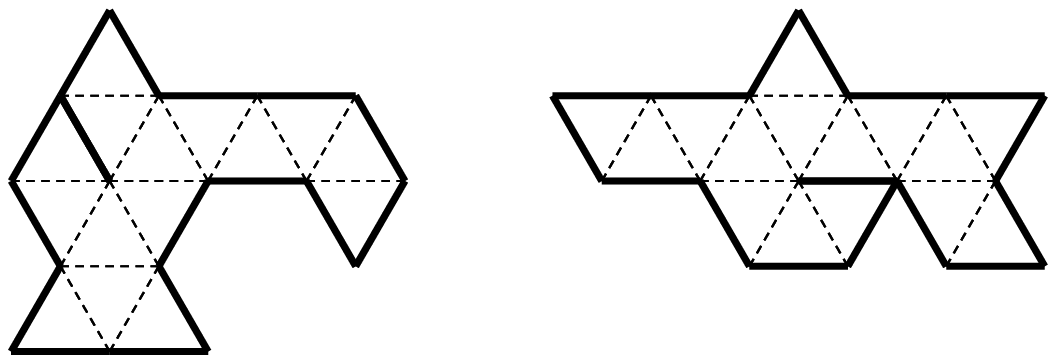}\hspace{0.08\linewidth}
\includegraphics[width=0.45\linewidth]{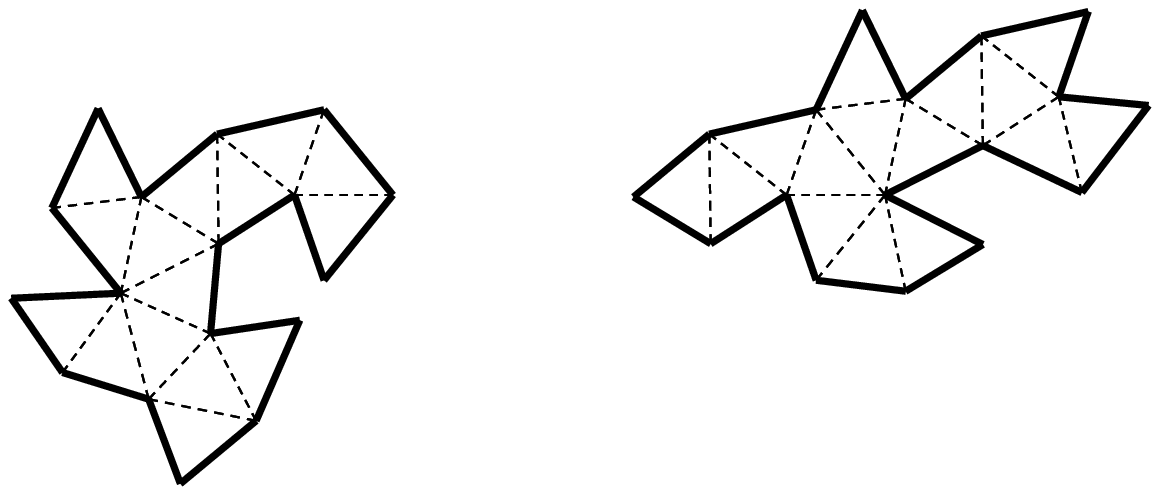}\caption{Pair $15_3$. Sunada triple $G=\PSL(4, 2)$, $G_i=\langle a_i, b_i, c_i\rangle$, $i=1,2$, with
$a_1=(0\ 14)(2\ 11)(4\ 7)(5\ 6)(8\ 10)(12\ 13)$, 
$b_1=(1\ 13)(2\ 7)(4\ 6)(8\ 9)$, 
$c_1=(0\ 12)(1\ 6)(3\ 5)(7\ 8)$, 
$a_2=(0\ 9)(2\ 5)(3\ 4)(6\ 8)(10\ 11)(12\ 13)$, 
$b_2=(0\ 10)(1\ 2)(6\ 9)(12\ 14)$, 
$c_2=(0\ 13)(1\ 11)(2\ 3)(7\ 10)$.
}
\label{15_3}
\end{center}
\end{figure}

\begin{figure}[h]
\begin{center}
\includegraphics[width=0.45\linewidth]{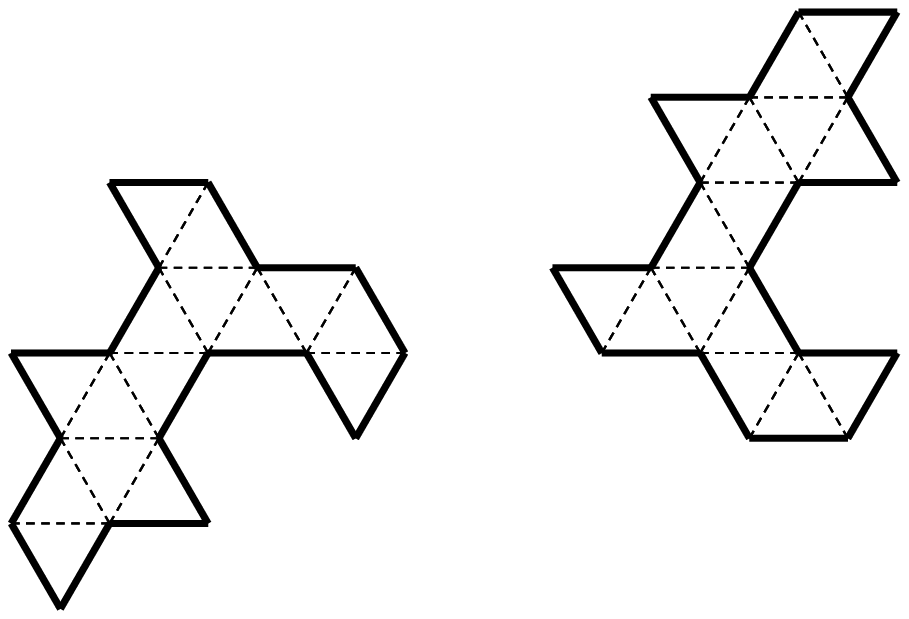}\hspace{0.08\linewidth}
\includegraphics[width=0.45\linewidth]{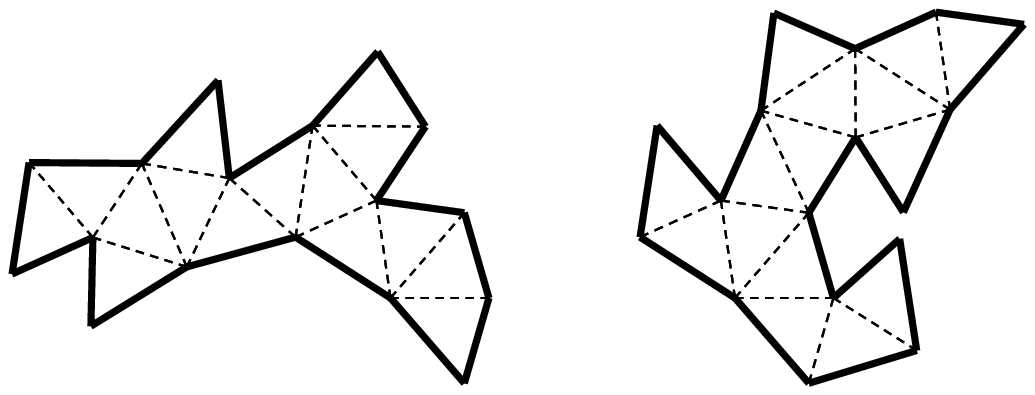}\caption{Pair $15_4$. Sunada triple $G=\PSL(4, 2)$, $G_i=\langle a_i, b_i, c_i\rangle$, $i=1,2$, with
$a_1=(0\ 14)(2\ 11)(4\ 7)(5\ 6)(8\ 10)(12\ 13)$, 
$b_1=(1\ 4)(2\ 8)(7\ 9)(8\ 9)$, 
$c_1=(0\ 12)(1\ 6)(3\ 5)(7\ 8)$, 
$a_2=(0\ 9)(2\ 5)(3\ 4)(6\ 8)(10\ 11)(12\ 13)$, 
$b_2=(6\ 9)(7\ 13)(12\ 14)$, 
$c_2=(0\ 13)(1\ 11)(2\ 3)(7\ 10)$.
}
\label{15_4}
\end{center}
\end{figure}

\begin{figure}[h]
\begin{center}
\includegraphics[width=0.45\linewidth]{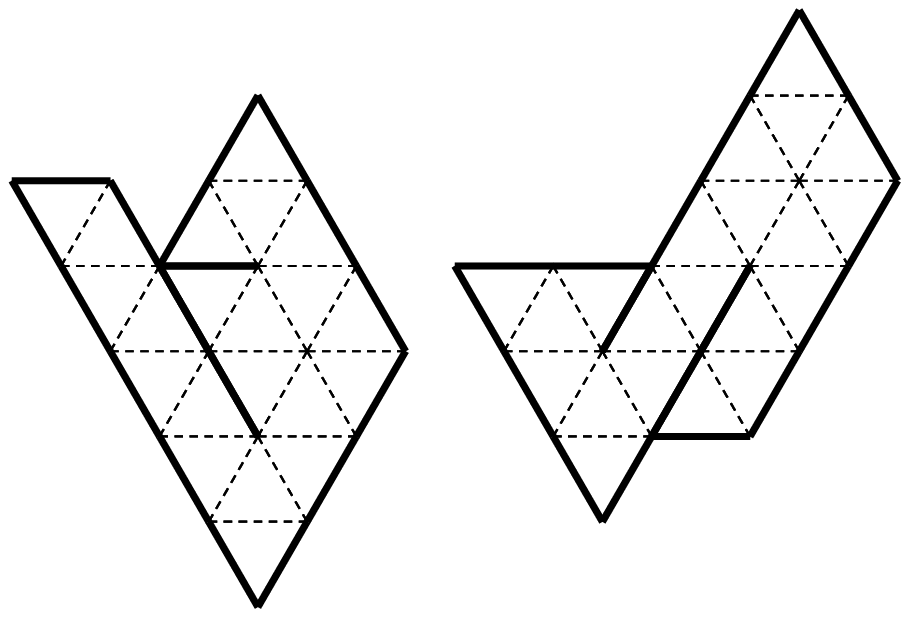}\hspace{0.08\linewidth}
\includegraphics[width=0.45\linewidth]{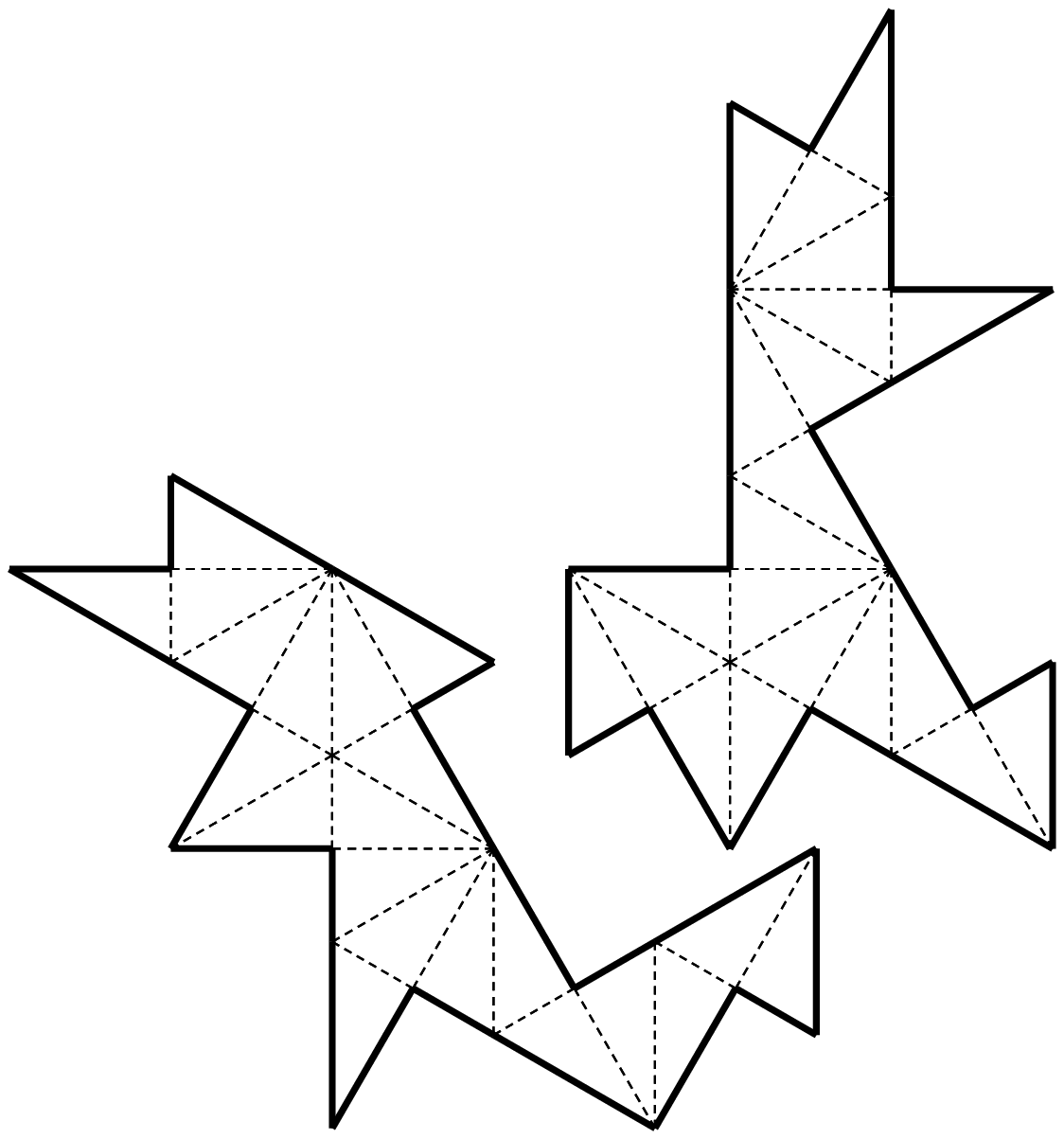}\caption{Pair $21_1$. Sunada triple $G=\PSL(3, 4)$, $G_i=\langle a_i, b_i, c_i\rangle$, $i=1,2$, with
$a_1=(2\ 7)(3\ 11)(5\ 12)(8\ 18)(13\ 14)(15\ 17)(16\ 20)$, 
$b_1=(0\ 17)(3\ 8)(4\ 12)(6\ 13)(9\ 19)(14\ 15)(16\ 18)$, 
$c_1=(1\ 8)(2\ 16)(4\ 11)(5\ 19)(7\ 14)(10\ 17)(13\ 20)$, 
$a_2=(0\ 1)(4\ 17)(7\ 12)(9\ 16)(10\ 20)(11\ 13)(15\ 19)$, 
$b_2=(0\ 20)(3\ 16)(6\ 11)(8\ 15)(9\ 19)(10\ 12)(14\ 18)$, 
$c_2=(1\ 8)(2\ 16)(4\ 11)(5\ 19)(7\ 14)(10\ 17)(13\ 20)$.
}
\label{21_1}
\end{center}
\end{figure}

The following gallery presents the  $17$ known families of isospectral pairs, 
as obtained by \textcite{BusConDoySem, OkaShu, Gir05}. All are based on a
Sunada triple $(G, G_1, G_2)$, where $G=\PSL(n + 1,q)$ is the
special linear automorphism group of a finite projective space of $(q^{n+1}-1)/(q-1)$
points, and $G_1, G_2$ are two subgroups, generated by 
$a_1, b_1, c_1$ and $a_2, b_2, c_2$ given below, respectively. These automorphisms
are collineations of order 2 of the
underlying finite projective space; $a_1, b_1, c_1$ act on points
while $a_2, b_2, c_2$ act on hyperplanes, numbered from $0$ to
$(q^{n+1}-1)/(q-1)-1$. The generators $a_i$ and $b_i$ allow to construct the
graphs (see section \ref{paperfoldingproof}) that specify the way in
which the tiles are glued together. Figures \ref{7_1} to \ref{21_1} give examples of pairs of isospectral
billiards obtained by applying the unfolding
rules on an equilateral triangle (left panel) or on a
scalene triangle (right panel).\\

Interestingly, the structure of pairs $13_6$ and $15_2$ forbids the construction of
any proper billiard, that is, structures where triangles do
not overlap. It is quite simple to convince oneself of this fact. 
In the case of the billiard $13_6$ (see Fig.~\ref{13_6}), the
initial triangle is unfolded six times around each of its corner.
Clearly, to have a non-overlapping billiard, each angle should be
less than $\pi/3$, which is impossible unless the initial triangle
is equilateral. 

For the billiard $15_2$ (see Fig.~\ref{15_2}), the initial triangle 
is unfolded six times around two of its corners, four times around
the third one, and thus two angles have to be less than $\pi/3$ and
one less than $\pi/2$. While it is possible to construct such a billiard,
it is impossible to get a pair of planar billiards. Indeed,  
the role of the angles is exchanged from one billiard to the other, which
leads to the condition that the three angles be less than $\pi/3$.
On the other hand, the presence of a loop in
the pair $21_1$ requires that one angle of the base triangle be $\pi/3$.


%\clearpage

%\clearpage
%\newpage

%%%%%%%%%%%%%%%%%%%%%%%%%%%%%%%%%%%%%%%%%%%%%%%%%%%%%%%%%%%%%%%%%%%%%%%%%%%%%%%%%%%%%%%%%%%%%%%%%%%%%%
\section{Spectral problems for Lie geometries}
%%%%%%%%%%%%%%%%%%%%%%%%%%%%%%%%%%%%%%%%%%%%%%%%%%%%%%%%%%%%%%%%%%%%%%%%%%%%%%%%%%%%%%%%%%%%%%%%%%%%%%
\label{appB}
\subsection{Generalized polygons}
Generalized polygons were introduced by \textcite{Ti}
in order to have a geometric interpretation of certain Chevalley groups of rank $2$. They are also the building bricks of (Tits) buildings,
the natural geometries for the groups with a BN-pair.

A group $G$ is said to have
a {\em BN-pair} $(B,N)$, where $B, N$ are subgroups of $G$, if the
following properties are satisfied: (BN1) $\langle B,N \rangle = G$;
(BN2) $H = B \cap N \unlhd N$ and $N/H = W$ is a Coxeter group (see, e.g., \cite{Tit:74}) 
with distinct generators $s_1,s_2,\ldots,s_n$; (BN3) $Bs_iBwB \subseteq BwB \cup Bs_iwB$ 
whenever $w \in W$ and $i\in\{1,2,\ldots,n\}$;
(BN4) $s_iBs_i \ne B$ for all $i\in\{1,2,\ldots,n\}$.
The subgroup $B$, respectively $W$, is a {\em Borel subgroup},
respectively the {\em Weyl group}, of $G$. The natural number $n$
is called the \emph{rank} of the BN-pair.\\

{\bf Example}.\quad
Suppose $\PG(1,q)$ is the projective line over the finite field $\mathbb{F}_q$; so $\PG(1,q)$ has $q + 1$ points.
Consider the natural action of $\PSL(2,q)$ on $\PG(1,q)$, and let $x$ and $y$ be distinct 
points of the projective line.
Set $B = \PSL(2,q)_x$ and $N = \PSL(2,q)_{\{x,y\}}$. Then $(B,N)$ is a BN-pair for $\PSL(2,q)$. 
Here $N/(B \cap N) = W$ is just the group of order $2$.\\

{\bf Example}.\quad
Consider the Desarguesian projective plane $\PG(2,q)$, and $\PSL(3,q)$ in its natural action on the latter plane.
Let $(x,L)$ be an incident point-line pair, and $\Delta$ a triangle (in the ordinary sense) that contains 
$x$ as a point and $L$ as a side. Set $B = \PSL(3,q)_{(x,L)}$ and $N = \PSL(3,q)_{\Delta}$; then $(B,N)$ is a BN-pair
for $\PSL(3,q)$ and $N/(B \cap N) = W$ is the dihedral group of order $6$.\\

See \cite{PT,HVM,TGQ,SFGQ} for standard references on the subject of generalized polygons. In this paper 
we consider only thick GPs.

Now let $G$ be a group with a BN-pair $(B,N)$ of rank $2$.
One can associate a generalized polygon $\mathcal{B}(G)$ with the group $G$ in the following way.
For this purpose, define $P_1 = \langle B,B^{s_1}\rangle$ and $P_2 = \langle B,B^{s_2}\rangle$.
\begin{itemize}
\item
Call the right cosets of $P_1$ ``points''.
\item
Call the right cosets of $P_2$ ``lines''.
\item
Call two  such  (distinct) cosets
``incident'' if their intersection is nonempty (so $P_1g$ is incident with $P_2h$, $g,h \in G$, if
$P_1g \cap P_2h \ne \emptyset$).
\end{itemize}
Then $\mathcal{B}(G)$ is a GP | say a generalized $n$-gon for some natural number $n$ | on which $G$ acts naturally as an automorphism group that permutes transitively the ordered $n$-gons (in the ordinary sense). 
\begin{quote}
{\bf Conjecture [J. Tits, \cite[\S 11.5.1]{Tit:74}]}.\quad
{\em If a finite thick generalized $n$-gon is such that the automorphism group permutes transitively the ordered $n$-gons (that is, if $\Delta$ is associated with a BN-pair), then $\Delta$ is isomorphic with the GP of an absolutely simple group over a finite field, or with the GP of a Ree group of type $^2\mathbf{F}_4$ over a finite field.}\\
\end{quote}
For more on the classification of BN-pairs of rank $2$, see \cite{KoeBN1,KoeBN2} and \cite{KoeBN4}.

\subsection{Duality principle}
Let $\Gamma = (\mP,\mB,\I)$ be a GP of order $(s,t)$. Then $\Gamma^D = (\mB,\mP,\I)$ clearly again is a GP, but now  of order
$(t,s)$. (The latter geometry is called the {\em point-line dual} of $\Gamma$.)
So any theorem which holds for a GP, has a dual interpretation; we call this principle ``duality principle''.

\subsection{Automorphisms and isomorphisms}
Let $\Gamma = (\mP,\mB,\I)$ and $\Gamma' = (\mP',\mB',\I')$ be GPs. Then an {\em isomorphism}  between $\Gamma$ and $\Gamma'$
is a pair $(\alpha,\beta)$ for which $\alpha$ is a bijection between $\mP$ and $\mP'$, $\beta$ is a bijection between $\mB$ and $\mB'$, and
$x \I L$ (in $\Gamma$) if and only if $x^{\alpha} \I L^{\beta}$. If there is an isomorphism between $\Gamma$ and $\Gamma'$, we say they are
``isomorphic'',  and write $\Gamma \cong \Gamma'$.

If $\Gamma = \Gamma'$ one speaks of an ``automorphism''. The set of all automorphisms of a GP forms a group, and the classical examples of
GPs are those examples that are associated with a Chevalley group (or, equivalently, with a ``classical'' BN-pair), cf.  \cite{HVM} for more details.

\subsection{Point spectra and order}
Let $\Gamma$ be a finite thick GP of order $(s,t)$, with associated collinearity matrix $\A$. Our first concern is to calculate $\spec(\A)$.

First, we recall the theorem of \textcite{FH}:
A finite thick generalized $n$-gon  exists if and only if $n \in \{3,4,6,8\}$.
We will do a case-by-case analysis according to this result.\\

{\bf Case $n = 3$.}

Recall that a generalized $3$-gon is the same as an axiomatic projective plane.
Now let $\Gamma$ be a finite projective plane of order $n$, $n \geq 2$, and put $n^2 + n + 1 = \vv$, its number of points.
Then $\A = \JJ_{\vv} - \II_{\vv}$, where $\JJ_{\vv}$ is the all $1$ $\vv\times \vv$-matrix, and $\II_{\vv}$ the $\vv\times\vv$-identity matrix.
It follows that
\begin{equation}
\spec(\A) = \{-1,\vv -1\}.
\end{equation}
So if the spectra of two finite projective planes coincide, their orders do as well.\\

{\bf Case $n = 4$.}

Let $\Gamma$ be a thick generalized $4$-gon, or also ''generalized quadrangle'' (GQ) of order $(s,t)$.
Then using the results of \textcite[1.2.2]{PT}, we have

\begin{equation}   \spec(\A) = \{ -t - 1, s - 1, s(t + 1)\}.                \end{equation}

Now let $\Gamma'$ also be a thick GQ, with the same spectrum, of order $(s',t')$.
There is only one negative eigenvalue, so $-t - 1 = -t' - 1$ and $t = t'$. Since $s - 1 < s(t + 1)$ ($s' - 1 < s'(t' + 1)$), it also follows that $s = s'$, and
hence $\Gamma$ and $\Gamma'$ have the same order.\\

{\bf Case $n = 6$.}

For this case, we need one more definition. A {\em distance regular graph} $G$ with diameter $d$ is a regular connected graph with valency $k$ for which there exist
natural numbers $b_0=k,b_1,\ldots,b_{d - 1};c_1 = 1,c_2,\ldots,c_d$  such that for each pair of vertices $x$ and $y$ at distance $j$, we have
\begin{itemize}
\item
$\vert \Gamma_{j - 1}(y)\cap \Gamma_1(x) \vert = c_j$, $1 \leq j \leq d$;
\item
$\vert \Gamma_{j + 1}(y)\cap \Gamma_1(x)\vert = b_j$, $0 \leq j \leq d - 1$.
\end{itemize}
Now define the constants $a_j = \vert\Gamma_j(y)\cap \Gamma_1(x)\vert$ for $0 \leq j\leq d$.
A result of \textcite{Gr} claims that the eigenvalues of the point graph of $G$ are the eigenvalues of the following ''intersection matrix'':
\begin{equation}
\left(\begin{array}{ccccc}
0&1&&&\\
k &a_1&c_2&&\\
&b_1&a_2&\cdots&\\
&\vdots&b_2&\cdots&\\
&&\vdots&&c_d\\
&&&b_{d-1}&a_d\\
\end{array}\right).\\
\end{equation}
It just so happens to be that the collinearity graph of a generalized hexagon (a generalized $6$-gon) 
is distance regular with a diameter of $3$. An easy exercise yields, for
a thick generalized hexagon of order $(s,t)$, the following intersection
matrix:
\begin{equation}  
\B =
\left(
\begin{array}{cccc}
0&1&0&0\\
s(t + 1)&s - 1& 1&0\\
0&st&s - 1&t + 1\\
0&0&st&(t + 1)(s - 1)
\end{array}\right).
\end{equation}
The determinant of $\B - x\II_4$  has the following roots:
\begin{equation}    
x = -t - 1,\ \ x = s(t + 1),\ \ x = s - 1 - \sqrt{st}, \ \ x = s - 1 +
\sqrt{st}.
\end{equation}
One observes that $-t - 1$ is strictly the smallest eigenvalue, while $s(t +
1)$ is the largest. 
It now easily follows that if a generalized hexagon of order $(s',t')$ has
the same spectrum as $\Gamma$, then it has the same order.\\

{\bf Case $n = 8$.}
Let $\Gamma$ be a thick generalized octagon ($8$-gon) of order $(s,t)$.
Again, the point graph is distance regular, now with a diameter of $4$. The intersection matrix is easily seen to be the following:
\begin{equation}  
\B =
\left(
\begin{array}{ccccc}
0&1&0&0&0\\
s(t + 1)&s - 1& 1&0&0\\
0&st&s - 1&1&0\\
0&0&st&s - 1&t + 1\\
0&0&0&st&(t + 1)(s - 1)
\end{array}\right),
\end{equation}
which has eigenvalues
\begin{eqnarray}  
x = -t-1,\ \ x = s - 1,\ \ x = s(t + 1),\nonumber\\
x = s - 1 - \sqrt{2st}, \ \ x =s - 1 + \sqrt{2st}.           
\end{eqnarray}
The third largest eigenvalue is $s - 1$, so if $\Gamma'$ is a thick
generalized octagon of order $(s',t')$ with the same spectrum, then $s  =
s'$. As $s(t + 1)$ is the largest eigenvalue of $\spec(\A)$, it follows that $t = t'$.
This ends the proof of Theorem \ref{KT2}.\eop

\subsection{Concluding remarks}
In this  section, we make some comments on generalized polygons that are characterized by their order.\\

\textsc{Projective planes}.\quad
For some small values, e.g. $n = 2$, it is known that there is a unique projective plane of order $n$ (up to isomorphism).
It is well-known, however, that as soon as $n$ is large enough and not a prime, nonisomorphic examples exist. On the other hand, for $p$ a prime, only one example is known, namely the classical example $\PG(2,p)$ arising from a BN-pair in $\PSL(2,p)$.\\

\textsc{Generalized quadrangles}.\quad
Many infinite classes of GQs are known, and several examples with small parameters are completely determined by their order.
We refer the interested reader to \cite[Chapter 6]{PT} for these examples. We make some comments
according to the known orders. Below, $q$ is always a prime power. We also
assume $s \leq t$ by reasons of duality. [Details and references can be
found in \cite[Chapter 3]{SFGQ}.]
\begin{itemize}
\item
$(s,t) = (q^2,q^3)$.\quad
Only one example is known (for each $q$), namely, the Hermitian quadrangle $\mathbf{H}(4,q^2)$.
\item
$(s,t) = (q - 1,q + 1)$.\quad
If $q \geq 8$ and $q$ is even, nonisomorphic examples are known for every $q$.
In the other cases, only unique examples are known.
\item
$(s,t) = (q,q)$.\quad
If $q$ is odd, nonisomorphic examples are known for every $q$. If $q \geq 8$ and $q$ is even, we have the same remark. The other values give unique examples.
\item
$(s,t) = (q,q^2)$.\quad
If $q \geq 5$, nonisomorphic examples are known for every $q$.
The examples of order $(2,4)$ and $(3,9)$ are unique.
\end{itemize}

\textsc{Generalized hexagons.}\quad
Up to duality, only two classes of generalized hexagons are known (both associated with classical groups): the {\em split Cayley hexagons $\mathbf{H}(q)$} of order $q$, $q$ a prime power,
and the {\em twisted triality hexagons $\mathbf{T}(q,q^3)$} of order $(q,q^3)$, cf. \cite[Chapter 2]{HVM}. 
We know that $\mathbf{H}(q) \cong \mathbf{H}(q)^D$ if and only if $q$ is a power of $3$ \cite{HVM}.
If $q$ is not a power of $3$, $\mathbf{H}(q) \not\cong \mathbf{H}(q)^D$, while both have the same spectrum.\\

\textsc{Generalized octagons}.\quad
Up to duality, the only known thick finite generalized octagons are the {\em Ree-Tits octagons} $\mathbf{O}(q)$, where $q$ is an odd power of $2$; they can be constructed from a BN-pair in the Ree groups of type
$^2\mathbf{F}_4$ \cite[Chapter 2]{HVM}. They have order $(q,q^2)$. \\

Finally, see the monograph \cite{CDS} for more information on graph spectra. 
\textcite{VDH} surveyed the known cases of graphs that are determined by their spectrum.
Some generalized quadrangles with small parameters are
mentioned that are uniquely determined by their spectrum.  Since such
examples must have the property that they 
are determined by their order, \textcite[Chapter 6]{PT} also  
yielded these examples.

%\clearpage
%\newpage

%%%%%%%%%%%%%%%%%%%%%%%%%%%%%%%%%%%%%%%%%%%%%%%%%%%%%%%%%%%%%%%%%%%%%%%%%%%%%%%%%%%%5
\section{Livsic cohomology}
%%%%%%%%%%%%%%%%%%%%%%%%%%%%%%%%%%%%%%%%%%%%%%%%%%%%%%%%%%%%%%%%%%%%%%%%%%%%%%%%%%%%5
\label{livsic}

In this appendix we describe a connection between isospectrality and cohomology.
Let $(M,g)$ be a Riemannian manifold without boundaries. The {\em length
  spectrum} is the discrete set 
\begin{equation} 
Lsp(M,g) = \{ L_{\gamma_1} < L_{\gamma_2} < \cdots  \}           
\end{equation}
of lengths of closed geodesics $\gamma_j$. \\

Denote by $(T^*M,\sum_jdx_j \wedge d\xi_j)$ the cotangent bundle of $M$
equipped with its natural symplectic form. 
Given the metric $g$, we define the {\em metric Hamiltonian} by
\begin{equation}  
H(x,\xi) = \vert \xi\vert = \sqrt{\sum_{ij = 1}^{n + 1}g^{ij}(x)\xi_i\xi_j},           
\end{equation}
and define the {\em energy surface} to be the unit sphere bundle
\begin{equation}  
S^*M = \{ (x,\xi) \parallel \vert \xi\vert = 1 \}.           
\end{equation}
The {\em geodesic flow} $G^t$ is the Hamiltonian flow
\begin{equation} 
G^t = \mbox{exp}t\Xi_H: T^*M \setminus 0 \mapsto T^*M \setminus 0,                    
\end{equation}
where $\Xi_H$ is the Hamiltonian vector field. Since it is homogeneous of degree $1$ with respect to the dilatation
$(x,\xi) \mapsto (x,r\xi)$, $r > 0$, one can restrict $G^t$ to $S^*M$. Its generator is also denoted by $\Xi$.\\

Livsic's cohomological problem asks whether a cocycle $f \in C^{\infty}(S^*M)$ satisfying
\begin{equation} 
\int_{\lambda} fds = 0     
\end{equation}
for every closed geodesic of the metric $g$ is necessarily a coboundary $f =
\Xi(g)$, where $\Xi$ is the generator of the geodesic flow $G^t$ and $g$ is
a function with a certain degree of regularity. 
Under a deformation $g_{\epsilon}$ of a metric $g = g_0$ preserving the
extended $Lsp(M,g)$ (including multiplicities), one has 
\begin{equation}  
\int_{\lambda}\dot{g}ds = 0, \ \ \forall \lambda.       
\end{equation}
When the cohomology is trivial, one can therefore write $\dot{g} = \Xi(f)$
for some $f$ with the given regularity. 
One does not expect the cohomology to be trivial in general settings, but
the results might be interesting for the length spectral deformation
problem.\\

%\newpage

%\addcontentsline{toc}{chapter}{\protect\bibname}

\bibliographystyle{apsrmp}

%\bibliography{biblio.bib}

\begin{thebibliography}{150}
\expandafter\ifx\csname natexlab\endcsname\relax\def\natexlab#1{#1}\fi
\expandafter\ifx\csname bibnamefont\endcsname\relax
  \def\bibnamefont#1{#1}\fi
\expandafter\ifx\csname bibfnamefont\endcsname\relax
  \def\bibfnamefont#1{#1}\fi
\expandafter\ifx\csname citenamefont\endcsname\relax
  \def\citenamefont#1{#1}\fi
\expandafter\ifx\csname url\endcsname\relax
  \def\url#1{\texttt{#1}}\fi
\expandafter\ifx\csname urlprefix\endcsname\relax\def\urlprefix{URL }\fi
\providecommand{\bibinfo}[2]{#2}
\providecommand{\eprint}[2][]{\url{#2}}



\bibitem[{\citenamefont{Aurich} \emph{et~al.}(1997)\citenamefont{Aurich,
  B\"acker, and Steiner}}]{AurBacSte97}
\bibinfo{author}{\bibnamefont{Aurich}, \bibfnamefont{R.}},
  \bibinfo{author}{\bibfnamefont{A.}~\bibnamefont{B\"acker}}, and
  \bibinfo{author}{\bibfnamefont{F.}~\bibnamefont{Steiner}},
  \bibinfo{year}{1997}, \bibinfo{journal}{Int. J. Mod. Phys. B}
  \textbf{\bibinfo{volume}{11}}, \bibinfo{pages}{805}.





\bibitem[{\citenamefont{Aurich and Steiner}(1990)}]{AurSte90}
\bibinfo{author}{\bibnamefont{Aurich}, \bibfnamefont{R.}}, and
  \bibinfo{author}{\bibfnamefont{F.}~\bibnamefont{Steiner}},
  \bibinfo{year}{1990}, \bibinfo{journal}{Physica D}
  \textbf{\bibinfo{volume}{43}}, \bibinfo{pages}{155}.


\bibitem[{\citenamefont{Balian and Bloch}(1974)}]{BalBlo74}
\bibinfo{author}{\bibnamefont{Balian}, \bibfnamefont{R.}}, and
  \bibinfo{author}{\bibfnamefont{C.}~\bibnamefont{Bloch}},
  \bibinfo{year}{1974}, \bibinfo{journal}{Ann. Phys. (N. Y. )}
  \textbf{\bibinfo{volume}{85}}, \bibinfo{pages}{514}.

\bibitem[{\citenamefont{Baltes and Hilf}(1976)}]{BalHil76}
\bibinfo{author}{\bibnamefont{Baltes}, \bibfnamefont{H.}}, and
  \bibinfo{author}{\bibfnamefont{E.~R.} \bibnamefont{Hilf}},
  \bibinfo{year}{1976}, \emph{\bibinfo{title}{Spectra of Finite Systems}}
  (\bibinfo{publisher}{Bibliographisches Institut},
  \bibinfo{address}{Mannheim}).

\bibitem[{\citenamefont{B\'erard}(1989)}]{Bera}
\bibinfo{author}{\bibnamefont{B\'erard}, \bibfnamefont{P.}},
  \bibinfo{year}{1989}, \bibinfo{journal}{Ast\'erisque}
  \textbf{\bibinfo{volume}{177-178}}, \bibinfo{pages}{127}.

\bibitem[{\citenamefont{B\'{e}rard}(1992)}]{Berard}
\bibinfo{author}{\bibnamefont{B\'{e}rard}, \bibfnamefont{P.}},
  \bibinfo{year}{1992}, \bibinfo{journal}{Math. Ann.}
  \textbf{\bibinfo{volume}{292}}, \bibinfo{pages}{547}.

\bibitem[{\citenamefont{B\'{e}rard}(1993)}]{Berard2}
\bibinfo{author}{\bibnamefont{B\'{e}rard}, \bibfnamefont{P.}},
  \bibinfo{year}{1993}, \bibinfo{journal}{J. London Math. Soc.}
  \textbf{\bibinfo{volume}{48}}, \bibinfo{pages}{565}.

\bibitem[{\citenamefont{B\'{e}rard and Besson}(1980)}]{BerBes}
\bibinfo{author}{\bibnamefont{B\'{e}rard}, \bibfnamefont{P.}}, and
  \bibinfo{author}{\bibfnamefont{G.}~\bibnamefont{Besson}},
  \bibinfo{year}{1980}, \bibinfo{journal}{Ann. Inst. Fourier, Grenoble}
  \textbf{\bibinfo{volume}{30}}, \bibinfo{pages}{237}.

\bibitem[{\citenamefont{Berry}(1981)}]{Berry}
\bibinfo{author}{\bibnamefont{Berry}, \bibfnamefont{M.~V.}},
  \bibinfo{year}{1981}, \bibinfo{journal}{Eur. J. Phys.}
  \textbf{\bibinfo{volume}{2}}, \bibinfo{pages}{91}.


\bibitem[{\citenamefont{Berry and Tabor}(1976)}]{BerTab76}
\bibinfo{author}{\bibnamefont{Berry}, \bibfnamefont{M.~V.}}, and
  \bibinfo{author}{\bibfnamefont{M.}~\bibnamefont{Tabor}},
  \bibinfo{year}{1976}, \bibinfo{journal}{Proc. Roy. Soc. London}
  \textbf{\bibinfo{volume}{349}}, \bibinfo{pages}{101}.

\bibitem[{\citenamefont{Berry and Tabor}(1977)}]{BerTab77a}
\bibinfo{author}{\bibnamefont{Berry}, \bibfnamefont{M.~V.}}, and
  \bibinfo{author}{\bibfnamefont{M.}~\bibnamefont{Tabor}},
  \bibinfo{year}{1977}, \bibinfo{journal}{Proc. Roy. Soc. London}
  \textbf{\bibinfo{volume}{356}}, \bibinfo{pages}{375}.

\bibitem[{\citenamefont{Berry and Wilkinson}(1984)}]{BerWil84}
\bibinfo{author}{\bibnamefont{Berry}, \bibfnamefont{M.~V.}}, and
  \bibinfo{author}{\bibfnamefont{M.}~\bibnamefont{Wilkinson}},
  \bibinfo{year}{1984}, \bibinfo{journal}{Proc. R. Soc. London}
  \textbf{\bibinfo{volume}{392}}, \bibinfo{pages}{15}.

\bibitem[{\citenamefont{Betcke and Trefethen}(2005)}]{BetTre}
\bibinfo{author}{\bibnamefont{Betcke}, \bibfnamefont{T.}}, and
  \bibinfo{author}{\bibfnamefont{L.~N.} \bibnamefont{Trefethen}},
  \bibinfo{year}{2005}, \bibinfo{journal}{SIAM Review}
  \textbf{\bibinfo{volume}{47}}(\bibinfo{number}{3}), \bibinfo{pages}{469}.

   
\bibitem[{\citenamefont{Blum} \emph{et~al.}(2002)\citenamefont{Blum,
  Gnutzmann, and Smilansky}}]{BluGnuSmi02}
\bibinfo{author}{\bibnamefont{Blum}, \bibfnamefont{G.}},
  \bibinfo{author}{\bibfnamefont{S.} \bibnamefont{Gnutzmann}}, and
  \bibinfo{author}{\bibfnamefont{U.}~\bibnamefont{Smilansky}},
  \bibinfo{year}{2002}, \bibinfo{journal}{Phys. Rev. Lett.}
  \textbf{\bibinfo{volume}{88}}, \bibinfo{pages}{114101}.



\bibitem[{\citenamefont{Bohigas} \emph{et~al.}(1984)\citenamefont{Bohigas,
  Giannoni, and Schmit}}]{BohGiaSch84}
\bibinfo{author}{\bibnamefont{Bohigas}, \bibfnamefont{O.}},
  \bibinfo{author}{\bibfnamefont{M.-J.} \bibnamefont{Giannoni}}, and
  \bibinfo{author}{\bibfnamefont{C.}~\bibnamefont{Schmit}},
  \bibinfo{year}{1984}, \bibinfo{journal}{Phys. Rev. Lett.}
  \textbf{\bibinfo{volume}{52}}, \bibinfo{pages}{1}.


\bibitem[{\citenamefont{Brooks}(1988)\citenamefont{Brooks}}]{Brooks88}
\bibinfo{author}{\bibnamefont{Brooks}, \bibfnamefont{R.}},
  \bibinfo{year}{1988}, \bibinfo{journal}{Amer. Math. Monthly}
  \textbf{\bibinfo{volume}{95}}, \bibinfo{pages}{823}.


\bibitem[{\citenamefont{Brouwer} \emph{et~al.}(1989)\citenamefont{Brouwer,
  Cohen, and Neumaier}}]{Gr}
\bibinfo{author}{\bibnamefont{Brouwer}, \bibfnamefont{A.~E.}},
  \bibinfo{author}{\bibfnamefont{A.~M.} \bibnamefont{Cohen}}, and
  \bibinfo{author}{\bibfnamefont{A.}~\bibnamefont{Neumaier}},
  \bibinfo{year}{1989}, \emph{\bibinfo{title}{Distance-Regular Graphs}}
 (\bibinfo{publisher}{Springer}, \bibinfo{address}{Heidelberg}).


\bibitem[{\citenamefont{Bruening} \emph{et~al.}(2008)\citenamefont{Bruening,
  Klawonn, and Puhle}}]{BruKlaPuh}
\bibinfo{author}{\bibnamefont{Bruening}, \bibfnamefont{J.}},
  \bibinfo{author}{\bibfnamefont{D.}~\bibnamefont{Klawonn}}, and
  \bibinfo{author}{\bibfnamefont{C.}~\bibnamefont{Puhle}},
  \bibinfo{year}{2008}, \bibinfo{journal}{J. Phys. A: Math. Theor.}
  \textbf{\bibinfo{volume}{40}}, \bibinfo{pages}{15143}.

\bibitem[{\citenamefont{Buser}(1988)}]{Buser}
\bibinfo{author}{\bibnamefont{Buser}, \bibfnamefont{P.}}, \bibinfo{year}{1988},
  \bibinfo{journal}{Geometry and Analysis on Manifolds (Katata/Kyoto, 1987),
  Lecture Notes in Math. 1339, Springer, Berlin} , \bibinfo{pages}{64}.

\bibitem[{\citenamefont{Buser} \emph{et~al.}(1994)\citenamefont{Buser, Conway,
  and Doyle}}]{BusConDoySem}
\bibinfo{author}{\bibnamefont{Buser}, \bibfnamefont{P.}},
  \bibinfo{author}{\bibfnamefont{J.}~\bibnamefont{Conway}}, and
  \bibinfo{author}{\bibfnamefont{P.}~\bibnamefont{Doyle}},
  \bibinfo{year}{1994}, \bibinfo{journal}{Int. Math. Res. Notices} 
\textbf{\bibinfo{volume}{9}}, \bibinfo{pages}{391}.

\bibitem[{\citenamefont{Chang and Deturck}(1989)}]{ChaDet}
\bibinfo{author}{\bibnamefont{Chang}, \bibfnamefont{P.-K.}}, and
  \bibinfo{author}{\bibfnamefont{D.}~\bibnamefont{Deturck}},
  \bibinfo{year}{1989}, \bibinfo{journal}{Proc. Amer. Math. Soc.}
  \textbf{\bibinfo{volume}{105}}(\bibinfo{number}{4}), \bibinfo{pages}{1033}.

\bibitem[{\citenamefont{Chapman}(1995)}]{Cha}
\bibinfo{author}{\bibnamefont{Chapman}, \bibfnamefont{S.~J.}},
  \bibinfo{year}{1995}, \bibinfo{journal}{Am. Math. Monthly}
  \textbf{\bibinfo{volume}{102}}(\bibinfo{number}{2}), \bibinfo{pages}{124}.

\bibitem[{\citenamefont{Chladni}(1802)}]{Chl}
\bibinfo{editor}{\bibnamefont{Chladni}, \bibfnamefont{E.}} (ed.),
  \bibinfo{year}{1802}, \emph{\bibinfo{title}{Die Akustik}}
  (\bibinfo{publisher}{Breitkopf und H\"artel}, \bibinfo{address}{Leipzig}).


\bibitem[{\citenamefont{Conway} \emph{et~al.}(1985)\citenamefont{Conway, Curtis, Norton, Parker, and Wilson}}]{ConCur}
\bibinfo{author}{\bibnamefont{Conway}, \bibfnamefont{J.~H.}}, and
  \bibinfo{author}{\bibfnamefont{R.~T.}~\bibnamefont{Curtis}}, and
  \bibinfo{author}{\bibfnamefont{S.~P.}~\bibnamefont{Norton}}, and
  \bibinfo{author}{\bibfnamefont{R.~A.}~\bibnamefont{Parker}}, and
  \bibinfo{author}{\bibfnamefont{R.~A.}~\bibnamefont{Wilson}}
  \bibinfo{year}{1985}, \emph{\bibinfo{title}{Atlas of Finite
Groups: Maximal Subgroups and Ordinary Characters for Simple Groups}} (\bibinfo{publisher}{Oxford University Press},
  \bibinfo{address}{Oxford}).



\bibitem[{\citenamefont{Conway and Sloane}(1992)}]{ConSlo}
\bibinfo{author}{\bibnamefont{Conway}, \bibfnamefont{J.~H.}}, and
  \bibinfo{author}{\bibfnamefont{N.~J.~A.} \bibnamefont{Sloane}},
  \bibinfo{year}{1992}, \bibinfo{journal}{Int. Math. Res. Notices}
  \textbf{\bibinfo{volume}{1992}}, \bibinfo{pages}{93}.


\bibitem[{\citenamefont{Courant and Hilbert}(1953)}]{CouHil}
\bibinfo{author}{\bibnamefont{Courant}, \bibfnamefont{R.}}, and
  \bibinfo{author}{\bibfnamefont{D.}~\bibnamefont{Hilbert}},
  \bibinfo{year}{1953}, \emph{\bibinfo{title}{Methods of Mathematical
  Physics}}, volume~\bibinfo{volume}{I} (\bibinfo{publisher}{Interscience},
  \bibinfo{address}{New York}).

\bibitem[{\citenamefont{Cvetkovic} \emph{et~al.}(1998)\citenamefont{Cvetkovic,
  Doob, and Sachs}}]{CDS}
\bibinfo{author}{\bibnamefont{Cvetkovic}, \bibfnamefont{D.~M.}},
  \bibinfo{author}{\bibfnamefont{M.}~\bibnamefont{Doob}}, and
  \bibinfo{author}{\bibfnamefont{H.}~\bibnamefont{Sachs}},
  \bibinfo{year}{1998}, \emph{\bibinfo{title}{Spectra of Graphs: Theory and
  Applications}}, \bibinfo{note}{3rd ed.} (\bibinfo{publisher}{Wiley}, \bibinfo{address}{New York})


\bibitem[{\citenamefont{Descloux and Tolley}(1983)}]{DesTol}
\bibinfo{author}{\bibnamefont{Descloux}, \bibfnamefont{J.}}, and
  \bibinfo{author}{\bibfnamefont{M.}~\bibnamefont{Tolley}},
  \bibinfo{year}{1983}, \bibinfo{journal}{Comput. Methods Appl. Mech. Engrg.}
  \textbf{\bibinfo{volume}{39}}, \bibinfo{pages}{37}.

\bibitem[{\citenamefont{Dhar} \emph{et~al.}(2003)\citenamefont{Dhar, Rao, UdayaShankar, and Sridhar}}]{DhaMadUdaSri}
\bibinfo{author}{\bibnamefont{Dhar}, \bibfnamefont{A.}},
  \bibinfo{author}{\bibfnamefont{D.M.}~\bibnamefont{Rao}},
  \bibinfo{author}{\bibfnamefont{N.}~\bibnamefont{UdayaShankar}}, and
  \bibinfo{author}{\bibfnamefont{S.}~\bibnamefont{Sridhar}},
  \bibinfo{year}{2003}, \bibinfo{journal}{Phys. Rev. E}
  \textbf{\bibinfo{volume}{68}}, \bibinfo{pages}{026208}.

\bibitem[{\citenamefont{Doyle and Rossetti}(2004)}]{DoyRoss04}
\bibinfo{author}{\bibnamefont{Doyle}, \bibfnamefont{P.~G.}} and
\bibinfo{author}{\bibnamefont{Rossetti}, \bibfnamefont{J.~P.}},
  \bibinfo{year}{2004},  \bibinfo{journal}{Geom. Topol.}
  \textbf{\bibinfo{volume}{8}}, \bibinfo{pages}{1227}.


\bibitem[{\citenamefont{Driscoll}(1997)}]{Dri}
\bibinfo{author}{\bibnamefont{Driscoll}, \bibfnamefont{T.~A.}},
  \bibinfo{year}{1997}, \bibinfo{journal}{SIAM Rev.}
  \textbf{\bibinfo{volume}{39}}(\bibinfo{number}{1}), \bibinfo{pages}{1}.

\bibitem[{\citenamefont{Driscoll and Gottlieb}(2003)}]{DriGot}
\bibinfo{author}{\bibnamefont{Driscoll}, \bibfnamefont{T.~A.}}, and
  \bibinfo{author}{\bibfnamefont{H.~P.~W.} \bibnamefont{Gottlieb}},
  \bibinfo{year}{2003}, \bibinfo{journal}{Phys. Rev. E}
  \textbf{\bibinfo{volume}{68}}, \bibinfo{pages}{016702}.

\bibitem[{\citenamefont{Earnest and Nipp}(1991)}]{EarNipp}
\bibinfo{author}{\bibnamefont{Earnest}, \bibfnamefont{A.~G.}}, and
  \bibinfo{author}{\bibfnamefont{G.}~\bibnamefont{Nipp}}, \bibinfo{year}{1991},
  \bibinfo{journal}{C. R. Math. Acad. Sci. Can.}
  \textbf{\bibinfo{volume}{13}}, \bibinfo{pages}{33}.

\bibitem[{\citenamefont{Even and Pieranski}(1999)}]{EvePie}
\bibinfo{author}{\bibnamefont{Even}, \bibfnamefont{C.}}, and
  \bibinfo{author}{\bibfnamefont{P.}~\bibnamefont{Pieranski}},
  \bibinfo{year}{1999}, \bibinfo{journal}{Europhys. Lett.}
  \textbf{\bibinfo{volume}{47}}, \bibinfo{pages}{531}.

\bibitem[{\citenamefont{Feit and Higman}(1964)}]{FH}
\bibinfo{author}{\bibnamefont{Feit}, \bibfnamefont{W.}}, and
  \bibinfo{author}{\bibfnamefont{D.}~\bibnamefont{Higman}},
  \bibinfo{year}{1964}, \bibinfo{journal}{J. Algebra}
  \textbf{\bibinfo{volume}{1}}, \bibinfo{pages}{114}.


\bibitem[{\citenamefont{Feynman and Hibbs}(1965)}]{FeyHib}
\bibinfo{author}{\bibnamefont{Feynman}, \bibfnamefont{R.~P.}} and
\bibinfo{author}{\bibnamefont{Hibbs}, \bibfnamefont{A.~R.}},
  \bibinfo{year}{1965}, \emph{\bibinfo{title}{Quantum Mechanics and Path Integrals}}
  (\bibinfo{publisher}{Mc. Graw-Hill},
  \bibinfo{address}{New York}).

\bibitem[{\citenamefont{Fox} \emph{et~al.}(1967)\citenamefont{Fox, Henrici, and
  Moler}}]{FoxHenMol}
\bibinfo{author}{\bibnamefont{Fox}, \bibfnamefont{L.}},
  \bibinfo{author}{\bibfnamefont{P.}~\bibnamefont{Henrici}}, and
  \bibinfo{author}{\bibfnamefont{C.}~\bibnamefont{Moler}},
  \bibinfo{year}{1967}, \bibinfo{journal}{SIAM J. Numer. Anal.}
  \textbf{\bibinfo{volume}{4}}, \bibinfo{pages}{89}.

\bibitem[{\citenamefont{Fulling and Kuchment}(2005)}]{FulKuc}
\bibinfo{author}{\bibnamefont{Fulling}, \bibfnamefont{S.~A.}}, and
  \bibinfo{author}{\bibfnamefont{P.}~\bibnamefont{Kuchment}},
  \bibinfo{year}{2005}, \bibinfo{journal}{Inverse Problems}
  \textbf{\bibinfo{volume}{21}}, \bibinfo{pages}{1391}.

\bibitem[{\citenamefont{Garabedian}(1998)}]{Garab}
\bibinfo{author}{\bibnamefont{Garabedian}, \bibfnamefont{P.~R.}},
  \bibinfo{year}{1998}, \emph{\bibinfo{title}{Partial Differential Equations}}
  (\bibinfo{publisher}{AMS Chelsea Publ., Amer. Math. Soc.},
  \bibinfo{address}{Providence, RI}).

\bibitem[{\citenamefont{Garabedian and Schiffer}(1952)}]{GS}
\bibinfo{author}{\bibnamefont{Garabedian}, \bibfnamefont{P.~R.}}, and
  \bibinfo{author}{\bibfnamefont{M.}~\bibnamefont{Schiffer}},
  \bibinfo{year}{1952}, \bibinfo{journal}{J. d'Anal. Math.}
  \textbf{\bibinfo{volume}{2}}, \bibinfo{pages}{281}.

\bibitem[{\citenamefont{Gassmann}(1926)}]{Ga}
\bibinfo{author}{\bibnamefont{Gassmann}, \bibfnamefont{F.}},
  \bibinfo{year}{1926}, \bibinfo{journal}{Math. Z.}
  \textbf{\bibinfo{volume}{25}}, \bibinfo{pages}{665}.

\bibitem[{\citenamefont{Georgeot and Prange}(1995)}]{GeoPra}
\bibinfo{author}{\bibnamefont{Georgeot}, \bibfnamefont{B.}}, and
  \bibinfo{author}{\bibfnamefont{R.~E.} \bibnamefont{Prange}},
  \bibinfo{year}{1995}, \bibinfo{journal}{Phys. Rev. Lett.}
  \textbf{\bibinfo{volume}{74}}(\bibinfo{number}{15}), \bibinfo{pages}{2851}.

\bibitem[{\citenamefont{Gerst}(1970)}]{Ge}
\bibinfo{author}{\bibnamefont{Gerst}, \bibfnamefont{I.}}, \bibinfo{year}{1970},
  \bibinfo{journal}{Acta Arith.} \textbf{\bibinfo{volume}{27}},
  \bibinfo{pages}{121}.

\bibitem[{\citenamefont{Giraud}(2004)}]{Gir04}
\bibinfo{author}{\bibnamefont{Giraud}, \bibfnamefont{O.}},
  \bibinfo{year}{2004}, \bibinfo{journal}{J. Phys. A: Math. Gen.}
  \textbf{\bibinfo{volume}{37}}(\bibinfo{number}{7}), \bibinfo{pages}{2751}.

\bibitem[{\citenamefont{Giraud}(2005{\natexlab{a}})}]{Giraud}
\bibinfo{author}{\bibnamefont{Giraud}, \bibfnamefont{O.}},
  \bibinfo{year}{2005}{\natexlab{a}}, \bibinfo{journal}{J. Phys. A}
  \textbf{\bibinfo{volume}{38}}, \bibinfo{pages}{L477}.

\bibitem[{\citenamefont{Giraud}(2005{\natexlab{b}})}]{Gir05}
\bibinfo{author}{\bibnamefont{Giraud}, \bibfnamefont{O.}},
  \bibinfo{year}{2005}{\natexlab{b}}, \bibinfo{journal}{J. Phys. A: Math. Gen.}
  \textbf{\bibinfo{volume}{38}}, \bibinfo{pages}{L477}.

\bibitem[{\citenamefont{Gnutzmann} \emph{et~al.}(2005)\citenamefont{Gnutzmann,
  Smilansky, and Sondergaard}}]{GnuSmiSon}
\bibinfo{author}{\bibnamefont{Gnutzmann}, \bibfnamefont{S.}},
  \bibinfo{author}{\bibfnamefont{U.}~\bibnamefont{Smilansky}}, and
  \bibinfo{author}{\bibfnamefont{N.}~\bibnamefont{Sondergaard}},
  \bibinfo{year}{2005}, \bibinfo{journal}{J. Phys. A: Math. Gen.}
  \textbf{\bibinfo{volume}{38}}, \bibinfo{pages}{8921}.

\bibitem[{\citenamefont{Gordon}
  \emph{et~al.}(1992{\natexlab{a}})\citenamefont{Gordon, Webb, and
  Wolpert}}]{GorWebWol}
\bibinfo{author}{\bibnamefont{Gordon}, \bibfnamefont{C.}},
  \bibinfo{author}{\bibfnamefont{D.}~\bibnamefont{Webb}}, and
  \bibinfo{author}{\bibfnamefont{S.}~\bibnamefont{Wolpert}},
  \bibinfo{year}{1992}{\natexlab{a}}, \bibinfo{journal}{Invent. math.}
  \textbf{\bibinfo{volume}{110}}, \bibinfo{pages}{1}.

\bibitem[{\citenamefont{Gordon}
  \emph{et~al.}(1992{\natexlab{b}})\citenamefont{Gordon, Webb, and
  Wolpert}}]{GorWebWol2}
\bibinfo{author}{\bibnamefont{Gordon}, \bibfnamefont{C.}},
  \bibinfo{author}{\bibfnamefont{D.}~\bibnamefont{Webb}}, and
  \bibinfo{author}{\bibfnamefont{S.}~\bibnamefont{Wolpert}},
  \bibinfo{year}{1992}{\natexlab{b}}, \bibinfo{journal}{Bull. Amer. Math. Soc.}
  \textbf{\bibinfo{volume}{27}}, \bibinfo{pages}{134}.

\bibitem[{\citenamefont{Gordon}(1986)}]{Gor}
\bibinfo{author}{\bibnamefont{Gordon}, \bibfnamefont{C.~S.}},
  \bibinfo{year}{1986}, \bibinfo{journal}{Contemp. Math.}
  \textbf{\bibinfo{volume}{51}}, \bibinfo{pages}{63}.

\bibitem[{\citenamefont{Gordon and Webb}(1994)}]{GorWeb}
\bibinfo{author}{\bibnamefont{Gordon}, \bibfnamefont{C.~S.}}, and
  \bibinfo{author}{\bibfnamefont{D.~L.} \bibnamefont{Webb}},
  \bibinfo{year}{1994}, \bibinfo{journal}{Proc. Amer. Math. Soc.}
  \textbf{\bibinfo{volume}{120}}, \bibinfo{pages}{981}.

\bibitem[{\citenamefont{Gorenstein}(1980)}]{Gor80}
\bibinfo{author}{\bibnamefont{Gorenstein}, \bibfnamefont{D.}},
  \bibinfo{year}{1980}, \emph{\bibinfo{title}{Finite Groups. Second edition}}
  (\bibinfo{publisher}{Chelsea Publishing Co.}, \bibinfo{address}{New York}).

\bibitem[{\citenamefont{Gottlieb}(2004)}]{Got}
\bibinfo{author}{\bibnamefont{Gottlieb}, \bibfnamefont{H.~P.~W.}},
  \bibinfo{year}{2004}, \bibinfo{journal}{Inverse problems}
  \textbf{\bibinfo{volume}{20}}, \bibinfo{pages}{155}.

\bibitem[{\citenamefont{Gottlieb and McManus}(1998)}]{GotMcM}
\bibinfo{author}{\bibnamefont{Gottlieb}, \bibfnamefont{H.~P.~W.}}, and
  \bibinfo{author}{\bibfnamefont{J.~P.} \bibnamefont{McManus}},
  \bibinfo{year}{1998}, \bibinfo{journal}{Journal of Sound and Vibration}
  \textbf{\bibinfo{volume}{212}}(\bibinfo{number}{2}), \bibinfo{pages}{253}.

\bibitem[{\citenamefont{Guhr} \emph{et~al.}(1998)\citenamefont{Guhr,
  M\"uller-Groeling, and Weidenm\"uller}}]{GuhMulWei98}
\bibinfo{author}{\bibnamefont{Guhr}, \bibfnamefont{T.}},
  \bibinfo{author}{\bibfnamefont{A.}~\bibnamefont{M\"uller-Groeling}}, and
  \bibinfo{author}{\bibfnamefont{H.}~\bibnamefont{Weidenm\"uller}},
  \bibinfo{year}{1998}, \bibinfo{journal}{Phys. Rep.}
  \textbf{\bibinfo{volume}{299}}, \bibinfo{pages}{189}.

\bibitem[{\citenamefont{Gutkin and Judge}(2000)}]{GutJud00}
\bibinfo{author}{\bibnamefont{Gutkin}, \bibfnamefont{E.}}, and
  \bibinfo{author}{\bibfnamefont{C.}~\bibnamefont{Judge}},
  \bibinfo{year}{2000}, \bibinfo{journal}{Duke Math. J.}
  \textbf{\bibinfo{volume}{103}}(\bibinfo{number}{2}), \bibinfo{pages}{191}.

\bibitem[{\citenamefont{Gutkin and Smilansky}(2001)}]{GutkSmil}
\bibinfo{author}{\bibnamefont{Gutkin}, \bibfnamefont{B.}}, and
  \bibinfo{author}{\bibfnamefont{U.}~\bibnamefont{Smilansky}},
  \bibinfo{year}{2001}, \bibinfo{journal}{J. Phys. A}
  \textbf{\bibinfo{volume}{34}}(\bibinfo{number}{31}), \bibinfo{pages}{6061}.



\bibitem[{\citenamefont{Gutzwiller}(1991)}]{Gut89}
\bibinfo{author}{\bibnamefont{Gutzwiller}, \bibfnamefont{M.~C.}},
  \bibinfo{year}{1991}, in \emph{\bibinfo{booktitle}{Chaos and Quantum
  Mechanics}}, edited by \bibinfo{editor}{\bibfnamefont{M.-J.}
  \bibnamefont{Giannoni}},
  \bibinfo{editor}{\bibfnamefont{A.}~\bibnamefont{Voros}}, and
  \bibinfo{editor}{\bibfnamefont{J.}~\bibnamefont{Zinn-Justin}}
  (\bibinfo{publisher}{North Holland}, \bibinfo{address}{Amsterdam}),
  \bibinfo{note}{les Houches Summer School Lectures, Session LII, 1989}.

\bibitem[{\citenamefont{Hannay and Thain}(2003)}]{HanTha}
\bibinfo{author}{\bibnamefont{Hannay}, \bibfnamefont{J.~H.}}, and
  \bibinfo{author}{\bibfnamefont{A.}~\bibnamefont{Thain}},
  \bibinfo{year}{2003}, \bibinfo{journal}{J. Phys. A: Math. Gen.}
  \textbf{\bibinfo{volume}{36}}, \bibinfo{pages}{4063}.


\bibitem[{\citenamefont{Hezari and Zelditch}(2009)}]{prep09}
\bibinfo{author}{\bibnamefont{Hezari}, \bibfnamefont{H.}},
  \bibinfo{author}{\bibfnamefont{S.}~\bibnamefont{Zelditch}},
  \bibinfo{year}{2009}, \bibinfo{journal}{Geom. Funct. Anal.}
  \textbf{\bibinfo{volume}{20}}, \bibinfo{pages}{160}.


\bibitem[{\citenamefont{Hirschfeld}(1998)}]{Hirsch}
\bibinfo{author}{\bibnamefont{Hirschfeld}, \bibfnamefont{J.~W.~P.}},
  \bibinfo{year}{1998}, \emph{\bibinfo{title}{Projective Geometries over Finite
  Fields. Second edition}} (\bibinfo{publisher}{The Clarendon Press, Oxford
  University Press}, \bibinfo{address}{New York}).

\bibitem[{\citenamefont{Hughes and Piper}(1973)}]{HP}
\bibinfo{author}{\bibnamefont{Hughes}, \bibfnamefont{D.~R.}}, and
  \bibinfo{author}{\bibfnamefont{F.~C.} \bibnamefont{Piper}},
  \bibinfo{year}{1973}, \emph{\bibinfo{title}{Projective Planes}}
  (\bibinfo{publisher}{Springer-Verlag}, \bibinfo{address}{New
  York/Heidelberg/Berlin}).



\bibitem[{\citenamefont{Iantchenko} \emph{et~al.}(2002)\citenamefont{Iantchenko,
  Sj\"ostrand, and Zworski}}]{ISZ02}
\bibinfo{author}{\bibnamefont{Iantchenko}, \bibfnamefont{A.}},
  \bibinfo{author}{\bibfnamefont{J.}~\bibnamefont{Sj\"ostrand}}, and
  \bibinfo{author}{\bibfnamefont{M.}~\bibnamefont{Zworski}},
  \bibinfo{year}{2002}, \bibinfo{journal}{Math. Res. Lett.}
  \textbf{\bibinfo{volume}{9}}, \bibinfo{pages}{337}.

\bibitem[{\citenamefont{Ikeda}(1980)}]{Ike}
\bibinfo{author}{\bibnamefont{Ikeda}, \bibfnamefont{A.}}, \bibinfo{year}{1980},
  \bibinfo{journal}{Ann. Sci. Ecole Normale Super.} \textbf{\bibinfo{volume}{13}},
  \bibinfo{pages}{303}.


\bibitem[{\citenamefont{Ivanov} \emph{et~al.}(1977)\citenamefont{Ivanov, Kotko,
  and Krein}}]{Krein}
\bibinfo{author}{\bibnamefont{Ivanov}, \bibfnamefont{L.}},
  \bibinfo{author}{\bibfnamefont{L.}~\bibnamefont{Kotko}}, and
  \bibinfo{author}{\bibfnamefont{S.}~\bibnamefont{Krein}},
  \bibinfo{year}{1977}, \emph{\bibinfo{title}{Boundary Value Problems in
  Variable Domains}}, volume~\bibinfo{volume}{19} (\bibinfo{publisher}{Math.
  Inst. of Lithuanian Acad. Sci.}, \bibinfo{address}{Vilnius}).

\bibitem[{\citenamefont{Jakobson} \emph{et~al.}(2006)\citenamefont{Jakobson,
  Levitin, Nadirashvili, and Polterovich}}]{JakLevNadPol}
\bibinfo{author}{\bibnamefont{Jakobson}, \bibfnamefont{D.}},
  \bibinfo{author}{\bibfnamefont{M.}~\bibnamefont{Levitin}},
  \bibinfo{author}{\bibfnamefont{N.}~\bibnamefont{Nadirashvili}}, and
  \bibinfo{author}{\bibfnamefont{I.}~\bibnamefont{Polterovich}},
  \bibinfo{year}{2006}, \bibinfo{journal}{J. Comp. Appl. Math.}
  \textbf{\bibinfo{volume}{194}}(\bibinfo{number}{1}), \bibinfo{pages}{141}.

\bibitem[{\citenamefont{Kac}(1966)}]{Kac}
\bibinfo{author}{\bibnamefont{Kac}, \bibfnamefont{M.}}, \bibinfo{year}{1966},
  \bibinfo{journal}{Am. Math. Monthly}
  \textbf{\bibinfo{volume}{73}}(\bibinfo{number}{4}), \bibinfo{pages}{1}.

\bibitem[{\citenamefont{Keller}(1962)}]{Kel62}
\bibinfo{author}{\bibnamefont{Keller}, \bibfnamefont{J.}},
  \bibinfo{year}{1962}, \bibinfo{journal}{J. Opt. Soc. Am.}
  \textbf{\bibinfo{volume}{52}}, \bibinfo{pages}{116}.

\bibitem[{\citenamefont{Keller and Rubinow}(1960)}]{KR}
\bibinfo{author}{\bibnamefont{Keller}, \bibfnamefont{J.~B.}}, and
  \bibinfo{author}{\bibfnamefont{S.~I.} \bibnamefont{Rubinow}},
  \bibinfo{year}{1960}, \bibinfo{journal}{Ann. Phys. (N.Y.)}
  \textbf{\bibinfo{volume}{9}}, \bibinfo{pages}{24}.

\bibitem[{\citenamefont{Knowles and McCarthy}(2004)}]{KnoMcC}
\bibinfo{author}{\bibnamefont{Knowles}, \bibfnamefont{I.~W.}}, and
  \bibinfo{author}{\bibfnamefont{M.~L.} \bibnamefont{McCarthy}},
  \bibinfo{year}{2004}, \bibinfo{journal}{J. Phys. A: Math. Gen.}
  \textbf{\bibinfo{volume}{37}}, \bibinfo{pages}{8103}.

\bibitem[{\citenamefont{Komatsu}(1976)}]{Ko}
\bibinfo{author}{\bibnamefont{Komatsu}, \bibfnamefont{K.}},
  \bibinfo{year}{1976}, \bibinfo{journal}{Kodai Math. Sem. Rep.}
  \textbf{\bibinfo{volume}{28}}, \bibinfo{pages}{78}.

\bibitem[{\citenamefont{Levitin} \emph{et~al.}(2006)\citenamefont{Levitin,
  Parnovski, and Polterovich}}]{LevParPol}
\bibinfo{author}{\bibnamefont{Levitin}, \bibfnamefont{M.}},
  \bibinfo{author}{\bibfnamefont{L.}~\bibnamefont{Parnovski}}, and
  \bibinfo{author}{\bibfnamefont{I.}~\bibnamefont{Polterovich}},
  \bibinfo{year}{2006}, \bibinfo{journal}{J. Phys. A: Math. Gen.}
  \textbf{\bibinfo{volume}{39}}, \bibinfo{pages}{2073}.



\bibitem[{\citenamefont{Mehra and Rechenberg}(2000)}]{MehRec00}
\bibinfo{author}{\bibnamefont{Mehra}, \bibfnamefont{J.}}, and
  \bibinfo{author}{\bibfnamefont{H.} \bibnamefont{Rechenberg}},
  \bibinfo{year}{2000},  \emph{\bibinfo{title}{The Historical Development of Quantum Theory, Volume 6: The Completion of Quantum Mechanics 1926-1941}}
  (\bibinfo{publisher}{Springer}, \bibinfo{address}{New York}).



\bibitem[{\citenamefont{Mehta}(1990)}]{Meh90}
\bibinfo{author}{\bibnamefont{Mehta}, \bibfnamefont{M.~L.}},
  \bibinfo{year}{1990}, \emph{\bibinfo{title}{Random Matrix Theory}}
  (\bibinfo{publisher}{Springer}, \bibinfo{address}{New York}).

\bibitem[{\citenamefont{Melrose}(1983)}]{Mel}
\bibinfo{author}{\bibnamefont{Melrose}, \bibfnamefont{R.}},
  \bibinfo{year}{1983}, \bibinfo{journal}{Math. Sci. Res. Inst.}
  \bibinfo{note}{Report No. 048-83}.

\bibitem[{\citenamefont{Melrose}(1996)}]{Me}
\bibinfo{author}{\bibnamefont{Melrose}, \bibfnamefont{R.~B.}},
  \bibinfo{year}{1996}, \bibinfo{journal}{Proc. Centre Math. Appl. Austral.
  Nat. Univ.} \textbf{\bibinfo{volume}{34}}, \bibinfo{pages}{137}.

\bibitem[{\citenamefont{Milnor}(1964)}]{milnor}
\bibinfo{author}{\bibnamefont{Milnor}, \bibfnamefont{J.}},
  \bibinfo{year}{1964}, \bibinfo{journal}{Proc. Natl. Acad. Sc. U.S.A}
  \textbf{\bibinfo{volume}{51}}, \bibinfo{pages}{542}.


\bibitem[{\citenamefont{Moon}
  \emph{et~al.}(2008)\citenamefont{Moon, Mattos, Foster, Zeltzer, Ko, and Manoharan}}]{Moon}
\bibinfo{author}{\bibnamefont{Moon}, \bibfnamefont{C.~R.}},
  \bibinfo{author}{\bibfnamefont{L.~S.}~\bibnamefont{Mattos}},
  \bibinfo{author}{\bibfnamefont{B.~K.}~\bibnamefont{Foster}}, 
  \bibinfo{author}{\bibfnamefont{G.}~\bibnamefont{Zeltzer}}, 
  \bibinfo{author}{\bibfnamefont{W.}~\bibnamefont{Ko}}, 
  \bibinfo{author}{\bibfnamefont{H.~C.}~\bibnamefont{Manoharan}},
  \bibinfo{year}{2008}, \bibinfo{journal}{Science}
  \textbf{\bibinfo{volume}{319}}, \bibinfo{pages}{782}.


\bibitem[{\citenamefont{Okada and Shudo}(2001)}]{OkaShu}
\bibinfo{author}{\bibnamefont{Okada}, \bibfnamefont{Y.}}, and
  \bibinfo{author}{\bibfnamefont{A.}~\bibnamefont{Shudo}},
  \bibinfo{year}{2001}, \bibinfo{journal}{J. Phys. A: Math. Gen.}
  \textbf{\bibinfo{volume}{34}}, \bibinfo{pages}{5911}.

\bibitem[{\citenamefont{Okada}
  \emph{et~al.}(2005{\natexlab{a}})\citenamefont{Okada, Shudo, Tasaki, and
  Harayama}}]{OkaShuTasHar}
\bibinfo{author}{\bibnamefont{Okada}, \bibfnamefont{Y.}},
  \bibinfo{author}{\bibfnamefont{A.}~\bibnamefont{Shudo}},
  \bibinfo{author}{\bibfnamefont{S.}~\bibnamefont{Tasaki}}, and
  \bibinfo{author}{\bibfnamefont{T.}~\bibnamefont{Harayama}},
  \bibinfo{year}{2005}{\natexlab{a}}, \bibinfo{journal}{J. Phys. A: Math. Gen.}
  \textbf{\bibinfo{volume}{38}}, \bibinfo{pages}{L163}.

\bibitem[{\citenamefont{Okada}
  \emph{et~al.}(2005{\natexlab{b}})\citenamefont{Okada, Shudo, Tasaki, and
  Harayama}}]{OkaShuTasHar05b}
\bibinfo{author}{\bibnamefont{Okada}, \bibfnamefont{Y.}},
  \bibinfo{author}{\bibfnamefont{A.}~\bibnamefont{Shudo}},
  \bibinfo{author}{\bibfnamefont{S.}~\bibnamefont{Tasaki}}, and
  \bibinfo{author}{\bibfnamefont{T.}~\bibnamefont{Harayama}},
  \bibinfo{year}{2005}{\natexlab{b}}, \bibinfo{journal}{J. Phys. A: Math. Gen.}
  \textbf{\bibinfo{volume}{38}}, \bibinfo{pages}{6675}.

\bibitem[{\citenamefont{Osgood}
  \emph{et~al.}(1988{\natexlab{a}})\citenamefont{Osgood, Phillips, and
  Sarnak}}]{OsgPhiSar}
\bibinfo{author}{\bibnamefont{Osgood}, \bibfnamefont{B.}},
  \bibinfo{author}{\bibfnamefont{R.}~\bibnamefont{Phillips}}, and
  \bibinfo{author}{\bibfnamefont{P.}~\bibnamefont{Sarnak}},
  \bibinfo{year}{1988}{\natexlab{a}}, \bibinfo{journal}{Proc. Natl. Acad. Sci.
  USA} \textbf{\bibinfo{volume}{85}}, \bibinfo{pages}{5359}.

\bibitem[{\citenamefont{Osgood}
  \emph{et~al.}(1988{\natexlab{b}})\citenamefont{Osgood, Phillips, and
  Sarnak}}]{OsgPhiSar2}
\bibinfo{author}{\bibnamefont{Osgood}, \bibfnamefont{B.}},
  \bibinfo{author}{\bibfnamefont{R.}~\bibnamefont{Phillips}}, and
  \bibinfo{author}{\bibfnamefont{P.}~\bibnamefont{Sarnak}},
  \bibinfo{year}{1988}{\natexlab{b}}, \bibinfo{journal}{J. Funct. Anal.}
  \textbf{\bibinfo{volume}{80}}, \bibinfo{pages}{212}.


\bibitem[{\citenamefont{Parzanchevski and Band}(2010)}]{ParzBand}
\bibinfo{author}{\bibnamefont{Parzanchevski}, \bibfnamefont{O.}},
  \bibinfo{author}{\bibfnamefont{R.}~\bibnamefont{Band}},
  \bibinfo{year}{2010}, \bibinfo{journal}{J. Geom. Anal.}
  \textbf{\bibinfo{volume}{20}}, \bibinfo{pages}{439}.



\bibitem[{\citenamefont{Pavloff and Schmit}(1995)}]{PavSch95}
\bibinfo{author}{\bibnamefont{Pavloff}, \bibfnamefont{N.}},
  \bibinfo{author}{\bibfnamefont{C.}~\bibnamefont{Schmit}},
  \bibinfo{year}{1995}, \bibinfo{journal}{Phys. Rev. Lett.}
  \textbf{\bibinfo{volume}{75}}, \bibinfo{pages}{61}.



\bibitem[{\citenamefont{Payne and Thas}(1984)}]{PT}
\bibinfo{author}{\bibnamefont{Payne}, \bibfnamefont{S.~E.}}, and
  \bibinfo{author}{\bibfnamefont{J.~A.} \bibnamefont{Thas}},
  \bibinfo{year}{1984}, \emph{\bibinfo{title}{Finite Generalized Quadrangles}},
  volume \bibinfo{volume}{110} (\bibinfo{publisher}{Pitman Advanced Publishing
  Program}, \bibinfo{address}{Boston/London/Melbourne}).

\bibitem[{\citenamefont{Pisani}(1996)}]{Pis96}
\bibinfo{author}{\bibnamefont{Pisani}, \bibfnamefont{C.}},
  \bibinfo{year}{1996}, \bibinfo{journal}{Ann. Phys.} \bibinfo{volume}{251} , \bibinfo{pages}{208}.

\bibitem[{\citenamefont{Pleijel}(1956)}]{Ple56}
\bibinfo{author}{\bibnamefont{Pleijel}, \bibfnamefont{A.}},
  \bibinfo{year}{1956}, \bibinfo{journal}{Commun. Pure Appl. Math.}
  \textbf{\bibinfo{volume}{9}}, \bibinfo{pages}{543}.

\bibitem[{\citenamefont{Porter}(1965)}]{Por65}
\bibinfo{author}{\bibnamefont{Porter}, \bibfnamefont{C.~E.}},
  \bibinfo{year}{1965}, \emph{\bibinfo{title}{Statistical Theories of Spectra:
  Fluctuations}} (\bibinfo{publisher}{Academic Press}, \bibinfo{address}{New
  York}).

\bibitem[{\citenamefont{Primack} \emph{et~al.}(1997)}]{PriSchSmi97}
\bibinfo{author}{\bibnamefont{Primack}, \bibfnamefont{H.}},
  \bibinfo{author}{\bibfnamefont{H.}~\bibnamefont{Schanz}}, and
  \bibinfo{author}{\bibfnamefont{U.}~\bibnamefont{Smilansky}}, and
  \bibinfo{author}{\bibfnamefont{I.}~\bibnamefont{Ussishkin}},
  \bibinfo{year}{1997}, \bibinfo{journal}{J. Phys. A: Math. Gen.}
  \textbf{\bibinfo{volume}{30}}, \bibinfo{pages}{6693}.



\bibitem[{\citenamefont{Protter}(1987)}]{Pro}
\bibinfo{author}{\bibnamefont{Protter}, \bibfnamefont{M.~H.}},
  \bibinfo{year}{1987}, \bibinfo{journal}{SIAM Review}
  \textbf{\bibinfo{volume}{29}}(\bibinfo{number}{2}), \bibinfo{pages}{185}.

\bibitem[{\citenamefont{Rauch}(1978)}]{Rau}
\bibinfo{author}{\bibnamefont{Rauch}, \bibfnamefont{J.}}, \bibinfo{year}{1978},
  \bibinfo{journal}{Amer. Math. Monthly} \textbf{\bibinfo{volume}{85}},
  \bibinfo{pages}{359}.



\bibitem[{\citenamefont{Richens and Berry}(1981)}]{BerRic81}
\bibinfo{author}{\bibnamefont{Richens}, \bibfnamefont{P.~J.}}, and
  \bibinfo{author}{\bibfnamefont{M.~V.} \bibnamefont{Berry}},
  \bibinfo{year}{1981}, \bibinfo{journal}{Physica D}
  \textbf{\bibinfo{volume}{2}}, \bibinfo{pages}{495}.


\bibitem[{\citenamefont{Riddel}(1979)}]{Rid79}
\bibinfo{author}{\bibnamefont{Riddel}, \bibfnamefont{R.~J.}},
  \bibinfo{year}{1979}, \bibinfo{journal}{J. Comput. Phys.}
  \textbf{\bibinfo{volume}{31}}, \bibinfo{pages}{21}.



\bibitem[{\citenamefont{Roth}(1984)}]{Roth}
\bibinfo{author}{\bibnamefont{Roth}, \bibfnamefont{J.~P.}},
  \bibinfo{year}{1984}, \bibinfo{journal}{Lect. Notes Math.}
  \textbf{\bibinfo{volume}{1096}}, \bibinfo{pages}{521}.



\bibitem[{\citenamefont{Schiemann}(1990)}]{Schiemann}
\bibinfo{author}{\bibnamefont{Schiemann}, \bibfnamefont{A.}},
  \bibinfo{year}{1990}, \bibinfo{journal}{Arch. Math.}
  \textbf{\bibinfo{volume}{54}}, \bibinfo{pages}{372}.

\bibitem[{\citenamefont{Scott}(1983)}]{Bull}
\bibinfo{author}{\bibnamefont{Scott}, \bibfnamefont{G.~P.}},
  \bibinfo{year}{1983}, \bibinfo{journal}{Bull. London Math. Soc.}
  \textbf{\bibinfo{volume}{15}}, \bibinfo{pages}{401}.

\bibitem[{\citenamefont{Segre}(1961)}]{Segre}
\bibinfo{author}{\bibnamefont{Segre}, \bibfnamefont{B.}},
  \bibinfo{year}{1961}, \emph{\bibinfo{title}{Lectures on Modern Geometry. With
  an appendix by Lucio Lombardo-Radice}}, volume~\bibinfo{volume}{7}
  (\bibinfo{publisher}{Edizioni Cremonese}, \bibinfo{address}{Rome}).


\bibitem[{\citenamefont{Sleeman and Hua}(2000)}]{SleHua}
\bibinfo{author}{\bibnamefont{Sleeman}, \bibfnamefont{B.~D.}}, and
  \bibinfo{author}{\bibfnamefont{C.}~\bibnamefont{Hua}}, \bibinfo{year}{2000},
  \bibinfo{journal}{Rev. Mat. Iberoam.} \textbf{\bibinfo{volume}{16}},
  \bibinfo{pages}{351}.

\bibitem[{\citenamefont{Smilansky and St\"ockmann}(2007)}]{SmiSto}
\bibinfo{author}{\bibnamefont{Smilansky}, \bibfnamefont{U.}}, and
  \bibinfo{author}{\bibfnamefont{H.-J.}~\bibnamefont{St\"ockmann}},
  \bibinfo{year}{2007}, \bibinfo{journal}{Eur. Phys. J. Special Topics}
  \textbf{\bibinfo{volume}{146}}, \bibinfo{pages}{341}.


\bibitem[{\citenamefont{Smithies}(1962)}]{Smi}
\bibinfo{author}{\bibnamefont{Smithies}, \bibfnamefont{F.}},
  \bibinfo{year}{1962}, \emph{\bibinfo{title}{Integral Equations}},
  volume~\bibinfo{volume}{49} (\bibinfo{publisher}{Cambridge University Press},
  \bibinfo{address}{Cambridge}).

\bibitem[{\citenamefont{Solomon}(2001)}]{Sol}
\bibinfo{author}{\bibnamefont{Solomon}, \bibfnamefont{R.}},
  \bibinfo{year}{2001}, \bibinfo{journal}{Bull. Amer. Math. Soc.}
  \textbf{\bibinfo{volume}{38}}, \bibinfo{pages}{315}.

\bibitem[{\citenamefont{Sridhar}(1991)}]{Sri}
\bibinfo{author}{\bibnamefont{Sridhar}, \bibfnamefont{S.}},
  \bibinfo{year}{1991}, \bibinfo{journal}{Phys. Rev. Lett.}
  \textbf{\bibinfo{volume}{67}}, \bibinfo{pages}{785}.

\bibitem[{\citenamefont{Sridhar and Heller}(1992)}]{SriHel}
\bibinfo{author}{\bibnamefont{Sridhar}, \bibfnamefont{S.}}, and
  \bibinfo{author}{\bibfnamefont{E.~J.} \bibnamefont{Heller}},
  \bibinfo{year}{1992}, \bibinfo{journal}{Phys. Rev. A}
  \textbf{\bibinfo{volume}{46}}, \bibinfo{pages}{R1728}.

\bibitem[{\citenamefont{Sridhar} \emph{et~al.}(1992)\citenamefont{Sridhar,
  Hogenboom, and Willemsen}}]{SriHogWil}
\bibinfo{author}{\bibnamefont{Sridhar}, \bibfnamefont{S.}},
  \bibinfo{author}{\bibfnamefont{D.~O.} \bibnamefont{Hogenboom}}, and
  \bibinfo{author}{\bibfnamefont{B.~A.} \bibnamefont{Willemsen}},
  \bibinfo{year}{1992}, \bibinfo{journal}{J. Stat. Phys.}
  \textbf{\bibinfo{volume}{68}}, \bibinfo{pages}{239}.

\bibitem[{\citenamefont{Sridhar and Kudrolli}(1994)}]{SriKud}
\bibinfo{author}{\bibnamefont{Sridhar}, \bibfnamefont{S.}}, and
  \bibinfo{author}{\bibfnamefont{A.}~\bibnamefont{Kudrolli}},
  \bibinfo{year}{1994}, \bibinfo{journal}{Phys. Rev. Lett.}
  \textbf{\bibinfo{volume}{72}}(\bibinfo{number}{14}), \bibinfo{pages}{2175}.

\bibitem[{\citenamefont{Stewardson and Waechter}(1971)}]{SteWae}
\bibinfo{author}{\bibnamefont{Stewardson}, \bibfnamefont{K.}}, and
  \bibinfo{author}{\bibfnamefont{R.~T.} \bibnamefont{Waechter}},
  \bibinfo{year}{1971}, \bibinfo{journal}{Proc. Cambridge
  Philos. Soc.} \textbf{\bibinfo{volume}{69}}, \bibinfo{pages}{353}.

\bibitem[{\citenamefont{Sunada}(1985)}]{Su}
\bibinfo{author}{\bibnamefont{Sunada}, \bibfnamefont{T.}},
  \bibinfo{year}{1985}, \bibinfo{journal}{Ann. Math.}
  \textbf{\bibinfo{volume}{121}}, \bibinfo{pages}{169}.

\bibitem[{\citenamefont{Tasaki} \emph{et~al.}(1997)\citenamefont{Tasaki,
  Harayama, and Shudo}}]{TasHarShu}
\bibinfo{author}{\bibnamefont{Tasaki}, \bibfnamefont{S.}},
  \bibinfo{author}{\bibfnamefont{T.}~\bibnamefont{Harayama}}, and
  \bibinfo{author}{\bibfnamefont{A.}~\bibnamefont{Shudo}},
  \bibinfo{year}{1997}, \bibinfo{journal}{Phys. Rev. E}
  \textbf{\bibinfo{volume}{56}}(\bibinfo{number}{1}), \bibinfo{pages}{R13}.

\bibitem[{\citenamefont{Thain}(2004)}]{Tha}
\bibinfo{author}{\bibnamefont{Thain}, \bibfnamefont{A.}}, \bibinfo{year}{2004},
  \bibinfo{journal}{Eur. J. Phys.} \textbf{\bibinfo{volume}{25}},
  \bibinfo{pages}{633}.

\bibitem[{\citenamefont{Thas} \emph{et~al.}(2006)\citenamefont{Thas, Thas, and
  {Van Maldeghem}}}]{TGQ}
\bibinfo{author}{\bibnamefont{Thas}, \bibfnamefont{J.~A.}},
  \bibinfo{author}{\bibfnamefont{K.}~\bibnamefont{Thas}}, and
  \bibinfo{author}{\bibfnamefont{H.}~\bibnamefont{{Van Maldeghem}}},
  \bibinfo{year}{2006}, \emph{\bibinfo{title}{Translation Generalized
  Quadrangles}}, volume~\bibinfo{volume}{26} (\bibinfo{publisher}{World
  Scientific}, \bibinfo{address}{Singapore}).

\bibitem[{\citenamefont{Thas}(2004)}]{SFGQ}
\bibinfo{author}{\bibnamefont{Thas}, \bibfnamefont{K.}}, \bibinfo{year}{2004},
  \emph{\bibinfo{title}{Symmetry in Finite Generalized Quadrangles}},
  volume~\bibinfo{volume}{1} (\bibinfo{publisher}{Birkh\"{a}user, Boston}).

\bibitem[{\citenamefont{Thas}(2006{\natexlab{a}})}]{KT}
\bibinfo{author}{\bibnamefont{Thas}, \bibfnamefont{K.}},
  \bibinfo{year}{2006}{\natexlab{a}}, \bibinfo{journal}{J. Phys. A: Math. Gen.}
  \textbf{\bibinfo{volume}{39}}, \bibinfo{pages}{L385}.

\bibitem[{\citenamefont{Thas}(2006{\natexlab{b}})}]{KT2}
\bibinfo{author}{\bibnamefont{Thas}, \bibfnamefont{K.}},
  \bibinfo{year}{2006}{\natexlab{b}}, \bibinfo{journal}{J. Phys. A: Math. Gen.}
  \textbf{\bibinfo{volume}{39}}, \bibinfo{pages}{13237}.


\bibitem[{\citenamefont{Thas}(2007{\natexlab{a}})}]{KT5}
\bibinfo{author}{\bibnamefont{Thas}, \bibfnamefont{K.}},
  \bibinfo{year}{2007}{\natexlab{a}},  \bibinfo{journal}{Inverse problems}
  \textbf{\bibinfo{volume}{23}}, \bibinfo{pages}{2021}.


\bibitem[{\citenamefont{Thas}(2007{\natexlab{b}})}]{KT3}
\bibinfo{author}{\bibnamefont{Thas}, \bibfnamefont{K.}},
  \bibinfo{year}{2007}{\natexlab{b}}, \bibinfo{journal}{J. Phys. A: Math.
  Theor.} \textbf{\bibinfo{volume}{40}}, \bibinfo{pages}{7233}.


\bibitem[{\citenamefont{Thas}(2009)}]{KoeBN1}
\bibinfo{author}{\bibnamefont{Thas}, \bibfnamefont{K.}},
  \bibinfo{year}{2009}, \bibinfo{journal}{Innov. Incidence Geom.}
 \textbf{\bibinfo{volume}{9}}, \bibinfo{pages}{189}.



\bibitem[{\citenamefont{Thas}(2010)}]{KoeBN2}
\bibinfo{author}{\bibnamefont{Thas}, \bibfnamefont{K.}},
  \bibinfo{year}{2010}, \bibinfo{note}{unpublished}.


\bibitem[{\citenamefont{Thas and {Van Maldeghem}}(2008)}]{KoeBN4}
\bibinfo{author}{\bibnamefont{Thas}, \bibfnamefont{K.}}, and
  \bibinfo{author}{\bibfnamefont{H.}~\bibnamefont{{Van Maldeghem}}},
  \bibinfo{year}{2008},  \bibinfo{journal}{Trans. Amer. Math. Soc.}
 \textbf{\bibinfo{volume}{360}}, \bibinfo{pages}{2327}.


\bibitem[{\citenamefont{Tits}(1959)}]{Ti}
\bibinfo{author}{\bibnamefont{Tits}, \bibfnamefont{J.}}, \bibinfo{year}{1959},
  \bibinfo{journal}{Inst. Hautes Etudes Sci. Publ. Math.}
  \textbf{\bibinfo{volume}{2}}, \bibinfo{pages}{13}.

\bibitem[{\citenamefont{Tits}(1974)}]{Tit:74}
\bibinfo{author}{\bibnamefont{Tits}, \bibfnamefont{J.}}, \bibinfo{year}{1974},
  \emph{\bibinfo{title}{Buildings of Spherical Type and Finite {BN}-Pairs}},
  volume \bibinfo{volume}{386} (\bibinfo{publisher}{Springer},
  \bibinfo{address}{Berlin}).

\bibitem[{\citenamefont{Urakawa}(1982)}]{Ura}
\bibinfo{author}{\bibnamefont{Urakawa}, \bibfnamefont{H.}},
  \bibinfo{year}{1982}, \bibinfo{journal}{Ann. Sci. Ecole Norm. Sup.}
  \textbf{\bibinfo{volume}{15}}, \bibinfo{pages}{441}.

\bibitem[{\citenamefont{Vaa} \emph{et~al.}(2005)\citenamefont{Vaa,
  Koch, and Bl\"umel}}]{VaaKocBlu}
\bibinfo{author}{\bibnamefont{Vaa}, \bibfnamefont{C.}}, and
  \bibinfo{author}{\bibfnamefont{P.~M.} \bibnamefont{Koch}}, and
  \bibinfo{author}{\bibfnamefont{R.} \bibnamefont{Bl\"umel}},
  \bibinfo{year}{2005}, \bibinfo{journal}{Phys. Rev. E}
  \textbf{\bibinfo{volume}{72}}, \bibinfo{pages}{056211}.

\bibitem[{\citenamefont{van Dam and Haemers}(2003)}]{VDH}
\bibinfo{author}{\bibnamefont{van Dam}, \bibfnamefont{E.~R.}}, and
  \bibinfo{author}{\bibfnamefont{W.~H.} \bibnamefont{Haemers}},
  \bibinfo{year}{2003}, \bibinfo{journal}{Linear Algebra Appl.}
  \textbf{\bibinfo{volume}{373}}, \bibinfo{pages}{241}.


\bibitem[{\citenamefont{{Van Maldeghem}}(1998)}]{HVM}
\bibinfo{author}{\bibnamefont{{Van Maldeghem}}, \bibfnamefont{H.}},
  \bibinfo{year}{1998}, \emph{\bibinfo{title}{Generalized Polygons}},
  volume~\bibinfo{volume}{93} (\bibinfo{publisher}{Birkh\"{a}user-Verlag},
  \bibinfo{address}{Basel}).


\bibitem[{\citenamefont{{Van Vleck}}(1928)}]{VVl}
\bibinfo{author}{\bibnamefont{{Van  Vleck}}, \bibfnamefont{J.H.}},
    \bibinfo{year}{1928}, \bibinfo{journal}{Proc. Natl. Acad. Sci. USA}
  \textbf{\bibinfo{volume}{14}}, \bibinfo{pages}{178}.



\bibitem[{\citenamefont{Vattay} \emph{et~al.}(1994)\citenamefont{Vattay,
  Wirzba, and Rosenqvist}}]{VatWirRos94}
\bibinfo{author}{\bibnamefont{Vattay}, \bibfnamefont{G.}},
  \bibinfo{author}{\bibfnamefont{A.}~\bibnamefont{Wirzba}}, and
  \bibinfo{author}{\bibfnamefont{P.E.}~\bibnamefont{Rosenqvist}},
  \bibinfo{year}{1994}, \bibinfo{journal}{Phys. Rev. Lett.}
  \textbf{\bibinfo{volume}{73}}(\bibinfo{number}{17}), \bibinfo{pages}{2304}.

\bibitem[{\citenamefont{Vign{\'e}ras}(1980)}]{Vig}
\bibinfo{author}{\bibnamefont{Vign{\'e}ras}, \bibfnamefont{M.~F.}},
  \bibinfo{year}{1980}, \bibinfo{journal}{Ann. Math.}
  \textbf{\bibinfo{volume}{112}}, \bibinfo{pages}{21}.

\bibitem[{\citenamefont{von Below}(2001)}]{vonBel}
\bibinfo{author}{\bibnamefont{von Below}, \bibfnamefont{J.}},
  \bibinfo{year}{2001}, \bibinfo{journal}{Lecture
Notes in Pure and Appl. Math.} \textbf{\bibinfo{volume}{219}}, \bibinfo{pages}{19}.


\bibitem[{\citenamefont{Vorobets}(1996)}]{Vor96}
\bibinfo{author}{\bibnamefont{Vorobets}, \bibfnamefont{Y.~B.}},
  \bibinfo{year}{1996}, \bibinfo{journal}{Russ. Math. Surveys}
  \textbf{\bibinfo{volume}{51}} (\bibinfo{number}{5}), \bibinfo{pages}{779}.



\bibitem[{\citenamefont{Vorobets and Stepin}(1998)}]{VoroStep}
\bibinfo{author}{\bibnamefont{Vorobets}, \bibfnamefont{Y.~B.}}, and
\bibinfo{author}{\bibnamefont{Stepin}, \bibfnamefont{A.~M.}},
  \bibinfo{year}{1998}, \bibinfo{journal}{Russian Mat. Zametki}
  \textbf{\bibinfo{volume}{63}}(\bibinfo{number}{5}), \bibinfo{pages}{660};
 translation in  \bibinfo{year}{1998}, \bibinfo{journal}{Math. Notes} 
\textbf{\bibinfo{volume}{63}} (\bibinfo{number}{5-6}), \bibinfo{pages}{582}.



\bibitem[{\citenamefont{Witt}(1941)}]{witt}
\bibinfo{author}{\bibnamefont{Witt}, \bibfnamefont{E.}}, \bibinfo{year}{1941},
  \bibinfo{journal}{Abh. Math. Semin. Univ. Hamburg}
  \textbf{\bibinfo{volume}{14}}, \bibinfo{pages}{323}.

\bibitem[{\citenamefont{Wolpert}(1978)}]{Wolp}
\bibinfo{author}{\bibnamefont{Wolpert}, \bibfnamefont{S.}},
  \bibinfo{year}{1978}, \bibinfo{journal}{Trans. Amer. Math. Soc.}
  \textbf{\bibinfo{volume}{244}}, \bibinfo{pages}{313}.

\bibitem[{\citenamefont{Wu} \emph{et~al.}(1995)\citenamefont{Wu, Sprung, and
  Martorell}}]{WuSprMar}
\bibinfo{author}{\bibnamefont{Wu}, \bibfnamefont{H.}},
  \bibinfo{author}{\bibfnamefont{D.~W.~L.} \bibnamefont{Sprung}}, and
  \bibinfo{author}{\bibfnamefont{J.}~\bibnamefont{Martorell}},
  \bibinfo{year}{1995}, \bibinfo{journal}{Phys. Rev. E}
  \textbf{\bibinfo{volume}{51}}(\bibinfo{number}{1}), \bibinfo{pages}{703}.

\bibitem[{\citenamefont{Zaremba}(1909)}]{Zar09}
\bibinfo{author}{\bibnamefont{Zaremba}, \bibfnamefont{S.}},
  \bibinfo{year}{1909}, \bibinfo{journal}{Bulletin international de
  l'Acad\'emie des sciences de Cracovie}, \bibinfo{note}{Feb.}, \bibinfo{pages}{125}.

\bibitem[{\citenamefont{Zelditch}(1998)}]{Zeld98}
\bibinfo{author}{\bibnamefont{Zelditch}, \bibfnamefont{S.}},
  \bibinfo{year}{1998}, \bibinfo{journal}{J. Diff. Geom.}
  \textbf{\bibinfo{volume}{49}}, \bibinfo{pages}{207}.

\bibitem[{\citenamefont{Zelditch}(1999)}]{Zeld99}
\bibinfo{author}{\bibnamefont{Zelditch}, \bibfnamefont{S.}},
  \bibinfo{year}{1999}, \bibinfo{journal}{Math. Res. Lett.}
  \textbf{\bibinfo{volume}{6}}, \bibinfo{pages}{457}.

\bibitem[{\citenamefont{Zelditch}(2000)}]{Zeld00}
\bibinfo{author}{\bibnamefont{Zelditch}, \bibfnamefont{S.}},
  \bibinfo{year}{2000}, \bibinfo{journal}{Geom. Funct. Anal.}
  \textbf{\bibinfo{volume}{10}}, \bibinfo{pages}{628}.

\bibitem[{\citenamefont{Zelditch}(2004{\natexlab{a}})}]{Zeld04}
\bibinfo{author}{\bibnamefont{Zelditch}, \bibfnamefont{S.}},
  \bibinfo{year}{2004}{\natexlab{a}}, \bibinfo{journal}{Comm. Math. Phys.}
  \textbf{\bibinfo{volume}{248}}, \bibinfo{pages}{357}.

\bibitem[{\citenamefont{Zelditch}(2004{\natexlab{b}})}]{Zeld}
\bibinfo{author}{\bibnamefont{Zelditch}, \bibfnamefont{S.}},
  \bibinfo{year}{2004}{\natexlab{b}}, \bibinfo{journal}{Surv. Differ. Geom.}
  \textbf{\bibinfo{volume}{IX}}, \bibinfo{pages}{401}.


\bibitem[{\citenamefont{Zelditch}(2004{\natexlab{c}})}]{Zeld04a}
\bibinfo{author}{\bibnamefont{Zelditch}, \bibfnamefont{S.}}, \bibinfo{year}{2004},
 in \emph{\bibinfo{title}{Geometric Methods in Inverse Problems and PDE Control}},
  (\bibinfo{publisher}{C. B. Croke, I . Lasiecka, G. Uhlmann, and
    M. S. Vogelius, eds.}), \bibinfo{journal}{IMA Vol. Math. Appl.}
  \textbf{\bibinfo{volume}{137}}, \bibinfo{address}{Springer-Verlag, New
    York}, \bibinfo{pages}{289-321}.



\bibitem[{\citenamefont{Zelditch}(2009)}]{Zeld09}
\bibinfo{author}{\bibnamefont{Zelditch}, \bibfnamefont{S.}},
  \bibinfo{year}{2009}, \bibinfo{journal}{Ann. Math.}
  \textbf{\bibinfo{volume}{170}}, \bibinfo{pages}{205}.


\bibitem[{\citenamefont{Zemlyakov and Katok}(1976)}]{ZemKat76}
\bibinfo{author}{\bibnamefont{Zemlyakov}, \bibfnamefont{A.~B.}}, and
  \bibinfo{author}{\bibnamefont{A.~N.} \bibnamefont{Katok}},
  \bibinfo{year}{1976}, \bibinfo{journal}{Math. Notes}
  \textbf{\bibinfo{volume}{18}}, \bibinfo{pages}{760}.

\end{thebibliography}

\end{document}